\documentclass[12pt]{iopart}

\usepackage{graphicx}
\usepackage{dcolumn}
\usepackage{bm}
\usepackage{color}
\usepackage{epstopdf}

\usepackage{amsmath}
\usepackage{amsfonts}
\usepackage{amssymb}
\usepackage{mathrsfs}
\usepackage{amsthm}
\usepackage{bbold}

\usepackage{subcaption}
\usepackage[labelfont=bf]{caption}
\usepackage{float}
\usepackage{upgreek}

\usepackage{hyperref}

\newtheorem{theorem}{Theorem}

\newcommand{\boldeta   }{\mbox{\boldmath$\eta$}}

\begin{document}

\title[Local sign stability and its implications for spectra of sparse random graphs]{Local sign stability and its implications for spectra of sparse random graphs and stability of ecosystems}

\author{Pietro Valigi}
\address{Department of Physics, Sapienza University of Rome, P.le Aldo Moro 5, Rome, Italy}
\ead{pietro.valigi@uniroma1.it}

\author{Izaak Neri}
\address{Department of Mathematics, King's College London, Strand, London, WC2R 2LS, UK}
\ead{izaak.neri@kcl.ac.uk} 

\author{Chiara Cammarota}
\address{Department of Physics, Sapienza University of Rome, P.le Aldo Moro 5, Rome, Italy}
\ead{chiara.cammarota@uniroma1.it}

\begin{abstract} 
We study the spectral properties of sparse random graphs with different topologies and type of interactions, and their implications on the stability of complex systems, with particular attention to ecosystems.  
Specifically, we focus on the behaviour of the leading eigenvalue in different type of random matrices (including interaction matrices and Jacobian-like matrices), relevant for the assessment of different types of dynamical stability.
By comparing the results on Erd\H{o}s-R\'{e}nyi and Husimi graphs with sign-antisymmetric interactions or mixed sign patterns, we introduce a sufficient criterion, called {\it strong local sign stability}, for stability not to be affected by system size, as traditionally implied by the complexity-stability trade-off in conventional models of random matrices. 
%This criterion generally holds for  graphs with local tree-like structures and antagonistic or unidirectional interactions.
The criterion requires sign-antisymmetric or unidirectional interactions and a local structure of the graph such that the number of cycles of finite length do not increase with the system size. Note that the last requirement is stronger than the classical local tree-like condition, which we associate to the less stringent definition of {\it local sign stability}, also defined in the paper.
In addition, for strong local sign stable graphs which show stability to linear perturbations irrespectively of system size, we observe that the leading eigenvalue can undergo a transition from being real to acquiring a nonnull imaginary part, which implies a dynamical transition from nonoscillatory to oscillatory linear response to perturbations. 
Lastly, we ascertain the discontinuous nature of this transition.
\end{abstract}
\maketitle

\section{Introduction}
Understanding the stability of dynamical systems is a fundamental question in various fields of science, ranging from ecology \cite{may1973stability, moore2012energetic} and economics \cite{quirk1965qualitative, maybe1969qualitative} to neuroscience \cite{sporns2016networks} and chemistry \cite{clarke1988stoichiometric, angeli2009tutorial}. In many cases, the stability analysis of a dynamical system can be reduced to a spectral problem involving a matrix, as discussed in Refs.~\cite{cesari2012asymptotic, hahn1967stability}.  Therefore, there  has been significant  interest in understanding how the statistical properties of matrix elements impact the spectral properties of the matrix, which in turn can shed light on the stability of the underlying dynamical system.

As early as the 1970s, using random matrices May has studied, for instance, the stability of fully connected ecosystems \cite{may1972will}.   Although this fueled significant interest \cite{Allesina2015}, it is only   recently that the  influence of  sparse network structure on dynamical stability has been studied.  Indeed, following pioneering work on the spectra of symmetric Erd\H{o}s-R\'{e}nyi graphs \cite{RodgBray1988, Reimer_sparse, cavity, susca2021cavity}, recent papers studied the spectra of random, directed graphs \cite{rogers2009cavity, metz2011spectra, bolle2013spectra, neri2016eigenvalue, Izaak_Metz_2019, neri2019spectral,  tarnowski2020universal, metz2020localization, tarnowski2022real} and the spectra of random graphs with predator-prey, mutualistic, or competitive interactions \cite{mambuca2022dynamical}.  

A surprising  finding of these more recent works  is that the spectra of sparse random graphs are   strongly affected by the sign patterns of their matrix entries, i.e., whether ${\rm sign}(A_{ij}A_{ji})=0$ (unidirectional interactions), ${\rm sign}(A_{ij}A_{ji})=-1$ (sign-antisymmetric interactions), or ${\rm sign}(A_{ij}A_{ji})=1$ (sign-symmetric interactions).  Notably,  the spectra of (infinitely large) sparse random graphs  are confined to a region in the complex plane with bounded real part when  ${\rm sign}(A_{ij}A_{ji})\in \left\{0,-1\right\}$ for all pairs $i,j$ of nodes~\cite{neri2019spectral, mambuca2022dynamical}, whereas the spectra of  (infinitely large) sparse random graphs encompass the full real axis when there exists a finite proportion of links with ${\rm sign}(A_{ij}A_{ji})= 1$~\cite{mambuca2022dynamical}.

 In the present paper, we provide a simple, sufficient criterion for the finiteness of the real part of the leading eigenvalue, which is the eigenvalue with the largest real part, of an infinitely large, sparse, random matrix. 
 To this aim, we rely on the concept of {\it sign stability}.

 Sign stability appeared first in studies on qualitative economics \cite{quirk1965qualitative, maybe1969qualitative} in the 1960s when economists were studying the impact of   qualitative properties of interaction matrices on the stability of economic systems, such as, the sign of the elements in the interaction matrices.    The importance of sign stability was soon realised for ecology \cite{may1973qualitative,jeffries1974qualitative, levins1975problems, solimano1982graph, logofet1982sign} and later it was also considered in chemistry \cite{clarke1975theorems}.  
 A matrix is sign stable if for any choice of the absolute values of the nonzero elements its leading eigenvalue is negative.
 %, and therefore bounded from above.
 In order for a matrix to be sign stable, it must  satisfy specific constraints on its topology and sign pattern \cite{quirk1965qualitative, jeffries1977matrix}. 
 Interestingly, tree graphs admit sign stable structures, for instance,  directed tree graphs and antagonistic tree graphs are sign stable.   On the other hand, in general, if cycles are present 
 the sign stability property may be lost.

 As sparse random graphs, e.g., sparse Erd\H{o}s-R\'{e}nyi graphs, contain cycles, they are not sign stable. However, Erd\H{o}s-R\'{e}nyi graphs are locally tree-like~\cite{mezard2001bethe, dembo2010gibbs, montanari2012weak}, and as tree graphs admit sign stable structures, we say that Erd\H{o}s-R\'{e}nyi graphs are {\it locally sign stable} if the signs of their interaction patterns correspond those of sign stable trees. A formal definition of local sign stability will be given in Sec.~\ref{sec:localsignstability} and will apply to a broader class of sparse random graphs. 
 %(see Sec.~\ref{sec:localsignstability} for a broader and formal definition of local sign stability). 
   We then 
   propose a stronger version of local sign stability, called in the following {\it strong local sign stability}, as sufficient condition for the finiteness of the real part of the leading eigenvalue.   
 Hence, we argue that strong local sign stability  allows us to predict the
  stability of large, sparse network structures and extends sign stability to sparse random graphs.

One important aspect of sign stability, which also applies under mild conditions, as discussed later, to (strong) local sign stability, is that it refers to all matrices with the same topology and sign pattern, independently from the absolute value of their nonzero elements.  
Sign stability is therefore a particularly robust type of stability, as it characterizes an infinitely large set of matrices.
This aspect  is particularly relevant in ecological applications for at least two reasons.  First, 
   most of the time, as discussed in the following section, to determine whether $\mathbf{M}$ is stable
  we need to determine its leading eigenvalue, which  requires knowledge of  the matrix entries $M_{ij}$.
     Unfortunately, in applications it is often the case that only partial information about the matrix $\mathbf{M}$ is available.      For example, in the context of ecology, it is relatively easy to determine both the    foodweb of the  trophic interactions between species, i.e.,  whether $M_{ij}=0$ or $M_{ij}\neq 0$,  and the type of the interactions, inter alia,  predator-prey (corresponding to sign-antisymmetric), mutualistic or competitive (corresponding to sign symmetric, respectively positive or negative) interactions.    On the other hand, it is significantly more difficult   to determine the strengths  $|M_{ij}|$ of the  trophic interactions between species~\cite{moore2012energetic, jacquet2016no}. 
 This raises the  question  whether stability can be determined from the sign pattern of the entries of the matrix~$\mathbf{M}$.
Second, different kinds of ecosystem stability (linear stability, structural stability, feasibility) can be studied by looking at the properties of different matrices obtained from the matrix~$\mathbf{A}$ of inter-species interactions without sign or topological alterations. 
In these cases, as explained in more details in the following section, if the topological properties and the sign pattern of the interaction network grant its strong local sign stability, the ecosystem can be declared at once feasible and stable, both with respect to small fluctuations and small changes in the external conditions.

A second interesting problem for sparse random graphs that we address is whether the leading eigenvalue is real-valued or whether it has a nonzero imaginary part. As it will be recalled, the presence of  pairs of conjugate complex leading eigenvalues has important consequences on the dynamical behaviour of the models associated as it gives rise to oscillatory dynamics in the vicinity of the fixed
point with frequency of oscillations inversely proportional to the absolute value of the imaginary part of the leading eigenvalue.  This aspect is especially relevant for strongly locally sign stable random graphs, as their leading eigenvalue is finite.
In particular for these cases, we will discuss how depending on the choice of model's parameter both situations can arise and we will describe the transition between the two corresponding dynamical phases.

The paper is structured as follows: we first review in Sec.~\ref{sec:stability_and_model} the problem of stability in ecology, including a general discussion on the dependence of stability on the system size, and we specify the random matrix models that we study in this paper. In Sec.~\ref{sec:sign_stability}, we review the properties of sign stability with particular attention to its occurrence in tree graphs, before presenting the main results of this paper in the Secs.~\ref{sec:localsignstability} and~\ref{sec:reentrances}.
In Sec.~\ref{sec:localsignstability}, we present the paper's main claim: we identify strong local sign stability as an important feature for the spectra of sparse random matrices and the stability of complex systems.
%, by determining the finiteness of the leading eigenvalue, %and we present .  Specifically, in this section we present the paper's main claim on strong local sign stability of sparse random graphs,
We also present several numerical results through direct diagonalisation of the models of sparse random matrices, as defined in Sec.~\ref{sec:stability_and_model}.
 In particular, we test the strong local sign stability criterion by showing that when its conditions hold the leading eigenvalue remains finite. We also show how significant the main conditions are by giving examples of ensembles that violate one of the conditions and show divergence of the leading eigenvalue.
In Sec.~\ref{sec:reentrances}, we determine the  imaginary part of the leading eigenvalue for  several type of sparse, random graphs, and in particular we identify a transition from a regime where the leading eigenvalue is real to a regime where the leading eigenvalue come in a pair of complex  eigenvalues.  We end the paper with a discussion in Sec.~\ref{sec:Discussion}, and a few Appendices with technical details.

\section{Stability in Ecology and Model Setup} \label{sec:stability_and_model}

In this section, we introduce the model setup of this paper.    In Secs.~\ref{Sec:StabGen} and \ref{sec:stabEco},  we review  the relation between, on one hand, the spectral properties of matrices and, on the other hand, the stability of linear dynamical systems and (nonlinear) ecosystems,  respectively. The reader not needing the basic mathematical background or not interested in the ecological applications can skip these two sections.   
In Sec.~\ref{sec:absolute-stability}, we review the concepts of absolute and size-dependent  stability, which play an important role in this paper.
Lastly, in Sec.~\ref{sec:RMTModels}, we define the random matrix models that we study in this paper, and in particular in  Sec.~\ref{sec:numerical_details}, we discuss the canonical model parameters that we use.    

\subsection{Absolute stability and size-dependent stability in linear dynamical systems}
\label{Sec:StabGen}
Let $\vec{x}(t)\in \mathbb{R}^N$,  with $t\geq0$ a time index, denote the evolution in time of the state of a system consisting of $N$ components. The simplest model for a dynamical system of $N$ interacting components is given by  a linear differential equation of the form 
\begin{equation}
 \frac{{\rm d}\vec{x}}{{\rm d}t} = \mathbf{M}\vec{x}, \label{eq:linear}
    \end{equation}
    where $\mathbf{M}\in \mathbb{R}^{N\times N}$  is an arbitrary matrix.

 The asymptotic state  $\vec{x}_{\infty} = \lim_{t\rightarrow \infty}\|\vec{x}(t)\|$ is determined by the eigenvalues  $\lambda_i(\mathbf{M})$ of the matrix $\mathbf{M}$.   When all  the $\lambda_i(\mathbf{M})$  have  negative real parts, then $\vec{x}_{\infty} =0$~\cite{cesari2012asymptotic},  and  we say that the matrix $\mathbf{M}$ is {\it stable}.  
 On the other hand, when there exists at least one eigenvalue with a positive real part, then $\vec{x}_{\infty}$ does not exist, as the norm of  $\|\vec{x}(t)\|$ diverges for large $t$, and we say that $\mathbf{M}$ is {\it unstable}.     If we order the eigenvalues such that  $\Re[\lambda_1] \geq \Re[\lambda_2]  \geq \ldots \geq \Re[\lambda_N]$, where $\Re(\cdot)$ denotes the real part of a complex number,  then $\mathbf{M}$ is stable if 
 \begin{equation}
     \Re[\lambda_1] <0.  \label{eq:stabCond}
 \end{equation}
In the intermediate regime for which 
 \begin{equation}
     \Re[\lambda_1] =0, 
 \end{equation}
 we say that the matrix is {\it marginally stable}.    Note that for marginally stable systems   $\|\vec{x}(t)\|$ may still diverge as a function of $t$ if the matrix $\mathbf{M}$ has degenerate eigenvalues~\cite{arnold1992ordinary, hirsch2012differential}.   
 
 Additionally, the transient dynamical behaviour after external perturbations is revealed by the imaginary part of the leading eigenvalue $\lambda_1({\bf M})$, which we denote by $\Im[\lambda_1({\bf M})]$. %, where $\Im(\cdot)$ denotes the imaginary part of a complex number.
 %  Additionally, the transient dynamical behaviour after external perturbations is revealed by the imaginary part of the leading eigenvalue $\Im[\lambda_1({\bf M})]$, where $\Im(\cdot)$ denotes the imaginary part of a complex number.
If $\Im[\lambda_1({\bf M})]=0$, then the transient is nonoscillatory, whereas a nonzero imaginary part implies oscillatory behaviour of $\vec{x}(t)$ in the vicinity of the origin.
The absolute value $|\Im[\lambda_1({\bf M})]|$ determines the frequency of oscillations of the slowest mode when the system is stable, and of the fastest unstable mode when the system is unstable.

Since in this paper we consider large complex systems,    following Ref.~\cite{mambuca2022dynamical}, we  introduce here two variants of linear stability in the limit of large $N$.   
Consider a sequence of matrices     $\mathbf{M}_N$  growing in size $N\in \mathbb{N}$.  In this case, we can distinguish two classes of matrix sequences, viz., those for which the real part of the leading eigenvalue converges to a  finite value, i.e., 
     \begin{equation}
        \lim_{N\rightarrow \infty} \Re[\lambda_1(\mathbf{M}_N)] \in \mathbb{R} ,\label{eq:potAbs1}
    \end{equation} 
    and those for which 
      \begin{equation}
        \lim_{N\rightarrow \infty} \Re[\lambda_1(\mathbf{M}_N)] = +\infty.  \label{eq:potAbs2}
    \end{equation} 
    
 Take as an example of the former the nondirected star graph $(\mathbf{M}_N)_{ij} = \delta_{i,1}(1-\delta_{j,1})+\delta_{j,1}(1-\delta_{i,1})$ (with $\lambda_1 = \sqrt{N-1}$, as text book calculation shows \cite{dorogovtsev2022nature})
and as an example of the latter, the directed star graph $(\mathbf{M}_N)_{ij} = \delta_{i,1}(1-\delta_{j,1})$ (with $\lambda_1=0$, which is  a simple linear algebra problem), where $\delta_{a,b}$ is the Kronecker delta function.     In the  former case, there exists a finite $d>0$ such that 
    \begin{equation}
        \lim_{N\rightarrow \infty} \Re[\lambda_1(  \mathbf{M}_N - d \mathbf{1}_N)] <0  , \label{eq:absStabPot}
    \end{equation}  
    where $\mathbf{1}_N$ stands for the identity matrix of size $N\times N$,
    and hence the sequence $\mathbf{M}_N - d \mathbf{1}_N$ is characterized by linear {\it absolute stability}.    In the latter case, any constant shift $d$ renders the ensemble stable up to a certain size $\overline{N}$, such that 
     \begin{equation}
         \Re[\lambda_1(  \mathbf{M}_N - d \mathbf{1}_{N})] <0  ,
    \end{equation}  
    for $N<\overline{N}$, while 
      \begin{equation}
         \Re[\lambda_1(  \mathbf{M}_N - d \mathbf{1}_{N})] > 0  ,
    \end{equation}  
    for $N>\overline{N}$, and we speak of {\it  size-dependent stability} of linear systems.  

Although linear systems simplify significantly the dynamics of complex systems, they can be insightful  for the study of complex systems, such as, ecological systems \cite{Allesina2015},  neural networks \cite{chaos, kadmon2015transition}, chemical interaction networks \cite{guo2021exploring}, and economic models \cite{moran2019may}, whenever the interest is to understand the  transient dynamics of  systems in the vicinity of a stable fixed point by linearizing the system of dynamical equations around the fixed point. We discuss the connection to more general nonlinear dynamics in more detail in the following section, where other types of stability properties for  dynamical systems are reviewed on an example of an ecological model and we will generalise absolute stability and size-dependent stability to those cases.

\subsection{Stability in Ecology}\label{sec:stabEco}
The possibility to predict and control the fate of ecosystems is of immediate concern for our lives, which strongly depend upon them. Therefore the concept of their stability in theoretical ecology has been investigated for decades leading to the classification of different types of stability of potential practical relevance.

We give a brief overview of the different notions of stability studied in the literature.  
Famously modelled by dynamical systems, ranging  from simple one- or two-species population evolution \cite{verhulst1845resherches, volterra1926fluctuations} to more recent studies about multi-species interactions   \cite{barbier2018generic}, ecosystem's stability with respect to small perturbations around putative fixed points has been  investigated  at length with {\it linear stability} analyses, \cite{may1972will, Allesina2015}, accompanied by considerations on their {\it global} or {\it local stability}. 
These approaches assume the existence of at least one equilibrium of ecological relevance, i.e., with nonnegative species' abundances. 
However it has been pointed out that the conditions for the existence of ecologically meaningful equilibria, called {\it feasibility}, are far from trivial \cite{rossberg2013food, dougoud2018feasibility}. 
Note that in feasible equilibria several species of the original pool can be extinct, and therefore do not enter in the final composition of the ecosystem.  Yet,  conditions for its {\it (un-)invadability}, corresponding with linear stability with respect to potential immigration of   species, must be explicitly looked at. 
A different kind of stability is represented by {\it structural stability}, which refers to the sensitivity of (feasible and linearly stable) equilibria (and of their stability and invadability) to changes in the ecological parameters.  The present paper discusses in the same  general context feasibility, structural and linear stability of model of ecosystems defined by several types of (sparse) interaction graphs.  

 The well-known complexity-stability trade-off in models of ecological systems generally affects their feasibility \cite{rossberg2013food, dougoud2018feasibility, roberts1974stability, meszena2006competitive}, linear stability \cite{may1972will, mccann2000diversity, mougi2012diversity, Allesina2015} and structural stability \cite{rossberg2013food, rossberg2017structural, rohr2014structural, biroli2018marginally, o2019metacommunity, lorenzana2022well}. 
Specifically, many models of ecosystems assembled from a pool of $S$ interacting species show a dependence of their stability properties on $S$. 
We will refer in short to this situation by calling it {\it size-dependent stability}.
Conversely, we define {\it absolute stability} the stability property of a system when it is not affected by its system size.
A formal definition of size-dependent stability in linear systems can be found in the previous section and its discussion in the general case is in Sec.~\ref{sec:absolute-stability}.  
Note that in many cases absolutely stable models can be trivially constructed from models with size-dependent stability by simply rescaling the inter-species interactions by $S$.
However, in absence of direct biological evidences of very weak inter-species interactions, such rescaling can appear unnatural and forced. Therefore in the following we will call absolutely stable only those models for which stationary points of the dynamics can be made available and can be stabilized without rescaling the interactions by the system size. 

The natural consequence of size-dependent stability is to severely constraint the possibility for these models to account for the emergence of a large biodiversity.
We will discuss how ecosystems whose interactions are structured according to sparse, locally tree-like, graphs, in some special cases, can benefit from absolute stability.
Such property will affect for different reasons all the different kinds of stability and feasibility, previously reviewed, allowing rich biodiversity to emerge.

To illustrate the mathematical implications of the different concepts of stability and the ways to address questions about them, 
we refer to a generalised Lotka-Volterra model where the $i$-th species' abundance $N_i\in \mathbb{R}^+$, with $i\in \left\{1,2,\ldots,S\right\}$, obeys  the following dynamical equation
\begin{equation}
    \frac{d N_i}{dt}=N_i\left[\left(r_i-\frac{N_i}{K_i}\right)-\sum^S_{j=1;j\neq i}\alpha_{ij}N_j\right]+\zeta=N_if_i(\{N_j\})+\zeta,
\end{equation}
where the   $\{\alpha_{ij}\}$ are the entries of the {\it interaction matrix} ${\bf A}$   and where $\zeta$ is the immigration rate (to be sent to zero before extracting the results, but useful to avoid considering ecosystems with trivial extinctions, and therefore to grant uninvadability).
The other parameters of the model, appearing in the first self-regulation term on the right hand side, are the growth rates in isolation $r_i$ and the carrying capacities $K_i/r_i$.
Depending on the choice of the graph structure, determined by the nonzero $\alpha_{ij}$ and of their sign and strength, different type of ecological models can be obtained and studied, from unstructured ecosystems to hierarchical food-webs, from predator-prey (sign-antisymmetric) types of interactions to mutualistic or competitive ones.

In this family of models, feasibility requires the existence of fixed points of the dynamics, therefore the existence of at least one non trivial meaningful solution $\vec{N}^*$, with elements $\{N_i^*\}$, to the set of equations $f_i(\{N_j\})=-\lim_{\zeta\rightarrow 0}\zeta/N_i$. 
All extinct species $i$ will have $N_i^*=0$ and $f_i(\{N_j^*\})<0$,
while surviving ones will be characterised by $N_i^*\neq0$ and $f_i(\{N_j^*\})=0$, therefore
\begin{equation}
    r_i=\frac{N_i^*}{K_i}+\sum^S_{j=1; j\neq i}\alpha_{ij}N_j^* \ .
    \label{eq:existence}
\end{equation}
The existence of a solution to the last equation is granted by the invertibility of the matrix 
${\bf B}$ with elements 
\begin{equation} 
    B_{ij}=\frac{\delta_{ij}}{K_i}+\alpha_{ij}
    \label{eq:Bij_def}
\end{equation}
 restricted to the $S^*$ surviving species:
\begin{equation}
    \vec{N}^*={\bf B}^{-1}\vec{r}\ .
\end{equation}
For ${\bf B}$ to be invertible, it is needed that none of its eigenvalue is null or, alternatively, that, in the infinitely large $S$ limit, the continuous part of the spectrum does not include the origin of the complex plane and none of the isolated eigenvalues is null.    

Feasibility also requires that all elements in $\vec{N}^*$ are non negative.
Explicit characterization of the probability to observe a feasible equilibrium are determined for ecosystems on dense unstructured graphs \cite{dougoud2018feasibility}, some special ecologically inspired structure of the graph \cite{rohr2014structural, grilli2017feasibility}, and can be studied numerically in some more contexts, but no result is known in general. 
Naturally, the requirement of having nonnegative abundances is more stringent than the condition for the existence of a solution to Eq.(\ref{eq:existence}), yet in the case previously studied the failing of the first condition closely anticipate the breaking of the second \cite{dougoud2018feasibility}.   Following this observation and in absence of a general rule able to asses full-fledged feasibility, we will consider the condition for the existence of a non trivial $\vec{N}^*$ as a good proxy for feasibility.

Interestingly, the matrix ${\bf B}$ is also directly relevant for structural stability, defined as the stability of the abundances of surviving species, $\vec{N}^*$, to small perturbations of the ecological parameters. In fact, as we show in \ref{app:structstab}, the susceptibility of $N_i^*$ to little variations $\xi_i$, $\eta_i$ and $\epsilon_{ij}$ of the three ecological parameters $r_i$, $K_i$ and $\alpha_{ij}$, respectively, is directly related to the inverse of ${\bf B}$:
\begin{align}
    r_i\rightarrow r_i+\xi_i \quad \Longrightarrow& \quad \frac{\partial N_i^*}{\partial \xi_j}=({\bf B}^{-1})_{ij} \ , \\
    K_i\rightarrow K_i+\eta_i \quad \Longrightarrow& \quad \frac{\partial N_i^*}{\partial \eta_k}\Bigg\rvert_{\vec{\eta}=0}=({\bf B}^{-1})_{ik} \frac{N_k^*}{K_k^2} \ , \\
    \alpha_{ij}\rightarrow \alpha_{ij}+\epsilon_{ij} \quad \Longrightarrow& \quad \frac{\partial N_i^*}{\partial \epsilon_{k\ell}}\Bigg\rvert_{\epsilon=0} = - ({\bf B}^{-1})_{ik} N_\ell^* \ .
\end{align}
Again, in all the three cases above, a singular behaviour emerges when the spectrum of ${\bf B}$ contains the origin of the complex plane hinting to a large susceptibility of the solution of $\vec{N}^*$ to ecological parameters.

Finally, the classical information on linear stability (stability with respect to dynamical fluctuations as induced by demographic noise, for instance), or Lyapunov stability concerning the domains of attraction of fixed point of the dynamics, is obtained by linearizing the system of dynamical equations around the fixed point $\vec{N}^*$ hence therein evaluating the {\it Jacobian} ${\bf J}$, {\it a.k.a.} the {\it community matrix}, with elements
\begin{equation}
    J_{ij}=\left.\frac{\partial [N_i f_i(\{N_j\})]}{\partial N_j}\right|_{\{N_j^*\}}=
    \delta_{ij}f_i(\{N_j^*\})+N_i^*\left.\frac{ \partial f_i(\{N_j\})}{\partial N_j}\right|_{\{N_j^*\}}
    \ .
\end{equation}
Note that contributions to the Jacobian coming from extinct species is diagonal and negative. The non trivial part comes from the $S^*$ surviving species and it gives rise to the $S^*\times S^*$ matrix 
\begin{equation}
    J_{ij}^*=-N_i^*B_{ij} \ ,
\end{equation}
with a non trivial stripy structure where the elements in each row are all rescaled by the same factor $N_i^*$. 
As discussed in the section about linear dynamical systems, linear stability requires 
that the real part of the leading eigenvalue $\lambda_1({\bf J})$
is negative,
and a nonzero imaginary part gives rise to oscillatory dynamics in the vicinity of the fixed point with frequency of oscillations inversely proportional to $|\Im[\lambda_1({\bf J})]|$.

The generalised Lotka-Volterra model discussed in this section provides an example of the structure of the matrices of interest when focusing on different facets of the stability of ecological systems. 
In these structures the interaction matrix ${\bf A}$ always plays an important role on the elements outside the diagonal, while the self regulation mechanism represented by the carrying capacities contributes to the non trivial diagonal.  Moreover, in the Jacobian, each row is multiplied by the abundance of the corresponding species at the fixed point.
Note that different examples of single species self-regulation mechanisms contained in the definition of $f_i$, such as those including the so called {\it Allee effect}~\cite{stephens1999allee} for instance, can lead to less straightforward connections between feasibility, structural and linear stability. 

A nowadays widespread approach to model ecosystems with large number of species is to account for the large variety of self-regulation  and interaction mechanisms by introducing random parameters, so that the matrices ${\bf A}$, ${\bf B}$, ${\bf J}$ may be represented by random matrices \cite{may1972will, Allesina2015, barbier2018generic, stone2018feasibility}. In the simple generalised Lotka-Volterra model considered above, this choice would require to introduce at least one probability distribution $p(\alpha)$ for the amplitudes of inter-species interactions $\alpha_{ij}$. When referring specifically to feasibility or structural stability determined by ${\bf B}$, the simplest setting would imply
assuming uniform growth rates across different species 
and unitary carrying capacities so that ${\bf B}_{\rm Id}={\bf A}+\mathbf{1}$. However, more generally it can be important to include in ${\bf B}$ the contribution of non trivial diagonal terms of a diagonal matrix ${\bf D}$, which are extracted from a second distribution $p_D(d)$ to describe the variability of carrying capacities 
%(and still uniform growth rates) 
${\bf B}={\bf A+D}$. 
Recall that the important stability trait of ${\bf B}$ or ${\bf B}_{\rm Id}$ is whether the spectrum contains or not the origin, which is answered by checking that the smallest real eigenvalue is  positive. Under this perspective it is completely equivalent to check whether the largest real eigenvalue of $-{\bf B}$ or $-{\bf B_{\rm Id}}$, or equivalently (as long as the spectra of ${\bf A}$ is symmetric around the origin) of ${\bf A}-\mathbf{1}$ or of ${\bf A}-{\bf D}$, is negative.
When focusing on ${\bf J}$, instead, the variability of the stationary abundances $N^*_i$ becomes a more relevant factor in the structure of the matrix. For simplicity, the carrying capacities are then set to the unity, $p_D(d)$ represents the distribution of the abundances, placed on the elements of a diagonal matrix ${\bf D}$, and we look at ${\bf J=}-{\bf D}{\bf B}_{\rm Id}$, or equivalently at ${\bf J=}{\bf D}({\bf A}-\mathbf{1})$, checking also in this case that the real part of its leading eigenvalue is negative.

Other key ingredients for model selection are the choice of the sign of interactions $\alpha_{ij}$ and $\alpha_{ji}$ \cite{mambuca2022dynamical, barbier2018generic, allesina2012} and the graph structure of inter-species interactions \cite{mambuca2022dynamical, marcus2022local}. The first aspect is related to which type of ecological behaviour determines the inter-species interactions: mutualistic (both positive), competitive (both negative), or predator-prey (of opposite sign). The first two cases will be also called sign-symmetric and the last sign-antisymmetric.
The interaction graph structure that can be considered spans from fully connected graphs to several types of sparse graphs. In this work we focus on the influence of short and long cycles on sparse graphs, therefore we will discuss and compare the results obtained on tree graphs (no cycles), Erd\H{o}s-R\'{e}nyi graphs (typical cycles with a length of the order $O(\log(S^*))$, and only a finite number of cycles of fixed length~\cite{bianconi2005loops}), and pure Husimi trees (cycles of fixed, short, length).

All the ecologically motivated characteristics of the random matrices ${\bf B}_{\rm Id}$, ${\bf B}$ and ${\bf J}$ highlighted and discussed in this section will be encoded in the different types of random matrix models introduced in the next section.
In that context the ecological notation is abandoned in favour of a more general random matrix notation where $N$ is the size of the matrix instead of $S$ or $S^*$, and a random diagonal matrix called ${\bf D}$ can either represent the diagonal matrix of inverse carrying capacities, or contain the elements of the vector of abundances $\vec{N}^*$. 

\subsection{Absolute stability and size-dependent stability for interaction-like and Jacobian-like ensembles} \label{sec:absolute-stability}

For all the ensembles of random matrices introduced in the previous section, and defined more generally in the next section, we are interested in  behaviour of their spectra when the matrix size is large, i.e., $N\gg 1$.
The stability of the corresponding dynamical system is assured when the spectrum of the associated interaction-like matrix does not include the origin of the complex plane and when the spectrum of the corresponding Jacobian-like matrix has leading eigenvalue with negative real part.

As  discussed in the previous sections, we  distinguish between models whose stability properties are not affected by their system size $N$, which we call {\it absolutely stable} models, and models whose stability is lost for $N$ larger than a finite size $\overline{N}$, and hence their stability is {\it size-dependent}.

Absolute stability, in terms of feasibility, linear, and structural stability, is granted when the real part of all eigenvalues of the corresponding relevant matrix is negative for all $N$, as discussed in Sec.~\ref{sec:stabEco}.  In other words, we require that $\Re[\lambda_i]<0$, $\forall i\in\left\{1,2,\ldots,N\right\}$ and  $\forall N\in \mathbb{N}$. 
Therefore, a necessary condition for absolute stability is that the spectra have a {\it real part bounded from above} in the large size limit, which implies that 
$\Re[\lambda_i]<a$, $\forall i\in\left\{1,2,\ldots,N\right\}$ and $\forall N$, where $a$ is   a finite constant. In such settings, absolute stability is obtained whenever, for all $N$, the matrix is  equipped with a diagonal that has  elements that are smaller than some finite $-d<0$, such that all eigenvalues have negative real part.

%Note that the stability of  densely connected models does not depend on their sign pattern. Hence, dense matrices are either size-dependent, or absolutely stable when their matrix entries  are properly rescaled by $N$, see, e.g., Refs,~\cite{Allesina2015, allesina2012, Cure2021}  
Previous works \cite{Allesina2015, allesina2012, Cure2021} have shown that the stability of densely connected models does not depend on their sign pattern. Hence, dense matrices are either size-dependent, or become absolutely stable when their matrix entries  are properly rescaled by $N$. Instead, we focus in this paper on sparsely connected models for which, interestingly, the sign pattern of interactions determines whether the matrix is absolutely stable or exhibits size-dependent stability~\cite{mambuca2022dynamical}.  Indeed, as we are going to discuss in detail in the next sections, for large, sparse, random matrices the existence  of a finite upper bound for the real part of all  eigenvalues may arise in specific settings  without the need of any {\it ad-hoc} global rescaling by system size, contrarily to the case of dense models.

\subsection{Random matrix models built from graphs} \label{sec:RMTModels}
We define the random matrix models that we study in this paper.
All the random matrix models are built from underlying graphs by using the weights of the edges as matrix entries. In particular entries opposite to the diagonal represent the weights of edges pointing to opposite directions. The matrix entry is zero if the corresponding directed edge is absent.
An edge is called nondirected when the two corresponding entries of the matrix are both nonzero.    

\subsubsection{The model structure.}We distinguish two type of random matrices defined on  sparse random graphs, namely, {\it interaction-like} matrices  ${\bf B}$, and {\it Jacobian-like}  matrices ${\bf J}$.    Both matrices ${\bf B}$ and  ${\bf J}$ are obtained from an {\it interaction} matrix ${\bf A}$ that is the adjacency matrix of a weighted graph, and which specifies the network of interactions between the system constituents.

Interaction-like matrices are the sum of the interaction matrix ${\bf A}$ and a diagonal matrix ${\bf D}$, i.e., 
\begin{equation} \label{eq:BDef}
{\bf B} :={\bf A}-{\bf D},
\end{equation}
where the entries $D_i\geq 0$ are independent and identically distributed random variables drawn from a distribution $p_D(d)$ with $d\in\mathbb{R}^+$, and the interaction matrix $\mathbf{A}$ will be defined in the next subsection.  
In the special case of $D_i=1$, for all $i$, we get
what we call the {\it shifted interaction} matrix
\begin{equation} \label{eq:BIDef}
{\bf B_{\rm Id}}:= {\bf A}-\mathbf{1},
\end{equation} 
where $\mathbf{1}$ is the identity matrix.

Instead, {\it Jacobian-like} matrices ${\bf J}$ are  defined as the product of ${\bf B_{\rm Id}}$ with $\mathbf{D}$,  viz.,
\begin{equation} \label{eq:JDef}
{\bf J}:=\mathbf{D}{\bf B_{\rm Id}},
\end{equation}
 where the entries $D_i\geq 0$ are as before independent and identically distributed random variables drawn from a distribution $p_D(d)$ with $d\in\mathbb{R}^+$ and finite second moment. 

\subsubsection{The interaction matrix $\mathbf{A}$.}
The random matrices ${\bf A}$ have elements 
\begin{equation}
A_{ij}:=C_{ij}\alpha_{ij},
\end{equation}
where $C_{ij}\in\left\{0,1\right\}$ are the entries of  the adjacency matrix of a nonweighted, nondirected, random graph,  and the $\alpha_{ij}\in \mathbb{R}$ are the weights of the edges of the graph, which represent the strengths of the interactions.  

The weights $\alpha_{ij}$ and $\alpha_{ji}$ are pairs of random variables extracted from a probability distribution $p_{\alpha}(u,l)$ that is symmetric under the exchange of its arguments, viz., %($p_p(u,l)=p_p(l,u)$): 
\begin{equation} \label{eq:pairs_distribution_def}
    p_{\alpha}(u,l):=p(|u|)p(|l|) \vec{\theta}(u) {\bf \Pi}(\pi^{\rm S},\pi^{\rm O})\ \vec{\theta}^{\ T}(l),
\end{equation}
where $\vec{\theta}(x)=\{\vartheta( x),\vartheta(-x)\}$, $\vartheta(x)$ is the Heaviside function,  
\begin{equation} \label{eq:pi_def}
{\bf \Pi}(\pi^{\rm S},\pi^{\rm O}):= 
\left(\begin{tabular}{cc}
    $\pi^{\rm S}(1-\pi^{\rm O})$ &  $0.5 \pi^{\rm O}$ \\
    $0.5 \pi^{\rm O}$ & $(1-\pi^{\rm S})(1-\pi^{\rm O})$
\end{tabular}\right) \ ,
\end{equation}
and $p(x)$ is a probability distribution  supported on $\mathbb{R}^+$ and with finite second moment; notice that we are particularly interested in models with unbounded support as their norm diverges in the infinite size limit, which is important for the findings in this paper.  
The constants $\pi^{\rm O}$ and $\pi^S$ determine the (anti)correlation between  the sign of  $\alpha_{ij}$ and  the sign of $\alpha_{ji}$ (their absolute values are  uncorrelated).  
For $\pi^{\rm O}=1$ elements opposite  to the main diagonal of $\mathbf{A}$ have opposite signs, i.e., $\alpha_{ij}\alpha_{ji}<0$, which we call {\it sign-antisymmetric} interactions. In this case, we speak of {\it antagonistic} model.
When $\pi^{\rm O}=0$, then elements opposite to the main diagonal   have the same sign, i.e., $\alpha_{ij}\alpha_{ji}>0$, and we speak of {\it sign-symmetric} interactions.   The elements are positive if  $\pi^{\rm S}=1$ and negative if $\pi^{\rm S}=0$, sometimes referred to as mutualistic and competitive interactions, respectively~\cite{allesina2012}.   For intermediate values of  $\pi^{\rm O}\in (0,1)$, we speak of a  
{\it mixture} models, as it contains a mixture of sign-antisymmetric and sign-symmetric interactions.

For the adjacency matrix $C_{ij}$  we focus in this paper on two models.  One is a random graph model that is locally tree-like, i.e., it has a small number of cycles of small length.   The second model  is  deterministic  and has many cycles of small length.    In this way, we will be able to address the effect of cycles on our results.  The models considered are: 
\begin{itemize}
\item {\it Erd\H{o}s-R\'{e}nyi graphs}:
There are two closely related variants of the Erd\H{o}s–R\'{e}nyi (ER) random graph model \cite{ER, bollobas2001random}. In the first model, a graph is chosen uniformly at random from the collection of all graphs which have $N$ nodes and $M$ edges. In the second model, the number of nodes $N$ is fixed and each edge connecting two of them exists with a probability $q$, which is fixed and independent from every other edge.
The Erd\H{o}s-R\'{e}nyi graphs we use are built according to the second model, that we denote as $G(N, q)$, setting $q=c/(N-1)$ where $c$ is it is the average number of edges on a single node, also called {\it connectivity} of the node.
In the limit $N\rightarrow\infty$, with $c$ fixed, Erd\H{o}s-R\'{e}nyi graphs are locally tree-like graph in the sense that with probability one the finite neighborhood of a randomly selected node is a tree, the typical cycles length $\ell$ grows like $\log N$~\cite{mezard2001bethe, dembo2010gibbs, montanari2012weak}, and they have only a finite number of cycles of fixed length~\cite{bianconi2005loops}). 

\item {\it Husimi trees}: Husimi trees are connected graphs for which no edge lies on more than one cycle~\cite{harary1953number}.   Loosely said, Husimi trees are trees built out of edges and cycles, such as,  triangles, quadrilaterals, pentagons, etc.   Husimi trees were introduced by Harary and Uhlenbeck \cite{harary1953number}, who recognised this graph structure in Husimi's virial expansion of the equation of state of a nonideal gas \cite{husimi1950note}, and whose terminology we adapt in this paper. If Husimi trees are built out of one type of cycle, then one speaks of pure Husimi trees, as in panel (c) of Fig.\ref{fig:sketch_sign_stable}, while otherwise they are mixed Husimi trees. If the cycles are triangles, then one speaks of a Husimi cactus. When the tree structure is  regular with coordination number $c$, a pure Husimi tree with cycles of length $\ell$ can be defined by using the notation $(c,\ell)$-pure Husimi tree. For instance, a $(c,1)$-pure Husimi tree is a Cayley tree or a Bethe lattice. Note that $(c,\ell)$-pure Husimi trees have a number of cycles of fixed length $\ell$ growing linearly with the system size.

\end{itemize}

As a recap, the models studied across this work can be identified from the matrix structure (interaction-like ${\bf B}$, shifted interaction ${\bf B_{\rm Id}}$, Jacobian-like ${\bf J}$), the choice of the distributions $p$ and $p_D$, the interactions sign pattern (antagonistic, mixture), the graph structure encoded in ${\bf C}$ (tree, $G(N, q)$ Erd\H{o}s-R\'{e}nyi graphs, $(c,\ell)$-pure Husimi tree).

\subsubsection{Canonical model parameters} \label{sec:numerical_details}

Here we list the  parameters that we use in the numerical results shown in the  following sections. Any variations on these will be reported in the figures captions.

As  anticipated, we deal with three matrix structures, viz., shifted interaction matrices ${\bf B_{\rm Id}}={\bf A}-\mathbf{1}$, interaction-like matrices ${\bf B}={\bf A}-{\bf D}$ and Jacobian-like matrices ${\bf J}=\mathbf{D}{\bf B_{\rm Id}}$.  In numerical examples, we need to specify the adjacency matrix $\mathbf{C}$, the distribution  $p_{\alpha}$ of weights $(\alpha_{ij},\alpha_{ji})$, and the distribution $p_D$ of diagonal entries $D_i$.

We consider two ensembles of adjacency matrices  ${\bf C}$, viz., sparse, Erd\H{o}s-R\'{e}nyi graphs and Husimi trees.   Erd\H{o}s-R\'{e}nyi graphs are drawn from the  $G(N, q)$ model with $q=c/(N-1)$.   Since we are interested in  sparse graphs, the connectivity is kept fixed at $c=2$, and thus does not  scale with the graph size $N$.  The Husimi trees  that  we employ are $(4, 4)$-pure Husimi tree, and hence all cycles have length $\ell=4$, as shown in panel (b) of Fig.~\ref{fig:sketch_not_sign_stable} for the case of a Husimi tree dressed with sign-antisymmetric interactions.

For the probability distribution  $p$  of the absolute values $|\alpha_{ij}|$ of the off-diagonal matrix entries $\alpha_{ij}$, which appears in Eq.~(\ref{eq:pairs_distribution_def}), we use a truncated Gaussian distribution.   In particular, we truncate a   Gaussian distribution with mean $\mu_{\scriptscriptstyle G}=1.0$ and variance $\sigma_{\scriptscriptstyle G}=0.6$ so that it is supported on the  positive part of the real line, viz., 
\begin{equation} \label{eq:truncated_gaussian}
    p(x)=\frac{2}{1+{\rm erf}\left(\frac{\mu_{\scriptscriptstyle {\rm G}}}{\sqrt{2\sigma_{\scriptscriptstyle {\rm G}}^2}}\right)} \vartheta(x)\mathcal{G}_{\mu_{\scriptscriptstyle {\rm G}},\sigma_{\scriptscriptstyle {\rm G}}}(x)
\end{equation}
where  $\mathcal{G}_{\mu_{\scriptscriptstyle G},\sigma_{\scriptscriptstyle G}}(x)$ is a Gaussian distribution with mean $\mu_{\scriptscriptstyle G}$ and variance $\sigma_{\scriptscriptstyle G}$, ${\rm erf}(x)$ is the error function, and $\vartheta(x)$ is  the Heaviside function. 

Notice that the first two moments of the truncated Gaussian distribution take the expressions
\begin{equation} \label{eq:tg_moments}
\begin{split}
    &\mu_{\scriptscriptstyle {\rm TG}} := \langle \, x \, \rangle_{\scriptscriptstyle {\rm TG}} = \mu_{\scriptscriptstyle {\rm G}} + \sqrt{\frac{2}{\pi}}\sigma_{\scriptscriptstyle {\rm G}} \frac{  {\rm exp}\left(-\frac{\mu_{\scriptscriptstyle {\rm G}}^2}{\sigma_{\scriptscriptstyle {\rm G}}^2}\right)}{1+{\rm erf}\left(\frac{\mu_{\scriptscriptstyle {\rm G}}}{\sqrt{2\sigma_{\scriptscriptstyle {\rm G}}^2}}\right)   } 
    \end{split}
\end{equation}
and
\begin{equation} 
\label{eq:tg_moments2}
\begin{split}
    &\langle \, x^2 \, \rangle_{\scriptscriptstyle {\rm TG}} = \mu_{\scriptscriptstyle {\rm G}}^2 + \sigma_{\scriptscriptstyle {\rm G}}^2 + \sqrt{\frac{2}{\pi}} \mu_{\scriptscriptstyle {\rm G}} \sigma_{\scriptscriptstyle {\rm G}}  \frac{{\rm exp}\left(-\frac{\mu_{\scriptscriptstyle {\rm G}}^2}{2\sigma_{\scriptscriptstyle {\rm G}}^2}\right)}{1+{\rm erf}\left(\frac{\mu_{\scriptscriptstyle {\rm G}}}{\sqrt{2\sigma_{\scriptscriptstyle {\rm G}}^2}}\right)   } \ ,
\end{split}
\end{equation}
where $\langle \cdot\rangle_{\scriptscriptstyle {\rm TG}}$ denotes the average with respect to the truncated Gaussian distribution. From Eqs.~(\ref{eq:pairs_distribution_def}), (\ref{eq:tg_moments}), and  (\ref{eq:tg_moments2}), the  first two cumulants of the distribution $p_\alpha(u,l)$ of the off-diagonal element pairs  follow readily as 
\begin{equation} \label{eq:pairs_distribution_moments}
\begin{split}
    &\mu := \langle \, u \, \rangle = \langle \, l \, \rangle = 0, 
\end{split}
\end{equation}
\begin{equation} \label{eq:pairs_distribution_moments2}
\begin{split}
    &\sigma^2 := \langle \, u^2 \, \rangle = \langle \, l^2 \, \rangle = \langle \, x^2 \, \rangle_{\scriptscriptstyle {\rm  TG}},
\end{split}
\end{equation}
and 
\begin{equation} \label{eq:pairs_distribution_moments3}
\begin{split}
    &\langle \, u \, l \, \rangle = (1-2\pi^{\rm O}) \; \mu_{\scriptscriptstyle {\rm TG}}^2.
\end{split}
\end{equation}
where $\langle \cdot\rangle$ denotes the average with respect to $p_\alpha(u,l)$. 

The matrix sign pattern is set by the choices of $\pi^O$ and $\pi^S$ as defined in Eq.~(\ref{eq:pi_def}). In our work we focus on antagonistic models with $\pi^O=1$ in which the interactions are  sign-antisymmetric and mixture models with $\pi^O=0.9$ and $\pi^S=0.5$, characterised by a majority of sign-antisymmetric interactions and a smaller portion of sign-symmetric ones.

In numerical examples we are using for $p_D$ a uniform distribution supported on $[d_{\rm min}, d_{\rm max}]$, i.e.,
\begin{equation} \label{eq:uniform_distr_def}
    p_D(d)= \frac{1}{d_{\rm max} - d_{\rm min}} \; \vartheta(x - d_{\rm min}) \; \vartheta(d_{\rm max} - x)
\end{equation}
where $d_{\rm min}$ and $d_{\rm max}$ are, respectively, the minimum and maximum values of the uniform distribution.   Notice that 
\begin{eqnarray} \label{eq:uniform_distr_extremes}
    d_{\rm min} = \mu_D - \sigma_D \sqrt{3}\quad  &  {\rm and} &\quad 
    d_{\rm max} = \mu_D + \sigma_D \sqrt{3},
\end{eqnarray}
where  $\mu_D$ and $\sigma_D$ are the mean and standard deviation of $p_D$, respectively.

We choose $\mu_D>0$ and $\sigma_D<\mu_D/\sqrt{3}$ such that $p_D(d)$ is supported on a subset of the positive real axis and, thereby, all the $D_i$ are positive.  More specifically, in Sec.~\ref{sec:localsignstability} we set $\mu_D=1$ and $\sigma_D=0.5$, whereas in Sec.~\ref{sec:reentrances} we set $\mu_D=1$ and $\sigma_D$ takes values equally spaced between $0$ and $0.30$.

We diagonalise matrices with   linear algebra routines of the Numpy submodule  \textit{linalg} of Python3.

\section{Sign stable  matrices} \label{sec:sign_stability}
We review sign stability of matrices  $\mathbf{M}\in \mathbb{R}^{N\times N}$, which plays an important role in this paper, as we extend this concept to random graphs in the next section.    
As anticipated in the Introduction,  sign stability refers to all matrices with the same topology and sign pattern, independently
from the value of their nonzero elements.
Therefore, sign stability  was introduced in  studies on qualitative economics in the 1960s \cite{maybe1969qualitative, quirk1965qualitative, logofet1982sign}, and  found a decade later applications in, among others, qualitative ecology \cite{may1973qualitative,jeffries1974qualitative, levins1975problems, solimano1982graph}, and  chemistry \cite{clarke1975theorems}.

We say that a matrix $\mathbf{M}$ is {\it sign stable} if any matrix $\mathbf{M}'$ with the same sign pattern  is stable. Note that sign stability is a stronger condition than stability, as it requires that the real part of the eigenvalues of all matrices $\mathbf{M}'$ in the equivalence class  
\begin{equation}
  \mathcal{M}(\mathbf{M}):= \left\{\mathbf{M}'\in \mathbb{R}^{N\times N}: {\rm sign}(M'_{ij}) = {\rm sign}(M_{ij})  \right\}  
\end{equation}
 are negative, where ${\rm sign}(x)\in \left\{-,0,+\right\}$.   Note that matrices $\mathbf{M}'$ in the equivalence class $ \mathcal{M}(\mathbf{M})$  can be generated from $\mathbf{M}$ through
 \begin{equation}
     M'_{ij} = y_{ij}M_{ij},
 \end{equation}
 where $y_{ij}>0$.  Also, note that $ \mathcal{M}(\mathbf{M}_1) = \mathcal{M}(\mathbf{M}_2)$, for all $\mathbf{M}_2\in \mathcal{M}(\mathbf{M}_1)$.

On first sight, sign stability may appear as a too strong condition to be useful.  However, as will become soon evident, there exist several, interesting examples of equivalence classes $\mathcal{M}$ that are sign stable.   Moreover, antagonistic, sign stable matrices have a tree structure, which will make them important for the spectral theory of random graphs that we discuss in the next section.  
In what follows we discuss necessary and sufficient conditions for the sign stability of the equivalence class $\mathcal{M}$ generated by the matrix $\mathbf{M}$.

In the case when $M_{ii}<0$ for all $i$, sufficient and necessary conditions for sign stability have been derived by Maybee and Quirk \cite{quirk1965qualitative}.    These are (see Theorem 3 in~\cite{quirk1965qualitative}): 
\begin{equation}
M_{ij}M_{ji}\leq 0, \quad \forall i\neq j,  \label{eq:cond1}
\end{equation}
and for all $m\geq 3$, 
\begin{eqnarray}
M_{i_mi_1}=0,  \nonumber\\ 
\quad  \quad \quad \forall   i_1\neq i_2\neq \cdots \neq i_m,  \quad {\rm such} \ {\rm that},
 \quad M_{i_1i_2}M_{i_2i_3}\ldots M_{i_{m-1}i_m}\neq 0. \label{eq:cond2} \end{eqnarray}
Condition (\ref{eq:cond1}) implies that edges are either directed or nondirected with sign-antisymmetric weights, and condition Eq.~(\ref{eq:cond2}) states that there are no directed cycles of length $m\geq 3$.   
 
In the case when $M_{ii}\leq 0$, the conditions  Eqs.~(\ref{eq:cond1})  and  (\ref{eq:cond2}) are sufficient and necessary conditions for marginal, sign stability (see Lemma 5 in~\cite{quirk1965qualitative}).   In Refs.~\cite{jeffries1974qualitative, jeffries1977matrix, yamada1987generic} the marginal stable case has been studied in more detail, in particular to determine the conditions for which $\lim_{t\rightarrow \infty}\|\vec{x}(t)\|$ diverges.     

\begin{figure}[h]
	\centering
	\begin{subfigure}[b]{0.32\textwidth}
		\centering
		\includegraphics[width=\textwidth]{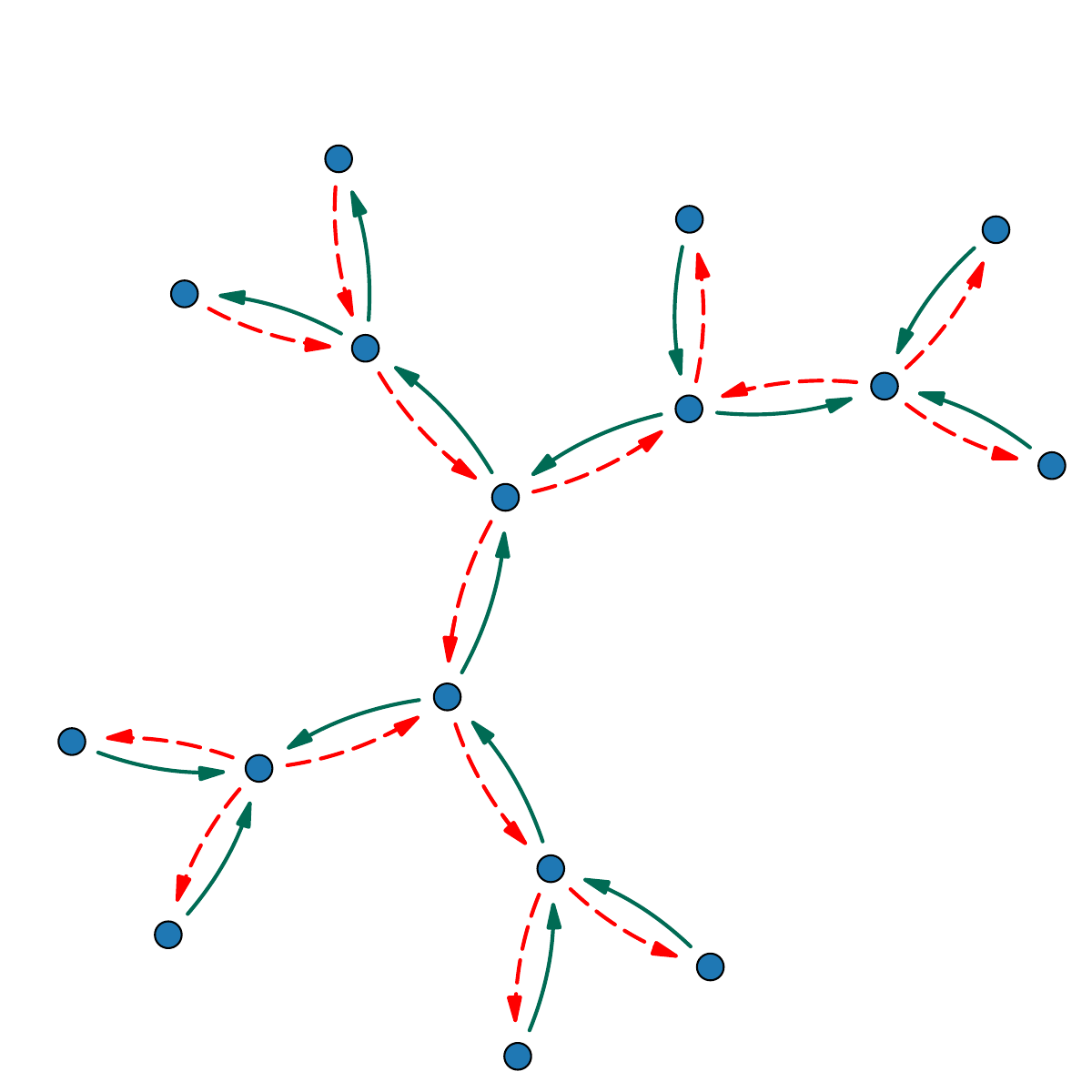}
		\caption{Antagonistic tree\\ \phantom{aaaaaaaaa}}
	\end{subfigure}
	\begin{subfigure}[b]{0.32\textwidth}
		\centering
		\includegraphics[width=\textwidth]{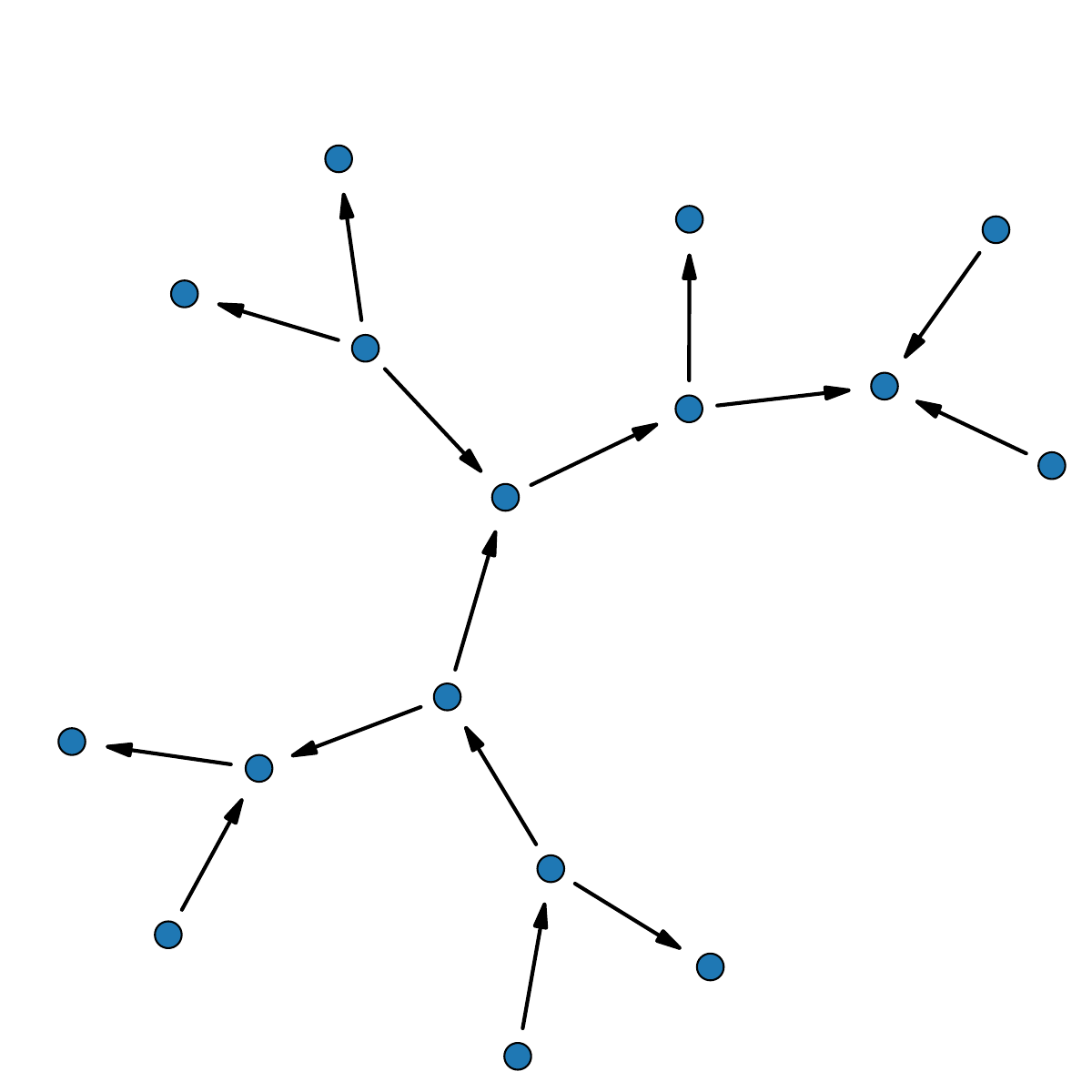}
		\caption{Oriented tree \\ \phantom{aaaaaaaaa}}
	\end{subfigure}
	\begin{subfigure}[b]{0.32\textwidth}
		\centering
		\includegraphics[width=\textwidth]{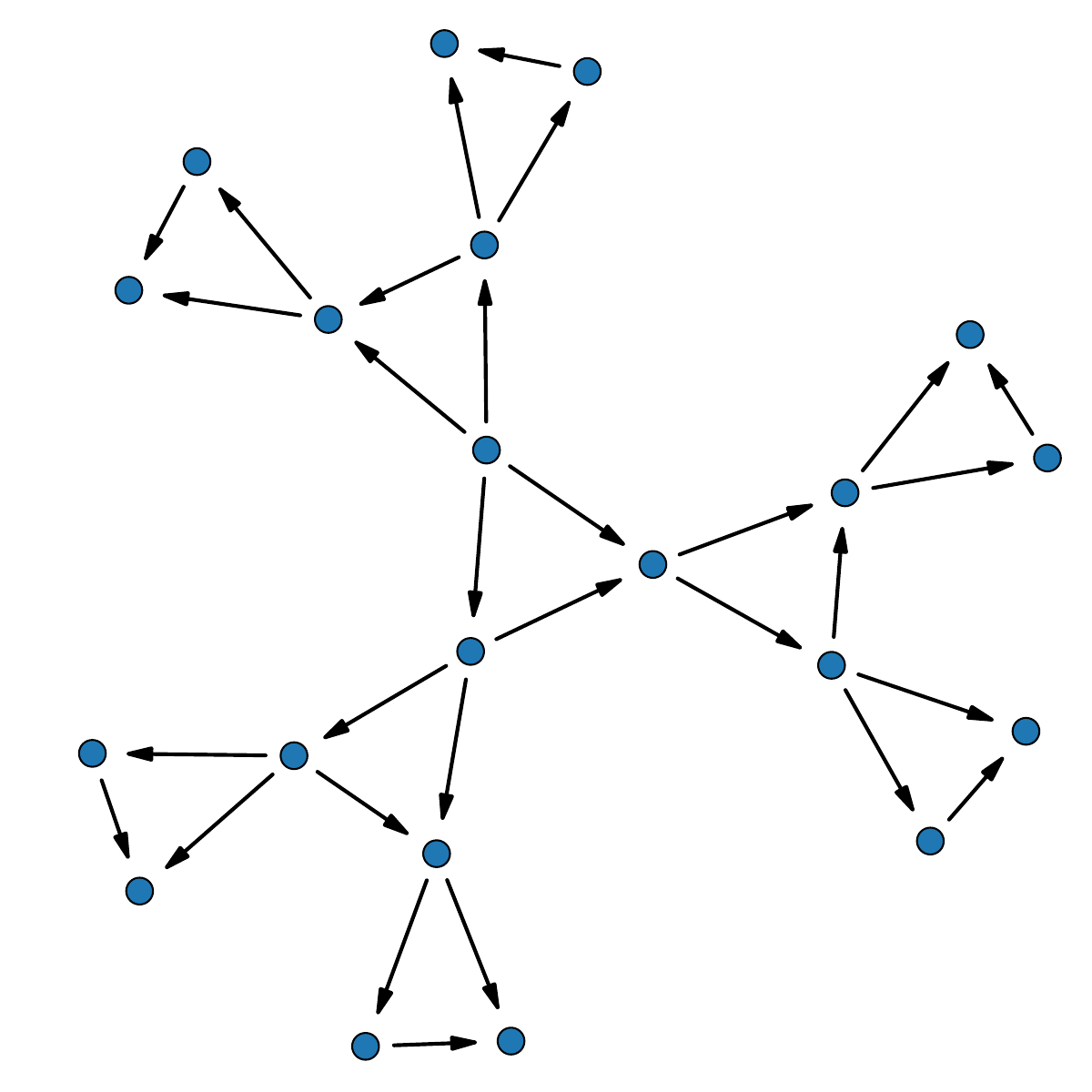}
		\caption{Feedforward oriented Husimi tree}
	\end{subfigure}
	\caption[]{Canonical examples of sign stable graphs. The arrows show the orientation of interactions. The color/style shows the sign of the weights associated with the edges.}
	\label{fig:sketch_sign_stable}
\end{figure}

\begin{figure}[H]
	\centering
	\begin{subfigure}[b]{0.32\textwidth}
		\centering
		\includegraphics[width=\textwidth]{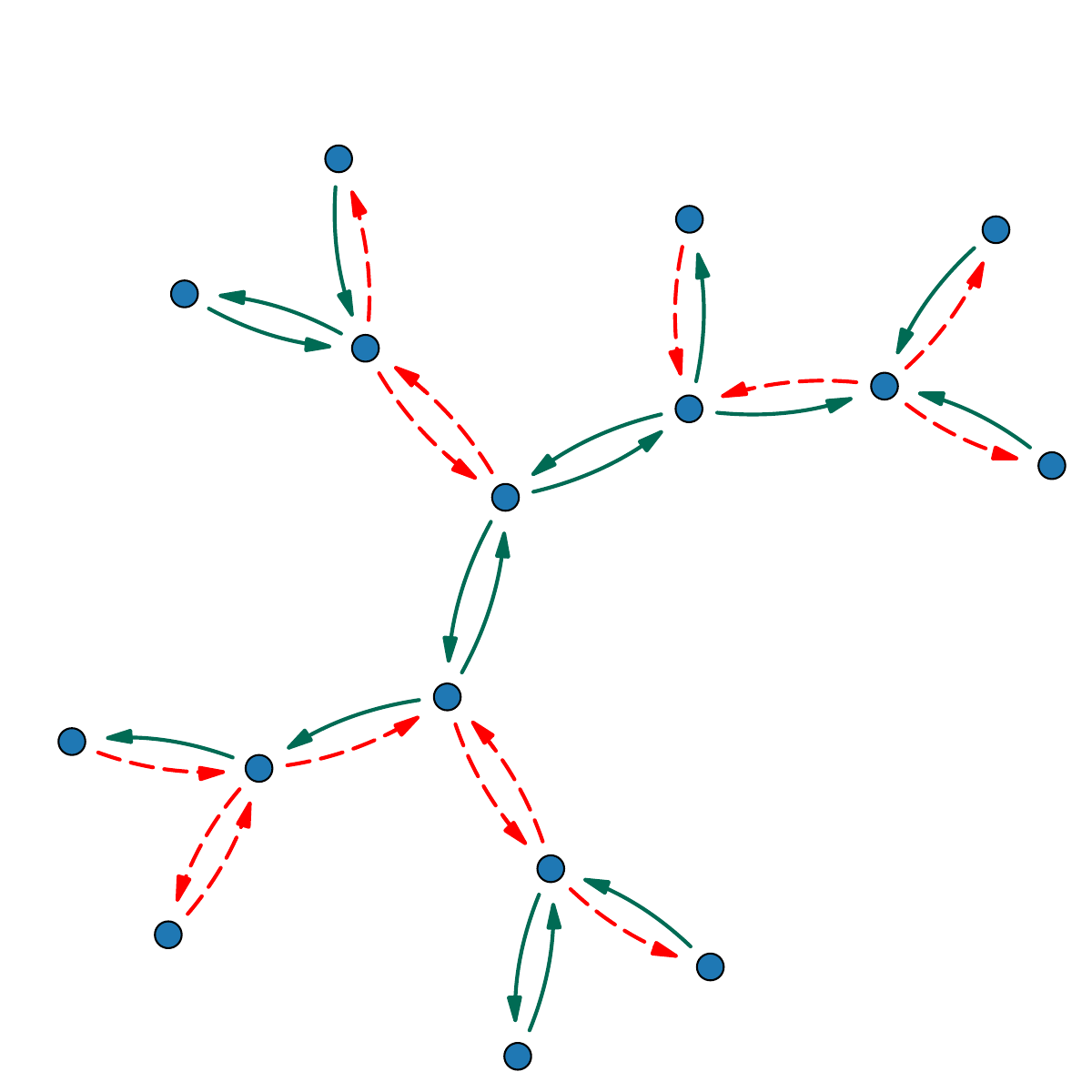}
		\caption{Mixture tree graphs}
	\end{subfigure}
        \hspace{8ex}
	\begin{subfigure}[b]{0.32\textwidth}
		\centering
		\includegraphics[width=\textwidth]{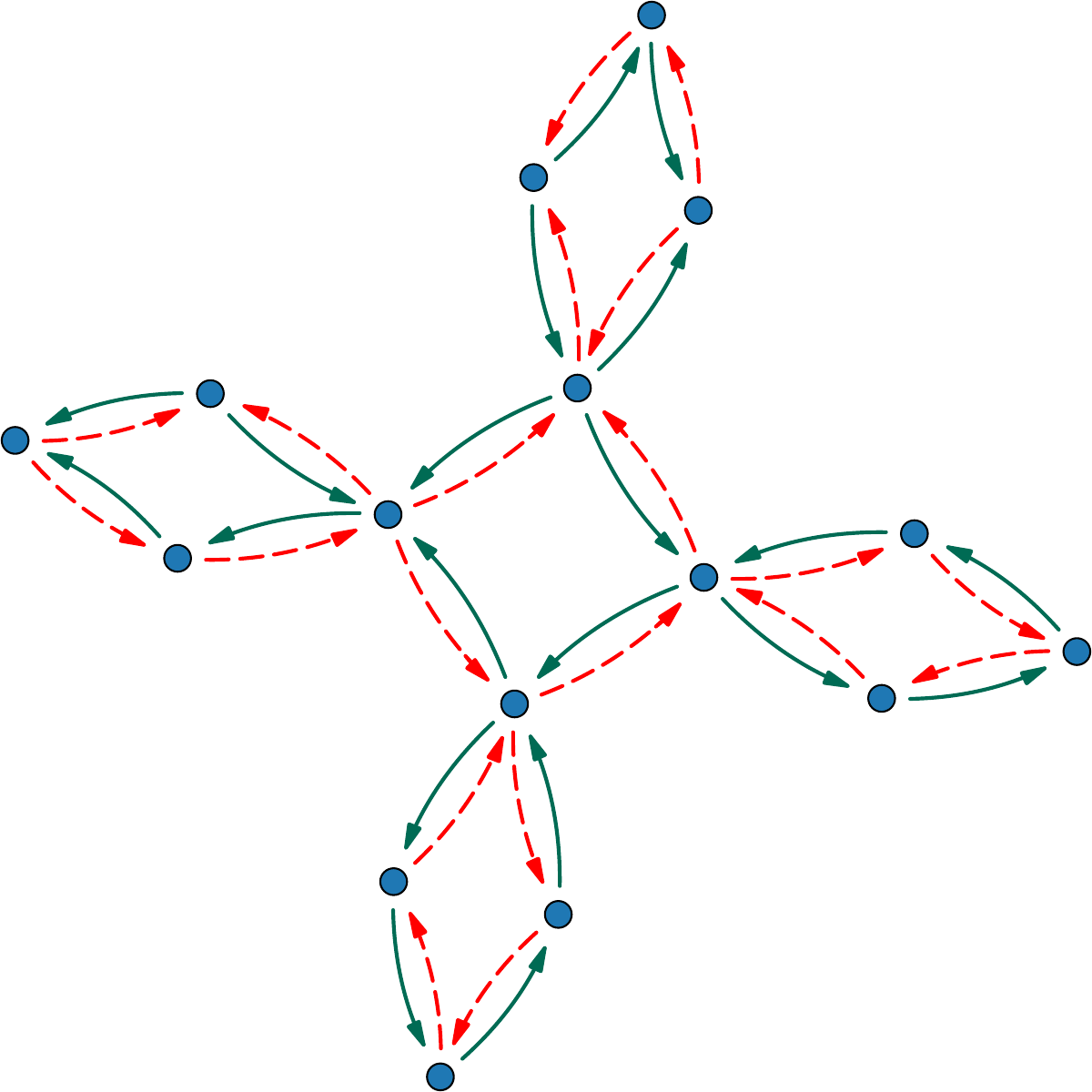}
		\caption{Antagonistic Husimi tree}
	\end{subfigure}
	\caption[]{Examples of graphs that are not sign stable due to the type of interactions {\bf (a)} or graph topology {\bf (b)}. The arrows show the orientation of interactions. The color/style shows the sign of the weights associated with the edges.}
	\label{fig:sketch_not_sign_stable}
\end{figure}

We will not discuss the derivation of the conditions Eqs.~(\ref{eq:cond1}) and (\ref{eq:cond2}), as these can be found in detail in Refs.~\cite{quirk1965qualitative, jeffries1977matrix}.   Nevertheless, we  mention three notable examples of sign stable matrices and show that they are sign stable:
%, and show explicitly that they are sign stable.    In particular, we consider the adjacency matrices of 
\begin{itemize}
    \item {\it  Nondirected antagonistic tree graphs}: this is a tree graph with sign-antisymmetric edges, i.e., $M_{ij}> 0 \Leftrightarrow M_{ji} < 0$ for all $i\neq j$, as illustrated in Panel (a) of Fig.~\ref{fig:sketch_sign_stable}.
    The condition (\ref{eq:cond1}) ensues from the  sign-antisymmetric nature of the edges, and the condition  (\ref{eq:cond2}) from  the absence
    of cycles in a  tree graph implies condition. 
    The sign stability of nondirected antagonistic tree graphs follows from the fact that (i) tree graphs with sign-antisymmetric interactions and negative diagonal elements have eigenvalues with negative real part, as we show in~\ref{app:signstab} and \ref{app:signstabdiag3}; (ii) the sign-antisymmetric nature of the interactions is preserved in the equivalence class $\mathcal{M}$ generated by an antagonistic tree graph. 
    Note that this example will be referred to extensively in the remainder of the paper. Importantly, simply lifting the constraint of sign-antisymmetric edges, as shown in Panel (a) of Fig.~\ref{fig:sketch_not_sign_stable}, will determine also in the case of a tree graph the failing of condition Eq.~(\ref{eq:cond1}). The corresponding matrices are not sign stable anymore. 

    \item {\it Oriented tree graphs}:  these are the adjacency matrices of  tree graphs with unidirectional edges, as sketched in Panel (b) of Fig.~\ref{fig:sketch_sign_stable}.  The unidirectionality of the edges implies
    \begin{equation}
    M_{ij}M_{ji} = 0, \quad \forall i\neq j,
    \end{equation} 
    and hence  Eq.~(\ref{eq:cond1}),  and the absence of cycles in a  tree graph implies condition Eq.~(\ref{eq:cond2}).   
    The sign stability of oriented tree graphs follows readily from the following two facts: (i) all eigenvalues of oriented tree graphs are equal to zero \cite{neri2019spectral, sachs1980spectra}, which is a direct consequence of the Coefficients Theorem for  Directed Graphs, see~\ref{app:signstab2}; (ii) the oriented tree property is preserved in the equivalence class $\mathcal{M}$ generated by an oriented tree graph.  
    
    \item {\it  Husimi trees built out of unidirectional feed-forward cycles}: these are Husimi trees built out of  motifs that are  feedforward cycles, as illustrated in Panel (c) of Fig.~\ref{fig:sketch_sign_stable}.   Adjacency matrices of such graphs are sign stable for exactly the same reason as the adjacency matrices of oriented tree graphs:  (i) $\lambda_j=0$ for all values of $j$, see~\ref{app:signstab2}; (ii) the orientedness and feedforward structure are preserved in the equivalence class  $\mathcal{M}$  generated by Husimi trees built out of unidirectional feed-forward cycles. Note that, if on this graph structure we consider instead antagonistic interactions, as shown in Panel (b) of Fig.~\ref{fig:sketch_not_sign_stable}, the condition Eq.~(\ref{eq:cond2}) will not be satisfied anymore as feedback loops are created, and the corresponding ensemble of matrices cannot be said to be sign stable. 
   
\end{itemize}

Note that not all sign stable matrices are tree graphs or Husimi trees. Let's comment on a couple of simple notable examples to illustrate how general is the concept of sign stability. 
An upper-diagonal matrix with negative entries on the diagonal is a sign stable matrix on any graph structure, even fully connected, as all eigenvalues correspond to the element on the diagonal.  
On the other hand, antisymmetric matrices for which  the entries satisfy
$M_{ij}=-M_{ji}$ and $M_{ii}=0$, are not sign-(marginally)stable, although having all imaginary eigenvalues, similarly to antagonistic tree graphs with zero diagonal entries (see~\ref{app:signstab}). The reason is that, at variance with antagonistic tree graphs with zero diagonal entries, most matrices in the corresponding equivalence class $\mathcal{M}$ of antisymmetric matrices are not marginally stable, because they are not antisymmetric and their eigenvalues can have nonnull (positive and negative) real parts. 
     
A further comment on the extension of sign stable ensembles is that adding negative terms to the diagonal of matrices in a sign stable random matrix ensemble does not affect their stability.    This leads to an asymmetry in the spectra of sign stable matrices, as the real part of the eigenvalues is bounded from above, but  can extend towards infinity on the negative real axis.

Finally, the two examples of graphs shown in Figure~\ref{fig:sketch_not_sign_stable} refer to ensembles which will not be characterised by sign stability due to the fact that either the condition Eq.~(\ref{eq:cond1}) or the condition Eq.~(\ref{eq:cond2}), respectively, are not satisfied.

\section{Implications of local sign stability on the spectra of random graphs} \label{sec:localsignstability}

In this section, we identify a useful  criterion, which we call {\it strong local sign stability}, for  the finiteness of the  average of the real part of the leading eigenvalue of  infinitely large, sparse, random graphs, i.e., $\lim_{N\rightarrow \infty}\langle \Re[\lambda_1]\rangle < \infty$; such random matrix models can be made absolutely stable through a constant shift of the diagonal entries, as we discussed in Sec.~\ref{sec:absolute-stability}.  

To introduce local sign stability and strong local sign stability, we first review the results of  Ref.~\cite{mambuca2022dynamical} on  the asymptotic behaviour of the real part of the leading eigenvalue of Erd\H{o}s-R\'{e}nyi graphs.   This paper shows that     the average, real part of the leading eigenvalue of an     Erd\H{o}s-R\'{e}nyi graphs with sign-antisymmetric  weights   converges to a finite limit as a function of $N$.     This result came as a surprise as the norm of the associated adjacency matrix diverges in the infinite size limit.   Moreover, it is sufficient to decorate the Erd\H{o}s-R\'{e}nyi graph with   a finite fraction of   sign-symmetric weights to have a  $\langle \Re[\lambda_1]\rangle$  that diverges as a function of $N$,   as expected for an   Erd\H{o}s-R\'{e}nyi  ensemble with diverging norm. 

In what follows, we  identify the property underlying the  finiteness of the real part of the leading eigenvalue in   Erd\H{o}s-R\'{e}nyi graphs with sign-antisymmetric weights, and we aim to extend this property so that it can be applied to other random graph ensembles. 
To understand the distinction between  Erd\H{o}s-R\'{e}nyi graphs with sign-antisymmetric and  sign-symmetric weights, we build on the locally tree-like property of  Erd\H{o}s-R\'{e}nyi graphs, see e.g.~Refs.~\cite{mezard2001bethe, dembo2010gibbs, montanari2012weak}.   We say that a random graph is locally tree-like when  for large enough values of $N$ the finite neighbourhood of a randomly selected node is almost certainly a tree. 
As discussed in Sec.~\ref{sec:sign_stability}, trees with sign-antisymmetric weights are sign stable, whereas this property does not hold for trees with sign-symmetric weights.    However, in general, random graphs have simple cycles and hence, given the condition in Eq.~(\ref{eq:cond2}), are not sign stable.  As a consequence, we can not rely directly on the concept of sign stability introduced in the previous section to understand the distinction between antagonistic and mixture Er\H{o}s-R\'{e}nyi graphs. 

This limitation leads us to introduce the weaker condition of local sign stability:    
%\begin{definition}[Local Sign Stability]
{\it let  $\mathbf{M}_N$ be a sequence of matrices built from weighted graphs  of $N$ nodes.   Let $\mathbf{M}^{(i)}_N(d)$ be the matrix of the weighted subgraph generated by a uniformly and randomly selected node $i$  and all of its nodes located within a distance $d$ of $i$.  We say that  $\mathbf{M}_N$  is locally sign stable if for all fixed $d$, the probability that    $\mathbf{M}^{(i)}_N(d)$ is  sign stable  converges to one as  a function of $N$.}
%\end{definition}

In the above definition we consider a matrix built from the graph as explained in Sec.~\ref{sec:RMTModels}.
In full generality this definition of local sign stability remains valid for graphs that are not locally tree-like as long as the short cycles satisfy the condition given by Eqs.~(\ref{eq:cond2}).  However, for nondirected graphs local sign stability requires locally tree-likeness, as cycles with nondirected edges cannot satisfy the Eqs.~(\ref{eq:cond2}).   

For nondirected graphs we make a step further and introduce the condition of strong local sign stability:  {\it let $\mathbf{M}_N$ be a sequence of matrices locally sign stable. We say that $\mathbf{M}_N$ is strongly locally sign stable if in addition the average number of cycles of fixed length does not asymptotically increase with $N$, which for nondirected graphs is  a more stringent requirement than the locally tree-likeness.}

%We say that a sequence of matrices is strongly locally sign stable if they are locally sign stable and %under the additional condition that
%the average number of loops of fixed length does not asymptotically increase with $N$, which for nondirected graphs is  a more stringent requirement than the locally tree-likeness.
%the weights are uncorrelated with the graph structure

Since (strong) locally sign stable matrices are not sign stable, they can have a leading eigenvalue with a positive real part. Nevertheless, finite neighbourhoods $\mathbf{M}^{(i)}_N(d)$ of large  tree-like antagonistic graphs are almost certainly sign stable. 
Thus, sign stability is broken by either cycles of length $\ln(N)$, which diverge for large enough size $N$, or by a small number of cycles of  finite length. 
Moreover, when the number of finite cycles is small but still growing with $N$~\cite{bianconi2005loops}, the leading eigenvalue can also still grow with $N$.
Conversely, here we claim that when the number of finite cycles remains finite, as required by the strong local sign stability condition,  
%Consequently, we claim that 
the real part of the leading eigenvalue can be positive, but  will not grow indefinitely when $N$ increases. %, as the contribution of cycles is not significant enough to make the leading eigenvalue diverge.  

We summarise this connection as follows.  
 Consider a  set   $\mathscr{M}$  of sequences $\left\{\mathbf{M}_N\right\}_{N\in \mathbb{N}}$ of  random graph models, as defined in Sec.~\ref{sec:RMTModels}.   Then,     
    \begin{equation} \label{eq:conjecture}
        \left\{\mathbf{M}_N\right\}_{N\in \mathbb{N}} \textrm{ is strongly locally   sign   stable} \Longrightarrow \lim_{N\to\infty} \langle \Re[\lambda_1(\mathbf{M}_N)]\rangle < \infty.
    \end{equation}
This 
connection holds  (trivially) when  the matrix norm $\|\mathbf{M}_N\|$ is bounded.     Indeed, the matrix norm is always larger or equal  than the real part of the leading eigenvalue, i.e.,  $\|\mathbf{M}_N\| \geq \Re[\lambda_1(\mathbf{M}_N)]$~\cite{horn2012matrix}.  
The interesting cases arise when  $\lim_{N\rightarrow \infty}\|\mathbf{M}_N\|=\infty$, e.g.,  for  the weighted Erd\H{o}s-R\'{e}nyi graphs that we consider in this paper.  When the matrix norm diverges with the system size, Eq.~\eqref{eq:conjecture} remains valid provided that the elements of the matrix have finite second moment.

Equation~(\ref{eq:conjecture}) characterises a sufficient condition, but strong LSS is not necessary to have finite $\Re[\lambda_1]$. 
Indeed, for instance, for antisymmetric matrices $\mathbf{M}_N$ it holds that  $\Re[\lambda_1]=0$, as all eigenvalues are immaginary, irrespectively of strong local sign stability. 
Nevertheless, strong LSS can be used as a condition to predict when the leading eigenvalue will have a finite real part and hence whether the corresponding dynamical system can be made absolutely stable or not.
To confirm the validity of this condition we  rely on  known results from the literature on spectra of graphs and we show new numerical results further corroborating the stated condition.  

First, let us consider known results on spectra of random graphs.   
For weighted random oriented graphs an explicit expression for the leading eigenvalue was derived in Refs.~\cite{neri2019spectral, tarnowski2020universal, metz2020localization}, which shows that the leading eigenvalue of large graphs is a growing function 
of the branching ratio $\langle K^{\rm in}K^{\rm out}\rangle/c$ and of the first or the second moment of the weights.
%of $\langle J^2\rangle_J\langle K^{\rm in}K^{\rm out}\rangle_K/c$ (where $\langle K^{\rm in}K^{\rm out}\rangle_K/c$ is the branching ratio of the oriented graph).
Interestingly, the average number of cycles of finite length $\ell$ also is a growing function of the branching ratio~\cite{neri2019spectral}, hence, for finite second moment of the distribution of weights, both diverge as soon as the branching ratio diverges with the system size.
Therefore, in this example strong local sign stability, granted by a finite number of small cycles, implies that the leading eigenvalue is  finite, provided that the distribution of weights has a well defined second moment.
%diverges as a function of $N$ when the branching ratio $\langle K^{\rm in}K^{\rm out}\rangle/c$ is infinite, and is finite for finite branching ratio,   which is in agreement with the connection established in Eq.~(\ref{eq:conjecture}). 
Another example of a strongly locally sign stable random graphs are Erd\H{o}s-R\'{e}nyi graphs (which have finite number of cycles of fixed length) with sign-antisymmetric weights: Ref.~\cite{mambuca2022dynamical} shows that the real part of the leading eigenvalue is finite as well in this case.   On the other hand, symmetric Erd\H{o}s-R\'{e}nyi graphs do not satisfy local sign stability condition Eq.~(\ref{eq:cond1}) and their spectra contain the whole real axis in the infinitely large limit, see e.g.~Refs.~\cite{RodgBray1988, Reimer_sparse, krivelevich2003largest, chung2004spectra}.   

Since examples already present in literature are limited, we now employ new numerical simulations to further verify Eq.~(\ref{eq:conjecture}) for different types of matrix structure. To illustrate the significance of the condition in Eq.~\eqref{eq:conjecture} we also explore two settings in which LSS does not hold and the leading eigenvalue grows with the system size. In particular we break strong LSS and LSS in two different ways by removing one of their two fundamental ingredients at a time: sign-antisymmetric entries and the locally tree-like structure. We compare spectral results for antagonistic (i.e., purely sign-antisymmetric) Erd\H{o}s-R\'{e}nyi graphs, which according to our definition are strongly locally sign stable, with two matrix ensembles that are not LSS, namely,  mixture Erd\H{o}s-R\'{e}nyi graphs (i.e., locally tree-like but with a small fraction of sign-symmetric links) and antagonistic Husimi trees (i.e., keeping the sign-antisymmetric links,  while dropping the locally tree-like structure). 
% Hence, the mixture ensemble has a network topology that is favourable, but its interaction pattern breaks strong local sign stability, while the antagonistic Husimi tree has a favourable interaction pattern, but a network topology  that breaks strong local sign stability.    

Since different stability criteria rely on different type of matrices, we consider three cases, as introduced in  Sec.~\ref{sec:stability_and_model}, namely,   shifted interaction matrices ${\bf B_{\rm Id}}$ in Sec.~\ref{subsec:lss_shifted}, interaction-like matrices $\mathbf{B}$ in Sec.~\ref{subsec:lss_interaction}, and Jacobian-like matrices $\mathbf{J}$ in Sec.~\ref{subsec:lss_jacobian}.     
%Note that the case of shifted interaction matrices was also studied in Ref.~\cite{mambuca2022dynamical}, but here we also study  the spectra of antagonistic Husimi trees that has  not been considered in Ref.~\cite{mambuca2022dynamical}.  Hence, here we  go beyond previous works by  breaking strong local sign stability through both the sign of the weights and through the network topology, and in addition, we go beyond previous works by considering matrices with diagonal disorder and Jacobian-like matrices.

\subsection{Shifted interaction matrices $\mathbf{B}_{\rm Id}$} \label{subsec:lss_shifted}

First, we illustrate the connection in Eq.~\eqref{eq:conjecture} on shifted interaction matrices, ${\bf B_{\rm Id}}={\bf A}-\mathbf{1}$, as defined in Sec.~\ref{sec:RMTModels}.
The $\mathbf{B}_{\rm Id}$ matrices are the simplest class of matrices that we consider, as  the diagonal entries are constant, and hence we can  focus  on the contribution of the off-diagonal matrix entries to the  spectrum. Note that this matrix structure was also studied in Ref.~\cite{mambuca2022dynamical}, but here for convenience we present again its results (obtained with a different weights distribution) also in comparison with 
%we also study 
the spectra of antagonistic Husimi trees that has not been considered before.

\begin{figure}[H]
	\centering
	%\captionsetup{justification=centering}
	\includegraphics[width=0.7\textwidth]{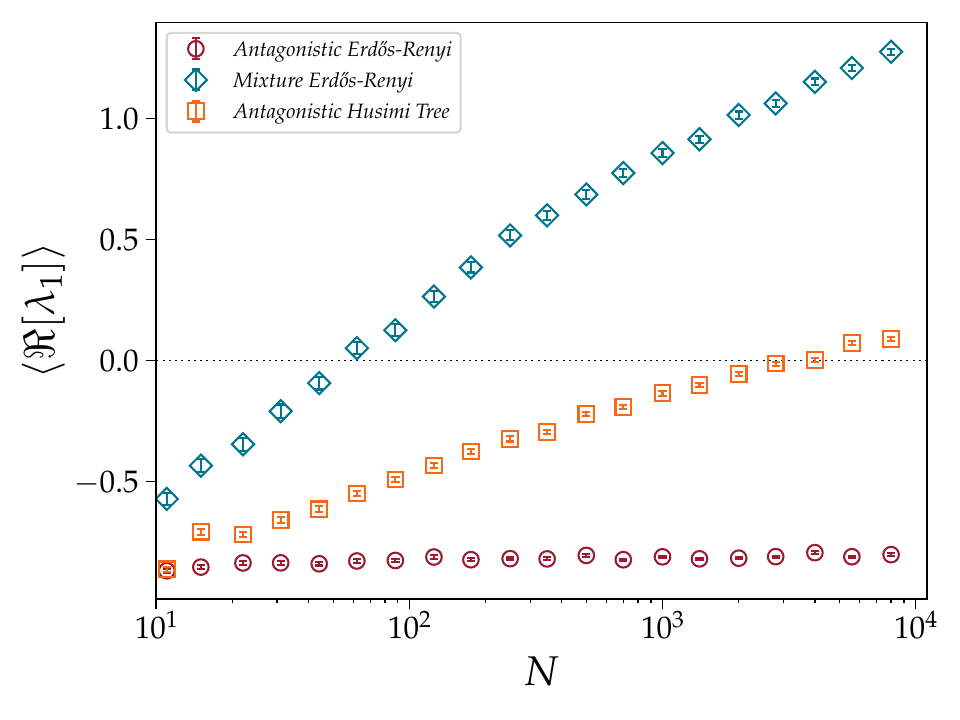}
	\caption[]{Average real part of the leading eigenvalue, $\langle \Re[\lambda_1]\rangle$, as a function of the matrix size $N$ for shifted interaction matrices $\mathbf{B}_{\rm Id}$. Results shown are   for antagonistic Erd\H{o}s-R\'{e}nyi graphs (red circles), mixture Erd\H{o}s-R\'{e}nyi graphs (blue diamonds) and antagonistic pure Husimi trees (orange squares). Markers are sample means obtained from numerically diagonalising $300$ matrix realisations, and error bars denote the error on the mean.    The  parameters used for the matrix ensembles  are detailed in Sec.~\ref{sec:numerical_details}.   }
	\label{fig:leading_eig_interaction}
\end{figure}

Figure~\ref{fig:leading_eig_interaction} plots the average value of the real part of the leading eigenvalue, $\langle  \Re[\lambda_1(\mathbf{B}_{\rm Id})]\rangle$, as a function of $N$ for the three ensembles under study, i.e.,  antagonistic Erd\H{o}s-R\'{e}nyi graphs, mixture Erd\H{o}s-R\'{e}nyi graphs, and antagonistic Husimi trees. These results, obtained by numerically diagonalising matrices, confirm that strong local sign stability implies that the leading eigenvalue does not diverge with the system size, while violating one of the conditions of LSS leads to its divergence. In fact, Fig.~\ref{fig:leading_eig_interaction} shows a qualitative difference in the behaviour of the real part  of the leading eigenvalue as a function of the matrix size~$N$: for mixture Erd\H{o}s-R\'{e}nyi  graphs and antagonistic Husimi trees, $\langle \Re[\lambda_1]\rangle$  increases monotonically  as a  function of $N$, while for  antagonistic Erd\H{o}s-R\'{e}nyi graphs, $\langle \Re[\lambda_1]\rangle$ quickly converges to a finite value.
Note that   the theoretical results in Ref.~\cite{mambuca2022dynamical} for the boundary of the spectrum of infinitely large graphs obtained with the cavity method support the numerical observation that in the antagonistic Erd\H{o}s-R\'{e}nyi case the leading eigenvalue has a real part which remains finite in the large $N$ limit. Therefore the saturation observed from direct diagonalization results in the antagonistic case turns out to be representative of the large $N$ behaviour.
Based on these grounds, in the following, we will rely on direct diagonalization results to extrapolate the large $N$ behaviour also for interaction-like and Jacobian-like matrices.

\begin{figure}[H]
	\centering
	%\captionsetup{justification=centering}
	\includegraphics[width=1.0\textwidth]{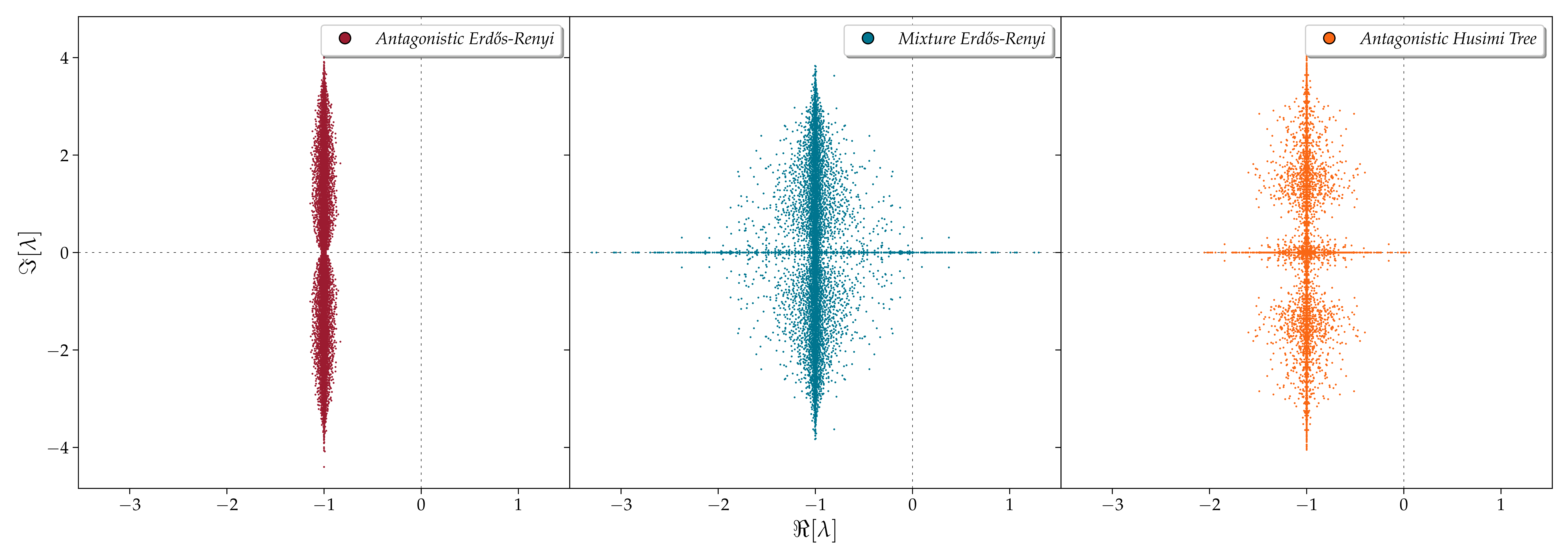}
	\caption[]{Spectra of  shifted interaction matrices $\mathbf{B}_{\rm Id}$  for antagonistic Erd\H{o}s-R\'{e}nyi (red, left), mixture Erd\H{o}s-R\'{e}nyi (blue, center) and antagonistic pure Husimi tree (orange, right).   Markers denote the eigenvalues of a single matrix $\mathbf{B}_{\rm Id}$ of size $N=8000$.   The  parameters used are those detailed  in Sec.~\ref{sec:numerical_details}.}
	\label{fig:interaction_spectra}
\end{figure}

So far, we have considered how strong local sign stability affects the leading eigenvalue of $\mathbf{B}_{\rm Id}$. Instead now, we investigate the effect of strong local sign stability on the full spectra of matrices $\mathbf{B}_{\rm Id}$, which are plotted in Fig.~\ref{fig:interaction_spectra}. 

Figure~\ref{fig:interaction_spectra} shows a qualitative difference between, on one hand, the spectra of antagonistic Erd\H{o}s-R\'{e}nyi graphs (red), and  on the other hand, the spectra of  mixture Erd\H{o}s-R\'{e}nyi graphs (blue) and antagonistic Husimi trees (orange). Indeed, in the latter two cases the spectrum develops long tails on the real axis, while in the former case the tails are absent.

Analysing how the spectra evolve as a function of the matrix size $N$, we have found that the  tails on the real axis   elongate  as the matrix size increases, populating larger and larger portions of the real axis (results not shown), which is in agreement with the results on the divergence of the leading eigenvalue in Fig.~\ref{fig:leading_eig_interaction}. On the other hand, the antagonistic Erd\H{o}s-R\'{e}nyi graph has a spectrum that remains confined in a part of the complex plane that has finite width along the real axis, even when the matrix size increases.   

Focusing on the imaginary parts of the spectra, we have found  that for all three ensembles under study  the spectra grow vertically as a function of $N$, covering an ever larger portion of the imaginary axis.   Notice that the latter result  is naively expected as   the matrix norm  diverges as a function of $N$, and,    since $\|\mathbf{M}\| \geq |\lambda_i(\mathbf{M})| \ \ \forall \, i$, there is no simple reason why the eigenvalue should be confined within a finite portion of the complex plane.     Hence for the antagonistic, Erd\H{o}s-R\'{e}nyi ensemble  the divergence of the norm materialises exclusively into the growth of the tails of the spectrum parallel to the imaginary axis.

In  light of what we outlined in Sec. \ref{sec:stability_and_model}, the observed qualitative difference  in the width of the spectrum on the real axis between ensembles that are strongly locally sign stable and those that are not, indicates that strong local sign stability may be an important characteristic of stability in  ecological models. In fact, both structural stability and feasibility require that  the origin of the complex plane is not part of the spectrum of interaction-like matrices, and  therefore, are not compatible with spectra that exhibit  tails covering the whole real axis. On the contrary, the spectrum of antagonistic Erd\H{o}s-R\'{e}nyi graphs contains a finite portion of the real axis and, therefore, as explained in  Sec.~\ref{sec:stability_and_model}, the   origin of the complex plane can be excluded from the spectrum after a finite shift of  the diagonal entries leading to absolutely stable models of ecosystems.  

Another  interesting feature that  we observe in Fig.~\ref{fig:interaction_spectra} is  a, so-called, \emph{reentrance effect} in the spectrum of   antagonistic Erd\H{o}s-R\'{e}nyi  graphs.    The reentrance effect implies that the width of the spectrum is small for eigenvalues with $\Im[\lambda]\approx 0$.  Increasing  $\Im[\lambda]$, the width of the spectrum increases until it reaches a maximum at $\Im[\lambda]=\Im[\lambda_1]$, after which the width of the spectrum decreases again to vanish at large values of $\Im[\lambda]$.   As a consequence,  the leading eigenvalue of antagonistic Erd\H{o}s-R\'{e}nyi  has typically a finite imaginary part, i.e., $\Im[\lambda_1] \neq 0$, and hence the leading eigenvalue comes in pairs with its complex conjugate. This reentrance effect, which was already observed  in Ref.~\cite{mambuca2022dynamical}, will be  discussed in-depth  in Sec.~\ref{sec:reentrances}. 

\subsection{Interaction-like matrices $\bf{B}$} \label{subsec:lss_interaction}
 
In the present section, we confirm the validity of Eq.~\eqref{eq:conjecture} also for interaction matrices with fluctuating (negative) diagonal entries, i.e., $\bf{B} = {\bf A}-{\bf D}$, with $D_i$s drawn independently from a distribution $p_D$ supported on a subset of the positive real axis.
%Therefore, to do not break the condition of local sign stability, the $D_i$s are drawn independently from a distribution $p_D$ supported on a subset of the positive real axis.
In particular we choose for simplicity a uniform distribution  $p_D(d)$ supported on $[d_{\rm min}, d_{\rm max}]$, with $d_{\rm min}>0$, even though the main results we obtain 
for the leading eigenvalue also holds for more general distributions $p_D$ as long as it is supported on $\mathbb{R}^+$.

\begin{figure}[H]
	\centering
	%\captionsetup{justification=centering}
	\includegraphics[width=0.7\textwidth]{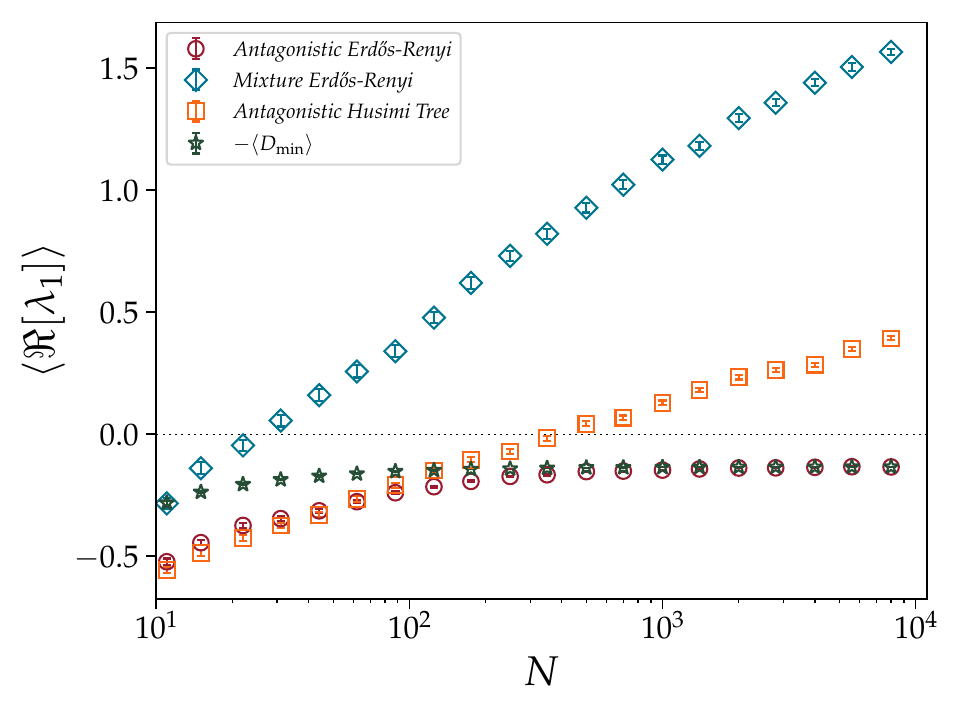}
	\caption[]{Average real part of the leading eigenvalue, $\langle \Re[\lambda_1]\rangle$, as a function of the matrix size $N$ for interaction-like matrices $\mathbf{B}$.   Results shown are   for antagonistic Erd\H{o}s-R\'{e}nyi graphs (red circles), mixture Erd\H{o}s-R\'{e}nyi graphs (blue diamonds) and antagonistic pure Husimi trees (orange squares). The green stars are the averages of the diagonal elements minima. Markers are sample means obtained from numerically diagonalising $300$ matrix realisations, and error bars denote the error on the mean.    The  parameters used for the matrix ensembles  are detailed in Sec.~\ref{sec:numerical_details}.}
	\label{fig:leading_eig_interaction_with_diagonal}
\end{figure}

Figure \ref{fig:leading_eig_interaction_with_diagonal} plots $\langle\Re[\lambda_1(\mathbf{B})]\rangle$ as a function of the matrix size $N$ in the three cases considered before in Fig.~\ref{fig:leading_eig_interaction}, albeit now with a distribution $p_D$ that has a nonzero variance.  The results of Fig.~\ref{fig:leading_eig_interaction_with_diagonal} are in correspondence with those  of Fig.~\ref{fig:leading_eig_interaction}, further establishing the connection between strong local sign stability and the  asymptotic finiteness of the leading eigenvalue.   Indeed,  for mixture,  Erd\H{o}s-R\'{e}nyi graphs  and antagonistic, Husimi tree  graphs  the real part of the leading eigenvalue is steadily growing with $N$, whereas for antagonistic, Erd\H{o}s-R\'{e}nyi graphs it converges to a finite value (as a function of $N$). 

A more detailed look at Fig.~\ref{fig:leading_eig_interaction_with_diagonal} reveals that for small values of $N<10^2$, the average leading eigenvalue, $\langle \Re[\lambda_1]\rangle$,   of the antagonistic Erd\H{o}s-R\'{e}nyi graph increases as a function of $N$, before it eventually saturates at its asymptotic value for $N\gtrsim 10^2$. The transient behaviour of $\langle\Re[\lambda_1(\mathbf{B})]\rangle$ at small values of  $N$ is different from the immediate convergence of $\langle\Re[\lambda_1(\mathbf{B})]\rangle$ in  Fig.~\ref{fig:leading_eig_interaction}. Moreover, according to Fig.~\ref{fig:leading_eig_interaction_with_diagonal}   the  asymptotic value is approximately equal to $-d_{\rm min}$, the largest possible value of the diagonal entries; notice that to consider finite size effects,  Fig.~\ref{fig:leading_eig_interaction_with_diagonal} shows in fact  $\langle -D_{\rm min}\rangle$ (green stars), with 
\begin{equation}
D_{\rm min} = {\rm min}_{j\in \left\{1,2,\ldots,N\right\}}D_j. \label{eq:Dmin}
\end{equation}  
This result is reminiscent of a related result for antagonistic tree graphs with fluctuating diagonal entries, which states that the leading eigenvalue of a interaction matrix associated with an antagonistic, tree graph is smaller or equal than $-D_{\rm min}$, see~\ref{sec:C3}.
Fig.~\ref{fig:leading_eig_interaction_with_diagonal} shows  that the same principle applies for antagonistic, Erd\H{o}s-R\'{e}nyi graphs, and moreover, for the specific parameters chosen it holds that the leading eigenvalue of a large antagonistic Erd\H{o}s-R\'{e}nyi graph is  approximately  equal to the largest possible diagonal element $-d_{\rm min}$.      

Note that $-d_{\rm min}$ does not always determine the leading eigenvalue of antagonistic Erd\H{o}s-R\'{e}nyi graphs.  For example, let us consider the limiting case of a trivial diagonal with no disorder, i.e., $p_D(d) = \delta(d-1)$, as  discussed in the previous Sec.~\ref{subsec:lss_shifted}. The results of  Fig.\ref{fig:leading_eig_interaction} show that $\langle \Re[\lambda_1]\rangle$ is larger than $-d_{\rm min} = -1$, and hence its value is not directly related to $d_{\rm min}$.    Thus, depending on the model parameters, the asymptotic behaviour of the leading eigenvalue of antagonistic Erd\H{o}s-R\'{e}nyi  graphs is either set to $-d_{\rm min}$ or it is determined by a complex interplay of various parameters.  

In Sec.~\ref{sec:reentrances}, we will study the transition between these two regimes in more detail.   Nevertheless, we emphasize that in both cases the real part of the leading eigenvalue converges to a finite value, marking a qualitative difference with respect to mixture Erd\H{o}s-R\'{e}nyi graphs and antagonistic Husimi trees.

At variance with shifted interaction matrices, we do not have theoretical results valid at infinite $N$ to confirm that the numerical results for the antagonistic case converge to a finite value. However, note that the real part of the leading eigenvalue for the largest $N$s observed converges to the upper boundary $-d_{\rm min}$ of the distribution of the elements on the diagonal (see in  Fig.~\ref{fig:leading_eig_interaction_with_diagonal} the trend of the average $-D_{\rm min}$ for comparison), strongly suggesting that the right boundary of the spectrum on the real axis in the antagonistic case is simply determined by the disorder on the diagonal and therefore by definition does not diverge with $N$.
% We stress again that  the numerical results for the antagonistic case can safely be considered representative of the large $N$ limit, as for the shifted interaction matrices they are consistent with theoretical results at infinite $N$ in Ref.~\cite{mambuca2022dynamical}. For the interaction-like matrices a small leap of faith is still needed but the coincidence of the real part of the leading eigenvalue for the largest $N$s observed with the upper boundary $-d_{\rm min}$ of the distribution of the elements on the diagonal strongly suggests that the right boundary of the spectrum on the real axis in the antagonistic case is simply determined by the disorder on the diagonal and therefore by definition does not diverge with $N$.

\begin{figure}[H]
	\centering
	%\captionsetup{justification=centering}
	\includegraphics[width=1.0\textwidth]{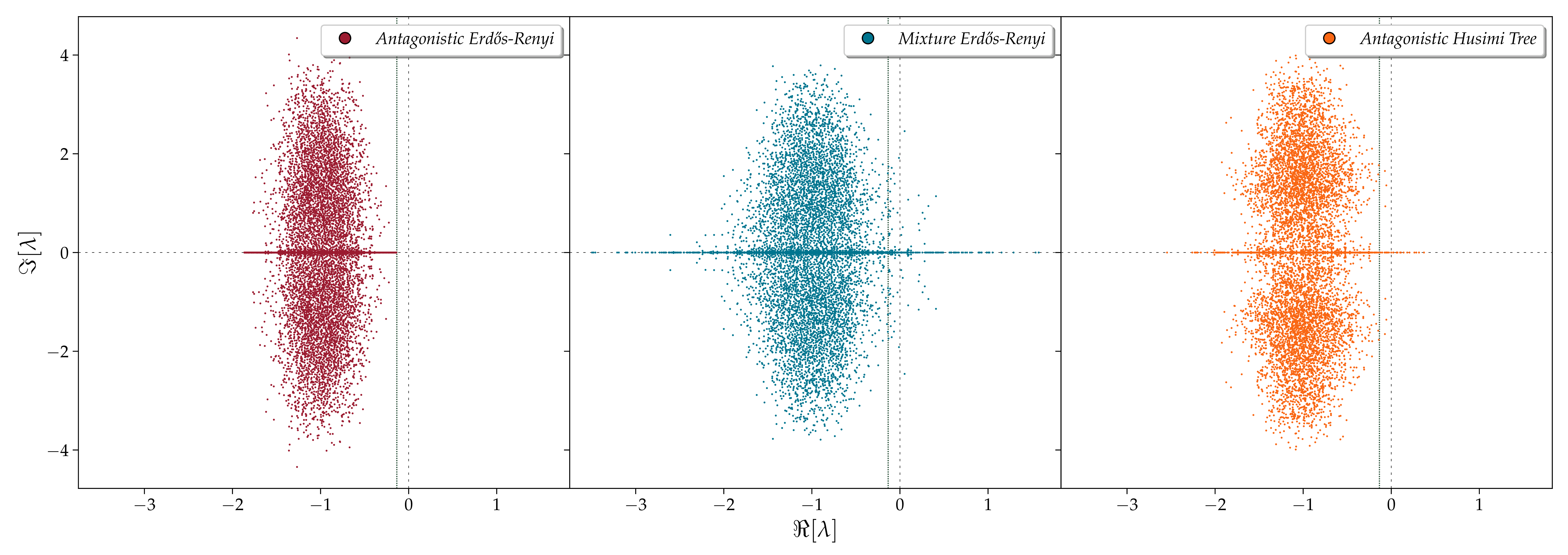}
	\caption[]{ Spectra of  Interaction-like  matrices $\mathbf{B}$  for antagonistic Erd\H{o}s-R\'{e}nyi (red, left), mixture Erd\H{o}s-R\'{e}nyi (blue, center) and antagonistic pure Husimi tree (orange, right).   Markers denote the eigenvalues of a single matrix $\mathbf{B}_{\rm Id}$ of size $N=8000$.  The green dotted line displays $-D_{\rm min}$, the maximum diagonal entry of $\mathbb{B}$.   The  parameters used are those detailed  in Sec.~\ref{sec:numerical_details}.}
	\label{fig:interaction_with_diagonal_spectra}
\end{figure}

Figure \ref{fig:interaction_with_diagonal_spectra} shows the full spectra of the matrices considered in Fig.~\ref{fig:leading_eig_interaction_with_diagonal}.     Comparing the spectra in Fig.~\ref{fig:interaction_with_diagonal_spectra} with those in Fig.~\ref{fig:interaction_spectra}, we observe again tails of eigenvalues on the real axis for the mixture Erd\H{o}s-R\'{e}nyi  and antagonistic $(4,4)$-pure Husimi tree  ensemble.    Note that in Fig.~\ref{fig:interaction_with_diagonal_spectra} we also observe a segment on the real axis in the spectrum of the  antagonistic Erd\H{o}s-R\'{e}nyi graph.   However, in the latter case,  the segment of eigenvalues  does not grow indefinitely as a function of $N$, and instead  it is confined to the interval $[-d_{\rm max},-d_{\rm min}]$, in agreement with the results in Fig.~\ref{fig:leading_eig_interaction_with_diagonal}.   Therefore, also for interaction-like matrices, strong LSS yields a finite segment  of eigenvalues on the real axis\footnote{Note that in general if the distribution $p_D$ is unbounded on $\mathbb{R}^+$, there will not be a finite segment of eigenvalues on the real axis, but the real part of the eigenvalues will still have a finite upper bound.} with direct consequences on the possibility to use strong LSS to predict structural stability and feasibility of models of ecosystems, irrespectively from system size, {\it i.e.},~absolute stability.  
%The tails on the real axis  show that also for disordered diagonal entries, strong local sign stability is related to structural stability and feasibility and consequently strong local sign stability is expected to be relevant for the stability of ecosystems.  

Another difference between Figs.~\ref{fig:interaction_spectra} and \ref{fig:interaction_with_diagonal_spectra} is that in the latter we do not observe  a reentrance effect in the spectrum of the antagonistic Erd\H{o}s-R\'{e}nyi graph.   We stress however that this is due to the choice of model parameters, and that in general interaction-like matrices can also exhibit reentrance effects.     In  Sec. \ref{sec:reentrances}, we will discuss in detail how reentrance effects appear in models with diagonal disorder, and how they depend on the model parameters.   

\subsection{Jacobian-like matrices $\mathbf{J}$} \label{subsec:lss_jacobian}
Lastly, we investigate the validity of Eq.(\ref{eq:conjecture}), on Jacobian-like matrices,  ${\bf J}=\mathbf{D}{\bf B_{\rm Id}}=\mathbf{D} \left( {\bf A}-\mathbf{1} \right)$,  that have a  distinctive stripy structure and negative diagonal, which is relevant for linear stability analysis in ecology as explained in Sec.~\ref{sec:stability_and_model}. In order to have a negative diagonal, we extract the entries of the diagonal matrix $\mathbf{D}$ from a uniform distribution $p_D(d)$ supported on $[d_{\rm min}, d_{\rm max}]$, with $d_{\rm min}>0$.      Further details on the various parameters can be found in Sec.~\ref{sec:numerical_details}.

\begin{figure}[H]
	\centering
	%\captionsetup{justification=centering}
	\includegraphics[width=0.7\textwidth]{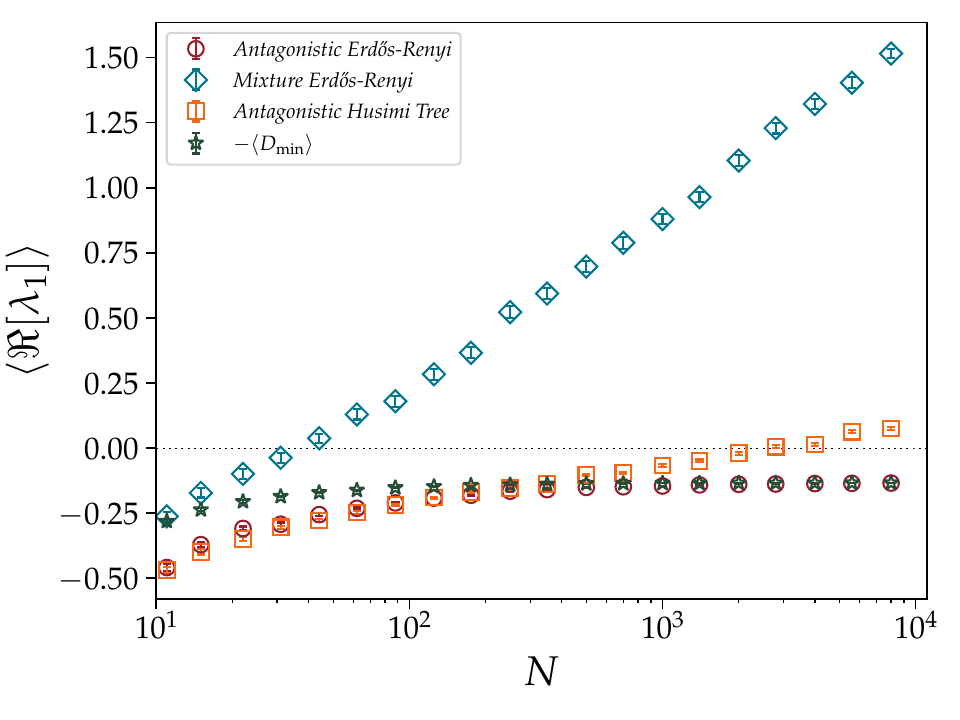}
	\caption[]{Average real part of the leading eigenvalue, $\langle \Re[\lambda_1]\rangle$, as a function of the matrix size $N$ for Jacobian-like matrices $\mathbf{J}$.   Results shown are   for antagonistic Erd\H{o}s-R\'{e}nyi graphs (red circles), mixture Erd\H{o}s-R\'{e}nyi graphs (blue diamonds) and antagonistic pure Husimi trees (orange squares).     The green stars are the averages of the largest element of the diagonal matrix $-D_{\rm min}$. Markers are sample means obtained from numerically diagonalising $300$ matrix realisations, and error bars denote the error on the mean.    The  parameters used for the matrix ensembles  are detailed in Sec.~\ref{sec:numerical_details}. }
	\label{fig:leading_eig_jacobian}  
\end{figure}

Figure~\ref{fig:leading_eig_jacobian} depicts the average,   real part of the leading eigenvalue,  $\langle \Re[\lambda_1(\mathbf{J})]\rangle$,  as a function of the matrix size $N$ in the three canonical models of interest, mirroring the analysis in Figs.~\ref{fig:leading_eig_interaction} and \ref{fig:leading_eig_interaction_with_diagonal}.    The numerical results confirm the connection between strong LSS and finiteness of leading eigenvalue, viz.,  $\langle\Re[\lambda_1(\mathbf{J})]\rangle$    rapidly converges to a finite value for antagonistic, Erd\H{o}s-R\'{e}nyi graphs, while   $\langle\Re[\lambda_1(\mathbf{J})]\rangle$ diverges as a function of  $N$ for  mixture, Erd\H{o}s-R\'{e}nyi graphs  and antagonistic, Husimi trees.    
In addition, in agreement with the results in Fig.~\ref{fig:leading_eig_interaction_with_diagonal},  Fig.~\ref{fig:leading_eig_jacobian} shows that for  antagonistic Erd\H{o}s-R\'{e}nyi graphs   the average leading eigenvalue saturates at a value that is approximately equal to $-D_{\rm min}$ (after a transient regime for small values of $N$).         For antagonistic tree graphs, \ref{sec:C3}   shows that the leading eigenvalue is smaller or equal than $-D_{\rm min}$. Fig.~\ref{fig:leading_eig_jacobian} shows that this principle also applies to Jacobian-like matrices defined on antagonistic Erd\H{o}s-R\'{e}nyi graphs.   
Taken together, as discussed previously for the interaction like matrices, for the model under study the right boundary of the spectrum on the real axis in the antagonistic case is  determined by the disorder on the diagonal and therefore by definition does not diverge with $N$. 

In~\ref{app:HusimiPlateau} we refine the results of Fig.~\ref{fig:leading_eig_jacobian}  for   $\langle\Re[\lambda_1(\mathbf{J})]\rangle$ as a function of $N$, by considering the  limit  $d_{\rm min}\rightarrow 0$.   In this limit, the antagonistic, Husimi tree exhibits strong transient effects, which we call the Husimi plateau. Nevertheless the results remain consistent with Fig.~\ref{fig:leading_eig_jacobian} as they eventually show a growth of $\langle\Re[\lambda_1(\mathbf{J})]\rangle$ with $N$. 
Note however that we do not expect that the conclusion about the finiteness of the boundary of the spectrum on the real axis at large $N$ in the antagonistic Erd\H{o}s-R\'{e}nyi case could be just the result of a finite size effect in correspondence of a long Husimi-like plateau, because the Husimi plateau only forms for fine-tuned values of $d_{\rm min}$, while in the antagonistic Erd\H{o}s-R\'{e}nyi case the leading eigenvalue seems to always converge to a finite value as we could not observe any divergence for any choice of the $d_{\rm min}$ studied.

The results in Fig.~\ref{fig:leading_eig_jacobian} have interesting implications  for the linear stability of ecosystems. Recalling the classical linear stability condition $\Re[\lambda_1(\mathbf{J})] < 0$,  the results in Fig.~\ref{fig:leading_eig_jacobian} imply that system size $N$ is not an important parameter for the linear stability of systems defined on strongly locally sign stable graphs, which therefore are absolutely stable.  
On the other hand, system size $N$ is an important parameter in the general case of models defined on graphs with sign-symmetric interactions or with a number of short cycles growing with $N$, as in the latter cases stability is  only attained for small enough values of $N$. 

\begin{figure}[H]
	\centering
	%\captionsetup{justification=centering}
	\includegraphics[width=1.0\textwidth]{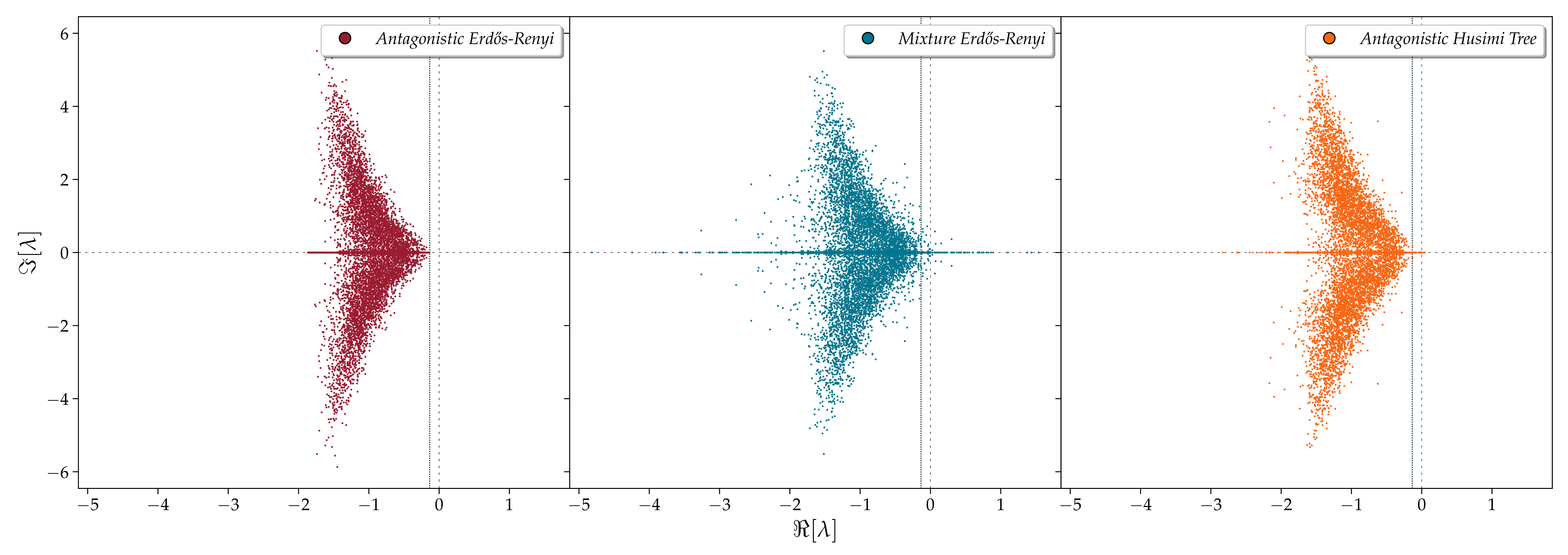}
	\caption[]{Spectra of a single realisation of Jacobian-like matrix $\mathbf{J}$ with size $N=8000$ for antagonistic Erd\H{o}s-R\'{e}nyi (red, left), mixture Erd\H{o}s-R\'{e}nyi (blue, center) and antagonistic pure Husimi tree (orange, right). The green dotted line displays $-D_{\rm min}$, the largest element of the diagonal matrix. The other parameters employed are discussed in Sec.~\ref{sec:numerical_details}.}
	\label{fig:jacobian_spectra}
\end{figure}

Figure \ref{fig:jacobian_spectra} shows the spectra of Jacobian-like matrices for the three canonical models under study.     The Jacobian-like spectra  have an arrow-like shape, which  resembles those already observed for the dense version of Jacobian-like matrices \cite{gibbs2018effect, stone2018feasibility}.  However, importantly in the sparse case a clear distinction is observed between,  on one hand, antagonistic Erd\H{o}s-R\'{e}nyi graphs, and  on the other hand,  mixture Erd\H{o}s-R\'{e}nyi graphs and antagonistic Husimi trees.    The latter two exhibit long tails on the real axis that increase with system size, while the former does not exhibit such tails.     Hence again, the divergence of the real part of the leading eigenvalue is due to tails  that develop on the real axis of the spectra of sparse random graphs, and     such tails are absent in strongly locally sign stable ensembles.    

Note that the spectrum of the   antagonistic, Erd\H{o}s-R\'{e}nyi  graph  in  the left panel of Fig.~\ref{fig:jacobian_spectra} does not exhibit a reentrance effect, similar to the spectrum of the interaction-like matrix in   the left panel of Fig.~\ref{fig:interaction_with_diagonal_spectra},  but different from the spectrum of the shifted interaction matrix in the left panel of Fig.~\ref{fig:interaction_spectra}.   We stress  that this is due to the choice of model parameters, and in fact Jacobian-like matrices can also exhibit reentrance effects.  In the next section we will investigate  how  reentrance in the spectra of Jacobian-like matrices is governed by an interplay between diagonal disorder and network structure.

\section{Discontinuous transition in the imaginary part of the leading eigenvalue} \label{sec:reentrances}

As shown in the left Panel of Fig.~\ref{fig:interaction_spectra}, the boundary of the spectrum of antagonistic Erd\H{o}s-R\'{e}nyi graphs exhibits a reentrance in correspondence of the real axis.  In this case, the leading eigenvalue comes in a pair of  complex conjugate values  with finite imaginary part.    From a dynamical systems point of view, the reentrance effect is interesting, as the imaginary part of the leading eigenvalue determines the frequency of oscillations of the slowest mode of relaxation towards the fixed point.   Hence,  when the leading eigenvalue has a nonzero imaginary part, then the leading, relaxation mode is oscillatory, while for leading eigenvalues that are real the leading mode is nonoscillatory. As a consequence, a transition from a phase in which the leading eigenvalue comes in a pair of two conjugate, complex values   to a phase in which it is real corresponds, from a dynamical perspective, to a transition from an oscillatory   to a nonoscillatory relaxation dynamics. For this reason we call it a {\it dynamical transition}.  

Note that in general reentrance effects can be interesting  both for  interaction-like matrices and Jacobian-like matrices. 
In the latter, they identify oscillatory dynamics in nonlinear systems of the generalised Lotka-Volterra type in the vicinity of a fixed point. In the former, they  may simply identify oscillatory dynamics of corresponding linear systems. 
However, for the generalised Lotka-Volterra model discussed in this paper, the dynamical behaviour is partially accessible by studying the Jacobian-like matrix only, and the reentrance effect visible in the spectrum of interaction-like matrix does not have any dynamical consequences. On the other hand, the emergence of a reentrance in the spectrum of the interaction-like matrix for the generalised Lotka-Volterra model can still represent an interesting piece of information in situations where the system is feasible/structurally stable, because the spectrum does not include the origin, although (complex) eigenvalues happen to have real parts of both positive and negative sign. In such cases the sign of the leading eigenvalue cannot directly determine structural stability or feasibility, as otherwise naively expected.

Antagonistic Erd\H{o}s-R\'{e}nyi graphs with constant diagonal entries, i.e., shifted interaction matrices,   exhibit a dynamical transition as a function of the mean degree $c$, as shown in Ref.~\cite{mambuca2022dynamical}.      Indeed, for large $c$ the boundary of the spectrum  resembles the elliptic law, and thereby the leading eigenvalue is typically real in the large size limit. On the other hand, for small $c$ the boundary of the spectrum shows a reentrance effect,  as shown in the left Panel of Fig.~\ref{fig:interaction_spectra}, and the leading eigenvalue is typically complex.    In addition, Ref.~\cite{mambuca2022dynamical} shows that the dynamical transition is a continuous transition in the following sense:  the imaginary part $\Im[\lambda^\star_1]$ of the typical value $\lambda^\star_1$ of the leading eigenvalue, for which we provide a mathematical definition later,  equals zero at the transition, and therefore the frequency of the oscillations near the transition is small.

In the present Section, we also study the dynamical transition in the leading eigenvalue, but now for  the  interaction-like and Jacobian-like matrices with sign-antisymmetric weights. 
Note that in   Figs.~\ref{fig:interaction_with_diagonal_spectra} and \ref{fig:jacobian_spectra} we have not observed  reentrance effects for interaction-like  and  Jacobian-like  matrices, and consequently their leading eigenvalue is  real.   Instead,  in Fig.~\ref{fig:interaction_spectra} we have observed a   reentrance in  the spectra of  shifted interaction matrices  with sign-antisymmetric weights, and consequently in this case the leading eigenvalue comes in a pair of two conjugate, nonreal eigenvalues. 
Since the latter random matrix ensemble can be obtained from the former ensembles in the limit of small $\sigma_D$, a transition, possibly related to the reentrance of the spectra in correspondence of the real axis, is to be expected at intermediate  values of $\sigma_D$. 

Interestingly, the dynamical transition we find in this section for interaction-like and Jacobian-like matrices occurs while the spectrum is still reentrant in correspondence of the real axis,  and hence it features a  discontinuous jump in the imaginary part $\Im[\lambda^\star_1]$ of the leading eigenvalue, at variance with the nature of the  transition studied in  Ref.~\cite{mambuca2022dynamical}. 
This Section is devoted to a careful study of this transition. In particular, we further discuss the definition of a suitable control parameter for the transition, both in the interaction-like and Jacobian-like matrices, involving $\sigma_D$ and other relevant parameters of the model. For both cases, in Sec.~\ref{sec:locTrans} we locate the transition point through a finite size scaling  analysis and in Sec.~\ref{sec:nature}, by some  additional finite size scaling studies   reported in~\ref{app:finite_size_scaling}, we also show that the transition takes place with a discontinuous jump. 

\subsection{Locating  the  transition point}
\label{sec:locTrans}

As anticipated by the results of the previous section, a change in the strength of the diagonal disorder, as quantified by  $\sigma_D$, can have a direct impact on the imaginary part of the leading eigenvalue of interaction-like and Jacobian-like matrices, determining a qualitative change in the relaxation  dynamics 
of a corresponding dynamical system. 

Concerning other model's parameter, in the interaction-like case, $\mu_D$ is only responsible of a global shift of the spectrum and cannot affect the imaginary part of the leading eigenvalue. 
A global rescaling of all matrix elements also affects trivially the spectrum. 
First non trivial changes in the spectrum emerge when $\sigma_D$ changes with respect to the scale of the off-diagonal elements, represented by their variance $\sigma$, and therefore the relevant control parameter for the dynamical transition must be $\sigma_D/\sigma$\footnote{Additional non trivial modifications of the spectra are introduced by changes in the relative importance of $\mu_G$ and $\sigma_G$ for fixed $\sigma$, as well as in general every change in the distribution of the diagonal and off-diagonal elements. Conversely, here we consider cases in which changes in $\sigma$ are only obtained by changing both $\mu_G$ and $\sigma_G$, with $\mu_G/\sigma_G$ fixed.}.

For Jacobian-like matrices the situation is more involved as their off-diagonal elements are the result of the product of pairs of random variables $D_i, A_{ij}$, with probability distribution $p_D$ and $p$. 
In this case, a global rescaling of the $D_i$ gives also a trivial global rescaling of all the elements of the matrix.
Any other modification of the parameters of $p_D$ and $p$ induces a non trivial modification of the probability distribution of the off-diagonal elements, which cannot be exactly recast in terms of the variation of a simple control parameter.
However, it is possible to derive a rough estimate of the scale of the off-diagonal elements via their variance $\sigma_{\rm O}=\sigma\sqrt{\mu_D^2+\sigma_D^2}$, which suggests that the relevant control parameter is approximately given by the ratio $\sigma_D/\sigma_{\rm O}=\sigma^{-1}/\sqrt{1+\mu_D^2/\sigma_D^2}$.

\begin{figure}[h]
	\centering
	\begin{subfigure}[b]{0.48\textwidth}
		\centering
		\includegraphics[width=\textwidth]{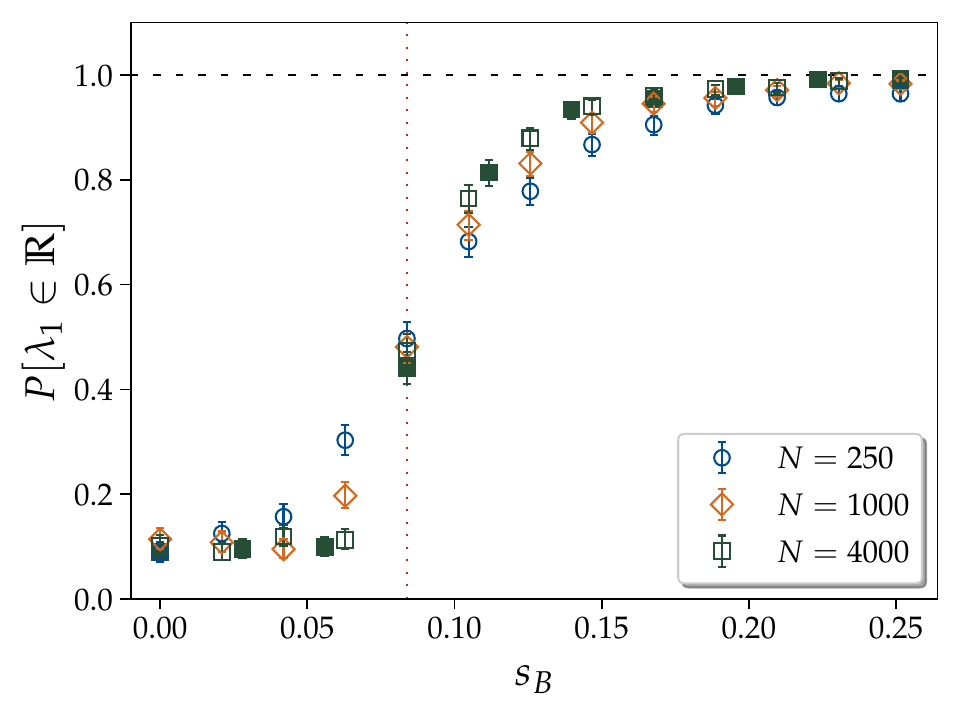}
        \caption{Interaction-like $\mathbf{B}$}
	\end{subfigure}
	\hfill
	\begin{subfigure}[b]{0.48\textwidth}
		\centering
		\includegraphics[width=\textwidth]{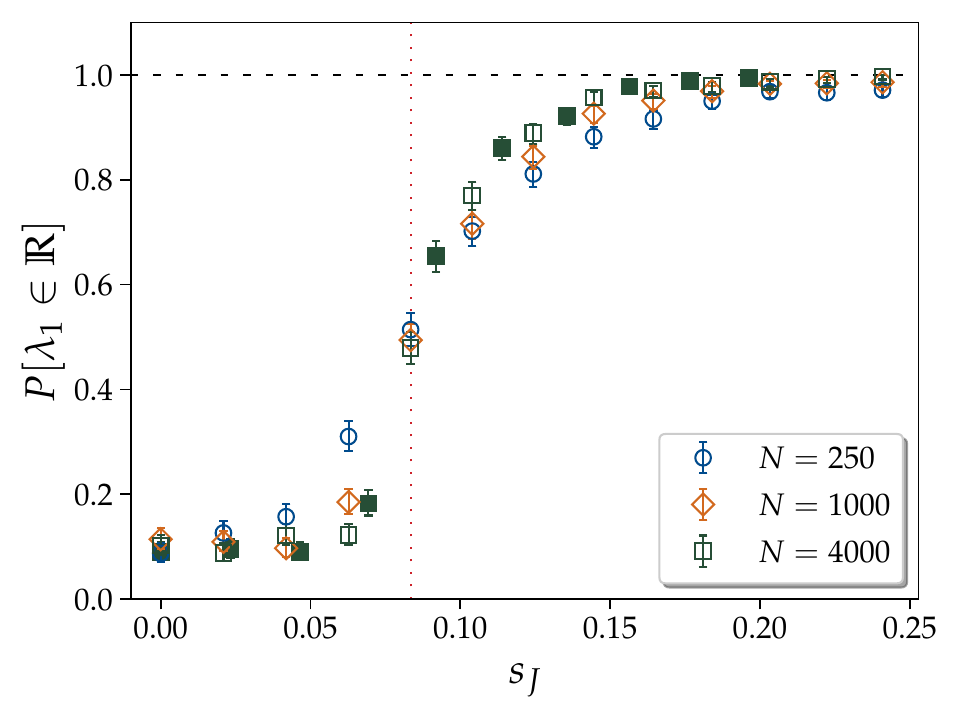}
		\caption{Jacobian-like $\mathbf{J}$}
	\end{subfigure}
	\caption[]{Probability $P[\lambda_1 \in \mathbb{R}]$ that the leading eigenvalue is real as a function of the control parameter, defined in Eq.~(\ref{eq:control_parameters}), for  interaction-like $\mathbf{B}$ (on the left) and Jacobian-like $\mathbf{J}$ (on the right) matrices of antagonistic  Erd\"{o}s-R\'{e}nyi graphs. The vertical dotted line shows the value $s_d \simeq 0.08$ at which  $P[\lambda_1 \in \mathbb{R}]\simeq 0.5$ and therefore indicates the transition from the complex phase to the real phase. Averages are performed over  $1000$ matrix realisations. The error bars depict the Wilson score interval at 95\% confidence, which is an asymmetric interval that does not exhibit  overshoot  phenomena and does not underestimate the uncertainty around the endpoints, see Refs.~\cite{wilson1927probable, Agresti1998ApproximateIB, brown2001interval}. The empty markers are obtained  by employing the parameters discussed in Sec.~\ref{sec:numerical_details} and for given values of $N$, and the filled ones are obtained for parameters  $\mu_G = 1.5$, $\sigma_G=0.9$, $\mu_D = 1.2$, ten values of $\sigma_D$ equally spaced in $[0, 0.45]$, and  for  $N=4000$.}
	\label{fig:reentrances_probability}
\end{figure}

We analyse in Fig.~\ref{fig:reentrances_probability} the probability  $P[\lambda_1 \in \mathbb{R}]$ that the  leading eigenvalue is real  as a function of the control parameters 
\begin{equation} \label{eq:control_parameters}
s_B :=\sigma_D/\sigma \quad  {\rm and} \quad s_{J}:=\sigma_D/\sigma_{\rm O},
\end{equation}
defined respectively for interaction-like and Jacobian-like matrices of antagonistic Erd\H{o}s-R\'{e}nyi graphs.   For small values of $s_B$ and $s_J$, $P[\lambda_1 \in \mathbb{R}]\approx 0.1$, and hence with high probability the leading eigenvalue has a nonzero imaginary part.
On the other hand, for large values of $s_B$ and $s_J$,    $P[\lambda_1 \in \mathbb{R}]\approx 1$, and hence the leading eigenvalue is typically real.
Because of the aforementioned  dynamical significance of $\Im[\lambda_1]$, we call the former the \emph{oscillatory phase} and the latter the  \emph{nonoscillatory phase}.

To show that the transition from an oscillatory to a nonoscillatory phase is a proper phase transition, we also plot in  Fig.~\ref{fig:reentrances_probability} the probability  $P[\lambda_1 \in \mathbb{R}]$ for different system sizes $N$.   Notably, the  transition becomes sharper as the system size increases, and the curves for different values of $N$ intersect at one single point, which we denote by $s_{\rm d}$, indicating where the {\it dynamical transition} takes place.
We also show results obtained at different values of $\sigma, \sigma_D, \mu_D$, which all collapse when plotted as function of $s_B$ and $s_J$. In particular, this evidence confirms that $s_J$ can be effectively used as relevant control parameter for the transition\footnote{Be warned that for Jacobian-like matrix, $s_J$ can be safely considered the  relevant control parameter %irrespectively of the choices of model parameters,
only as long as the modifications in the model parameters $\sigma, \mu_D, \sigma_D$ do not give rise to a significant change in the shape of the distribution of the off-diagonal elements when the transition takes place, as it is the case for the two series of data for the largest system size in the right panel of Fig.~\ref{fig:reentrances_probability}.}.

Note that in the oscillatory phase   $P[\lambda_1 \in \mathbb{R}]\approx 0.1$ for large values of $N$, and hence  there is a small nonzero probability that the leading eigenvalue is real.   Hence, the imaginary part of the leading eigenvalue is not a self-averaging quantity (i.e., it does not converge to a deterministic number in the infinite size limit), which is in correspondence with the numerical results from Ref.~\cite{mambuca2022dynamical}. 

\subsection{Characterising the discontinuity of the transition}\label{sec:nature}

In this Section, we determine whether the dynamical transition occurs with a  continuous or discontinuous variation of the imaginary part of the leading eigenvalue.   

\begin{figure}[h]
	\centering
	\begin{subfigure}[b]{0.48\textwidth}
		\centering
		\includegraphics[width=\textwidth]{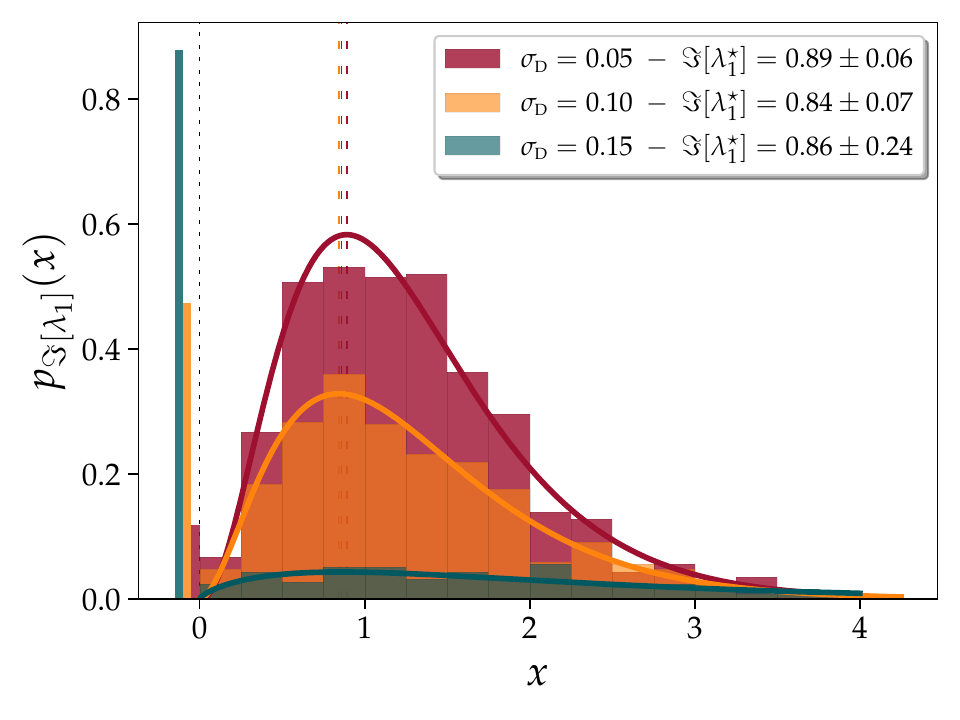}
        \caption{Interaction-like $\mathbf{B}$}
	\end{subfigure}
	\hfill
	\begin{subfigure}[b]{0.48\textwidth}
		\centering
		\includegraphics[width=\textwidth]{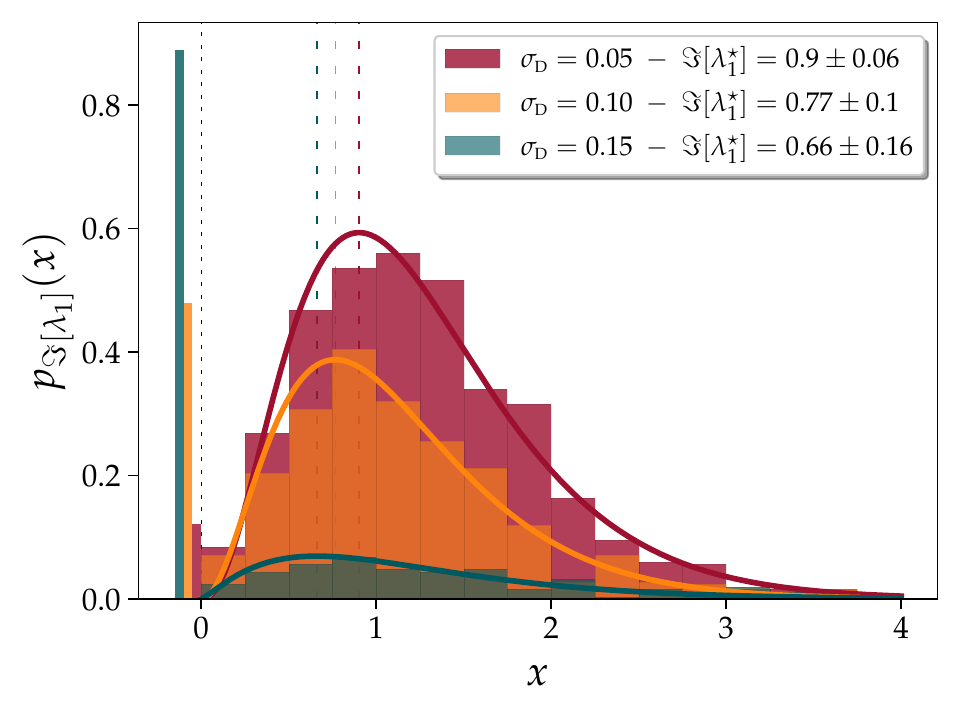}
		\caption{Jacobian-like $\mathbf{J}$}
	\end{subfigure}
	\caption[]{Histograms of the  imaginary part  of the leading eigenvalues of interaction-like $\mathbf{B}$ (on the left) and Jacobian-like $\mathbf{J}$ (on the right) matrices of antagonistic  Erd\"{o}s-R\'{e}nyi graphs for $N=4000$ and different values of the strength of the diagonal disorder: $\sigma_D = 0.05$ (corresponding to $s_B \simeq 0.04$ and $s_J \simeq 0.04$) in red, $\sigma_D = 0.10$ (corresponding to $s_B \simeq 0.08$ and $s_J \simeq 0.08$) in orange, and $\sigma_D = 0.15$ (corresponding to $s_B \simeq 0.13$ and $s_J \simeq 0.12$) in green;  the other parameters employed are as discussed in Sec.~\ref{sec:numerical_details}.  The solid lines are  $\gamma$-distributions fitted to the histograms (see main text for detailed explanations), and the dashed vertical lines denote the   mode of these fitted distributions, which provide the estimates for $\Im[\lambda_1^{\star}]$. The delta peaks in zero denote the probability $P[\lambda_1\in\mathbb{R}]$ that an eigenvalue is real.}
	\label{fig:reentrances_hist}
\end{figure}

Figure~\ref{fig:reentrances_hist}  displays the distribution $p_{\Im[\lambda_1]}(x)$ of the imaginary parts of the leading eigenvalue  obtained by generating a large sample of  interaction-like matrices $\mathbf{B}$ (left Panel)  and Jacobian-like matrices  $\mathbf{J}$ (right Panel);  notice that for each pair of conjugate leading eigenvalues with nonzero imaginary part we only focus on the one with  $\Im[\lambda_1]>0$. 
From Fig.~\ref{fig:reentrances_hist}, we observe that the distribution $p_{\Im[\lambda_1]}(x)$ consists of two parts, viz.,   a  delta distribution at zero 
%of height $P[\lambda_1 \in \mathbb{R}]$, determining 
carrying the fraction $P[\lambda_1 \in \mathbb{R}]$
of matrix realisations that have a real-valued leading eigenvalue, and  a continuous distribution  corresponding with eigenvalues that have nonzero imaginary part: 
\begin{equation}
p_{\Im[\lambda_1]}(x) = P[\lambda_1 \in \mathbb{R}] \delta(x) + (1- P[\lambda_1 \in \mathbb{R}] ) \tilde{p}_{\Im[\lambda_1]}(x). \label{eq:pInm}
\end{equation}
We define 
\begin{equation}
\Im[\lambda^\star_1] :=  {\rm argmax}_{x\in\mathbb{R}^+}\:\tilde{p}_{\Im[\lambda_1]}(x), 
\end{equation}
as the typical value of the imaginary part of the leading eigenvalue.

As the control parameter changes from $s<s_{\rm d}$, to $s\approx s_{\rm d}$ and to $s>s_{\rm d}$ (obtained for $\sigma_D=0.05$, $\sigma_D=0.10$ and $\sigma_D=0.15$ in correspondence of $\sigma_{\rm G}=0.6$ and $\mu_{\rm G}=1.0$) it can be observed that the maximum of the continuous part of the distribution, indicated by the vertical dashed line, is mostly independent on the control parameter both for interaction-like and Jacobian-like matrices. The change in the control parameters mainly affects the weight  
 $P[\lambda_1 \in \mathbb{R}]$ carried by the two parts of the distribution $p_{\Im[\lambda_1]}(x)$, as it is usual with discontinuous phase transitions and at variance with the continuous dynamical transition found in Ref.~\cite{mambuca2022dynamical} for shifted interaction matrices as a function of $c$. Since the numerical results in Fig.~\ref{fig:reentrances_hist}  are obtained at finite $N$, we perform in \ref{app:finite_size_scaling} a detailed,  finite size scaling analysis of the distribution $p_{\Im[\lambda_1]}(x)$, which confirms that the discontinuity in $\Im[\lambda^\star_1]$  at the transition is to be expected also in the large $N$ limit.  

Following the approach implemented in \cite{mambuca2022dynamical}, to get an estimate of the typical value, $\Im[\lambda_1^{\star}]$, of the portion of the distribution corresponding to nonzero imaginary part we take the mode of a $\gamma$-distribution 
$   \gamma(x;\alpha, \beta)~=~\frac{\beta^\alpha x^{\alpha-1} e^{-\beta x}}{\Gamma(\alpha)}
$
fitted on the histogram of $\Im[\lambda_1]>0$. Here $\Gamma(\alpha)$ is the gamma function with  parameters $\alpha, \beta \in \mathbb{R}^+$  real and positive.

\begin{figure}[h]
	\centering
	\begin{subfigure}[b]{0.48\textwidth}
		\centering
		\includegraphics[width=\textwidth]{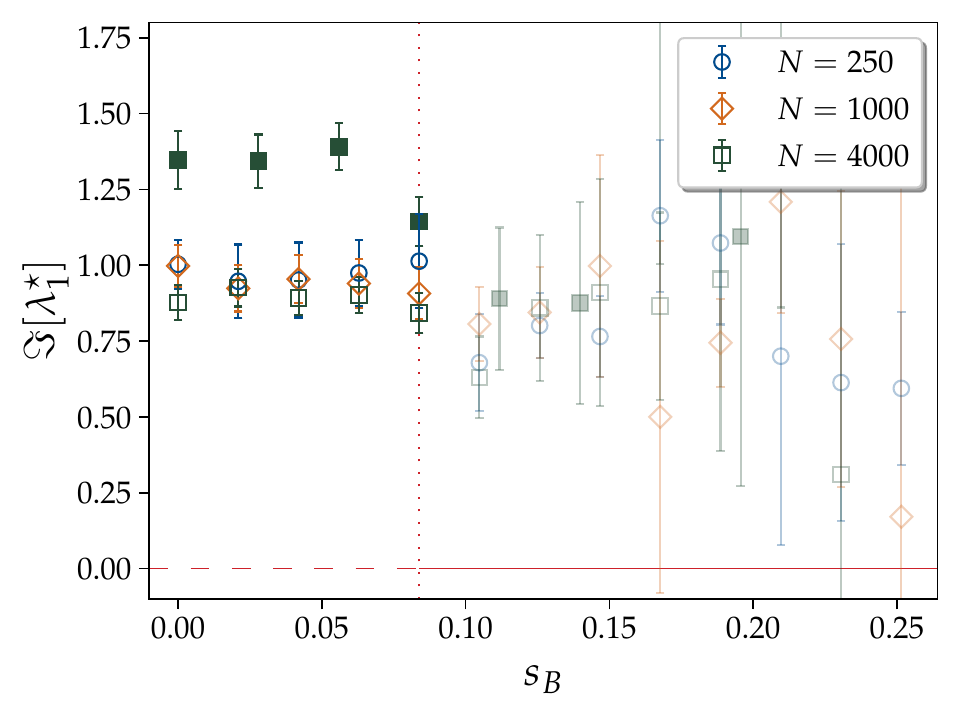}
		\caption{Interaction-like $\mathbf{B}$}
	\end{subfigure}
	\hfill
	\begin{subfigure}[b]{0.48\textwidth}
		\centering
		\includegraphics[width=\textwidth]{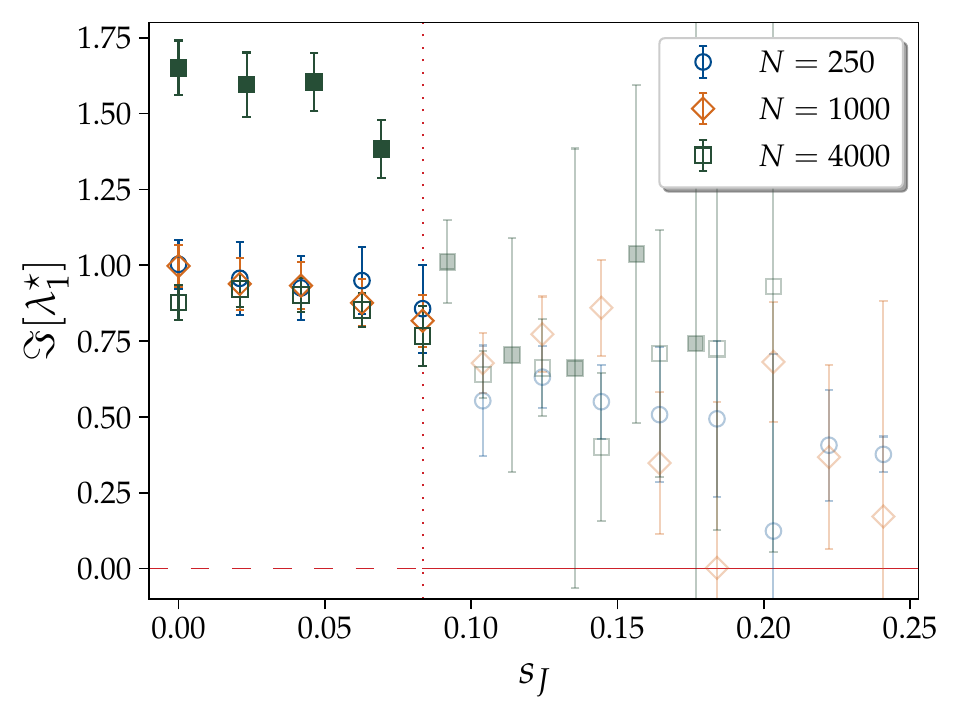}
		\caption{Jacobian-like $\mathbf{J}$}
	\end{subfigure}
        \caption[]{Imaginary part $\Im[\lambda_1^\star]$ of the typical, nonreal, leading eigenvalue as a function of the control parameters, as defined in Eq.~(\ref{eq:control_parameters}), for interaction-like $\mathbf{B}$ (on the left) and Jacobian-like $\mathbf{J}$ (on the right) matrices of antagonistic  Erd\"{o}s-R\'{e}nyi graphs.  The values of   $\Im[\lambda_1^\star]$ are estimated as shown in Fig.~\ref{fig:reentrances_hist}, viz., as the mode of a  $\gamma$-distribution fitted to a histogram built out of $1000$ matrix  realisations. The  error bars are computed by standard error propagation from the fitting parameters.   The  parameters employed for empty and filled markers 
        are the same as those used 
        %for unfilled and filled markers 
        in Fig.~\ref{fig:reentrances_probability}, respectively. The vertical dotted line depicts the transition value $s_d$. Note that the markers after the transition are denoted in a lighter shade because they represent data which are merely the result of finite size effects: in the nonoscillatory phase the frequency of complex leading eigenvalues decreases with system size. After the transition the eigenvalues are instead typically real, as represented by the 
        solid red line ($\Im[\lambda_1^\star]=0$).
        Its left dashed continuation shows rare real leading eigenvalues also present in the oscillatory phase. 
        }
	\label{fig:reentrances_imaginary_part}
\end{figure}
The typical value $\Im[\lambda_1^{\star}]$ for different sizes $N$ is plotted in Fig.~\ref{fig:reentrances_imaginary_part} as a function of the control parameter for the interaction-like and Jacobian-like case and with the same settings as in Fig.~\ref{fig:reentrances_probability}.
As it can be observed again, $\Im[\lambda_1^{\star}]$ is an almost constant  function of the control parameter and its value at the transition point is positive until it vanishes at $s_{\rm d}$, showing the discontinuous nature of the transition and the associated finite size effects. 
For further evidence, the lighter markers show that  $\Im[\lambda_1^{\star}]$ is nonzero also in the nonoscillatory phase (the dynamical transition point is indicated by the vertical dotted line) when the leading eigenvalue is  typically real.    
Finally, notice that while the location of the transition is only controlled by $s_B$ or $s_J$, respectively, the value of $\Im[\lambda_1^{\star}]$ depends on the particular choice of the model parameters. This value reflects the global rescaling of the matrix as discussed at the beginning of Sec.~\ref{sec:locTrans}.

\begin{figure}[h]
	\centering
	%\captionsetup{justification=centering}
	\includegraphics[width=1.0\textwidth]{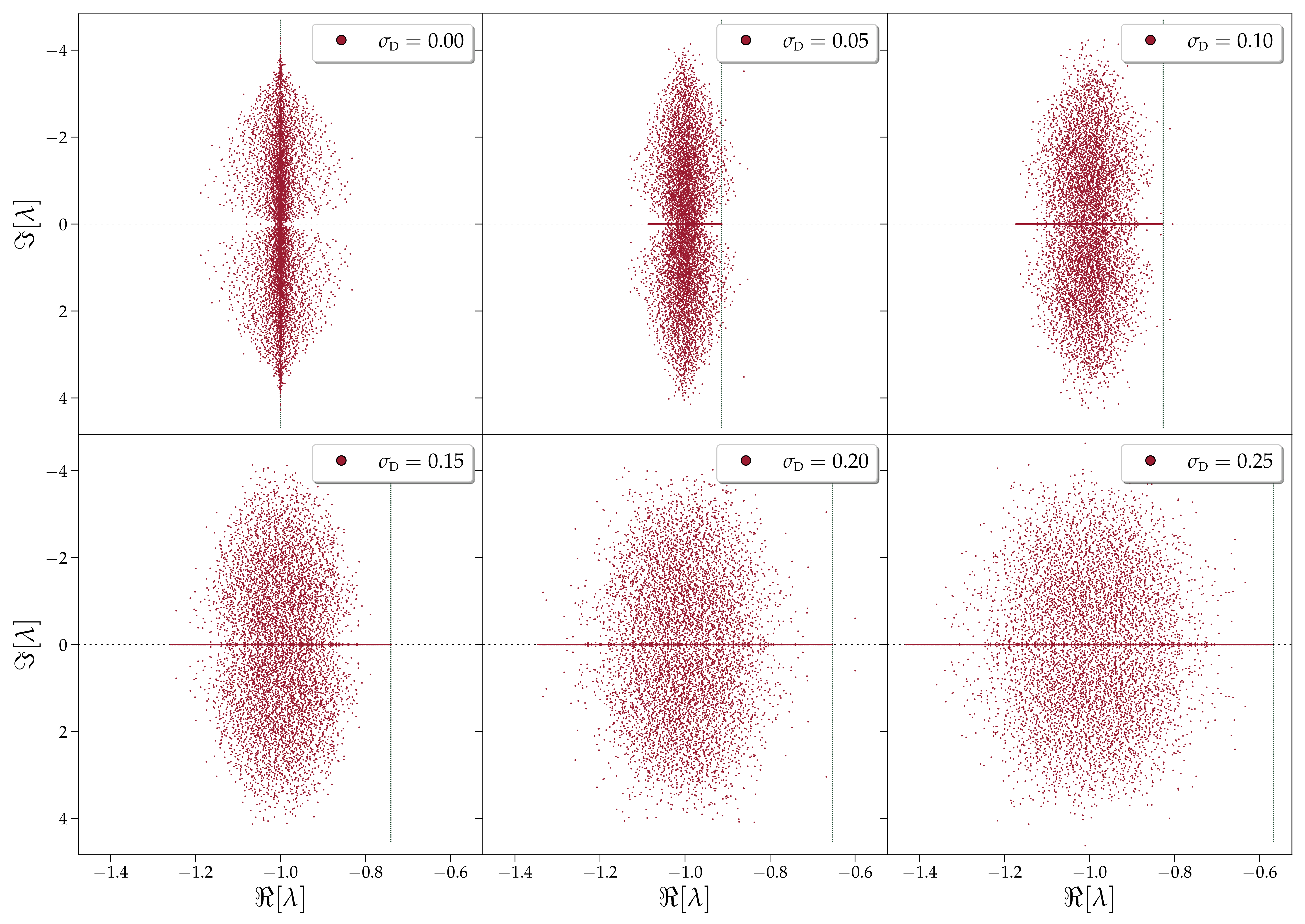}
	\caption[]{Spectra of single realisations of  antagonistic, Erd\H{o}s-R\'{e}nyi, interaction-like matrices $\mathbf{B}= {\bf A}-{\bf D}$ of size $N=8000$. The diagonal entries of $\mathbf{D}$  are drawn from  a uniform distribution with $\sigma_D$ as  depicted in the legend, corresponding to values of  $s_B=0$, $s_B \simeq 0.4$, $s_B \simeq 0.8$, $s_B \simeq 0.13$, $s_B \simeq 0.17$ and $s_B \simeq 0.21$ (left to right, up to bottom). The green dotted line displays  $-D_{\rm min}$. The other parameters employed are as discussed in Sec.~\ref{sec:numerical_details}.}
	\label{fig:reentrances_spectra_interaction_with_diagonal}
\end{figure}

\begin{figure}[h]
	\centering
	%\captionsetup{justification=centering}
	\includegraphics[width=1.0\textwidth]{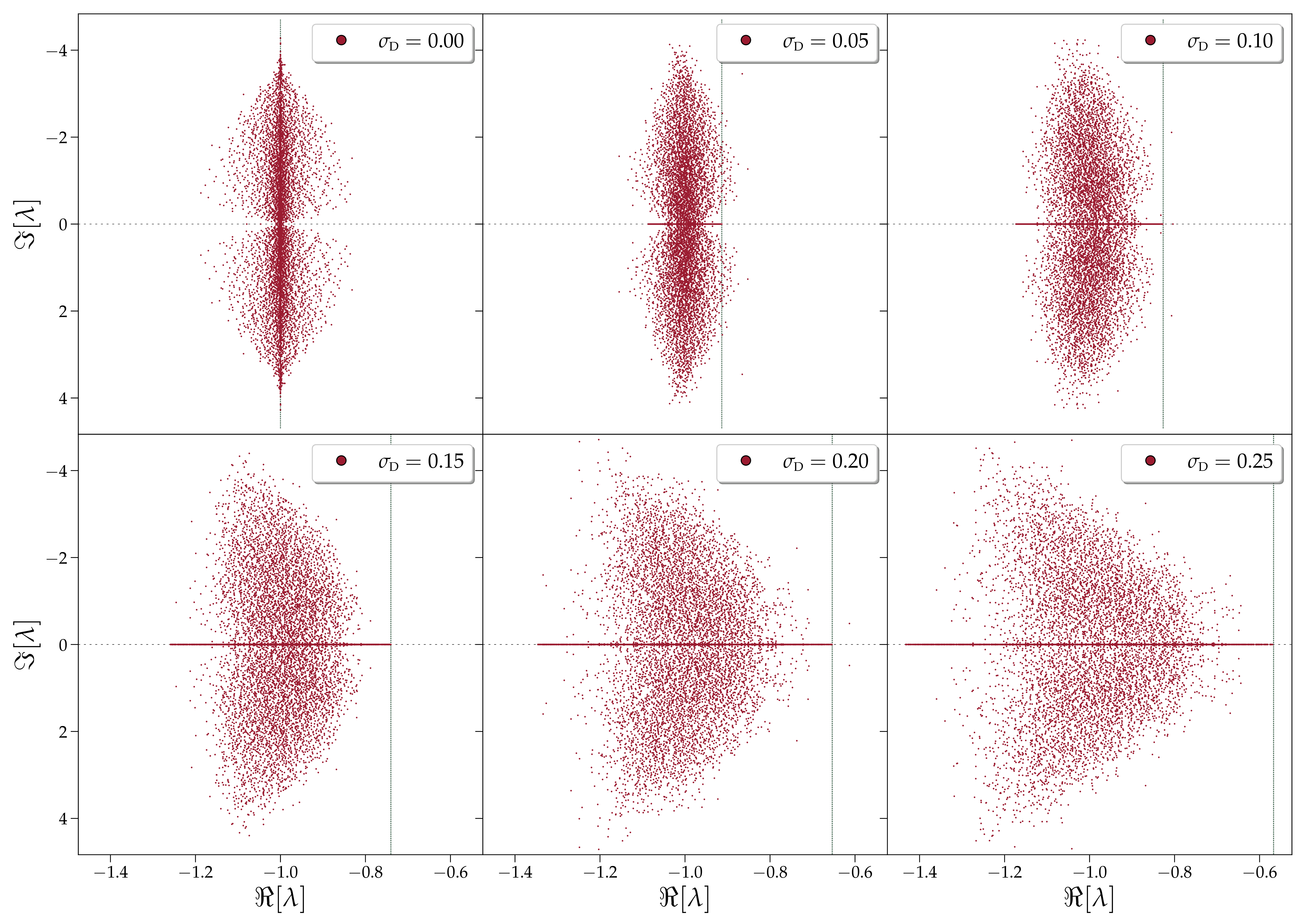}
	\caption[]{Spectra of  single realisations of antagonistic, Erd\H{o}s-R\'{e}nyi, Jacobian-like matrices $\mathbf{J} = -\mathbf{D}{\bf B_{\rm Id}}$ of size $N=8000$. The diagonal entries of $\mathbf{D}$ are drawn from a   uniform distribution with $\sigma_D$ as depicted in the figure legends,  corresponding to values  $s_J=0$, $s_J \simeq 0.4$, $s_J \simeq 0.8$, $s_J \simeq 0.12$, $s_J \simeq 0.16$ and $s_J \simeq 0.20$ (left to right, up to bottom). The green dotted line displays $-D_{\rm min}$. The other parameters employed are as discussed in Sec.~\ref{sec:numerical_details}.}
	\label{fig:reentrances_spectra_jacobian}
\end{figure}

Lastly,  to develop a better understanding about the mechanism of the discontinuity in the dynamical transition of interaction-like  and Jacobian-like matrices,  we investigate their spectra  in Figs.~\ref{fig:reentrances_spectra_interaction_with_diagonal} and~\ref{fig:reentrances_spectra_jacobian} at different values of the control parameters.      Let us first discuss the spectra of  antagonistic, interaction-like matrices in Fig.~\ref{fig:reentrances_spectra_interaction_with_diagonal}.    We observe that the spectrum contains two parts, viz., a cloud of eigenvalues that have a nonzero imaginary part and a segment of real-valued eigenvalues. 
In particular,
the width of the segment of real eigenvalues is approximately equal to  $[-D_{\rm max},-D_{\rm min}]$, and hence the segment width increases as a function of $\sigma_D$ (or as a function of $s_B$, for fixed $\sigma$). 
The discontinuous nature observed for the dynamical transition is originated from the competition between the width of the cloud of complex eigenvalues and the width of the segment on the real axis.
It can be observed that for $s_B<s_{\rm d}$, the leading eigenvalue belongs to the cloud of complex eigenvalues, and since the shape of this cloud is reentrant the imaginary part of the leading eigenvalue in this regime is  nonzero.   On the other hand, for   $s_B>s_{\rm d}$, the leading eigenvalue belongs to the segment of real eigenvalues, and hence has null imaginary part.   Since the eigenvalue cloud is still reentrant at the point  $s_B = s_{\rm d}$ when the segment width overtakes the width of the cloud, the transition is discontinuous. 
 %one which is real and one that exhibits %reentrance.     
This behaviour is different from the continuous transition driven by the connectivity $c$ at $s_B=0$, as discussed  in Ref.~\cite{mambuca2022dynamical}.
In this case, the transition in the  imaginary part of the leading eigenvalue is the result of a gradual reshaping of the spectrum resulting in the progressive disappearance of the reentrance in correspondence of the real axis at large $c$.  

For 
the spectra of Jacobian matrices, shown 
in Fig.~\ref{fig:reentrances_spectra_jacobian}, 
the qualitative picture is %here
similar to what we have discussed for Fig.~\ref{fig:reentrances_spectra_interaction_with_diagonal} for interaction-like matrices, viz., the spectrum consists of two parts, one being a cloud of complex eigenvalues  reentrant in correspondence of the real axis %for small values of $s_J$, 
and the other is a segment of real eigenvalues approximately supported on $[-D_{\rm max},D_{\rm min}]$.   Again, a discontinuous transition on the imaginary part of the leading eigenvalue takes place as
the cloud of eigenvalue is reentrant when the width of the segment overtakes the width of the cloud of eigenvalues.  

In both cases, in the nonoscillatory phase, $\Re[\lambda_1]\approx -D_{\rm min}$, as the support of the real eigenvalues is well approximated by $[-D_{\rm max},-D_{\rm min}]$, which is consistent with the results in Figs.~\ref{fig:leading_eig_interaction_with_diagonal} and~\ref{fig:leading_eig_jacobian}.

Moreover, in both cases, it emerges that the reentrance of the cloud of complex eigenvalue in correspondence of the real axis also becomes slightly less pronounced when $\sigma_D$ increases, which hints to the possibility that, for specific settings, the disappearance of the reentrance could take place before the largest real eigenvalues become the leading one.
In such situation a continuous transition of the imaginary part of the leading eigenvalue is to be expected, instead of the discontinuous transition observed here.
In general, by modifying the parameters contained in the distributions $p$ and $p_D$ and the connectivity of the graph, it should be possible to obtain a phase diagram containing a line of discontinuous transition ending in a continuous transition point. 
This interesting endeavour is left for future work.
Finally, there is one notable distinction between Figs.~\ref{fig:reentrances_spectra_interaction_with_diagonal} and \ref{fig:reentrances_spectra_jacobian}, namely,  for large values of $\sigma_D$ the spectra of interaction-like matrices have a  round shape, albeit not circular, while the spectra of Jacobian-like matrices has a characteristic arrow shape, which has also been observed before for dense matrices, see Ref.~\cite{gibbs2018effect, stone2018feasibility}.  
This different behaviour can produce a qualitative difference in the phase diagram containing discontinuous and continuous dynamical transitions of interaction-like and Jacobian-like matrices, which will be also interesting to study.

\section{Discussion}\label{sec:Discussion}

In this paper, we have focused on two interesting features of the spectra of sparse random graphs. First, we have analyzed how the leading eigenvalue of random graphs depends on the system size, taking into account the sign pattern of matrix entries and the graph topology. Second, we  have studied how the imaginary part of the leading eigenvalue transitions from zero to a nonzero value as a function of the model parameters in the large system limit. These features are specific to sparse network topologies and do not appear in highly connected graphs.
   
More specifically, this paper presents a simple, general criterion for predicting the asymptotic behavior of the leading eigenvalue of infinitely large, sparse, random graphs, based on the concept of strong local sign stability.
Although in some previous examples (Refs.~\cite{neri2019spectral,mambuca2022dynamical}) the importance of the sign pattern and network topology on the stability of 
some matrices was emerging, here we also systematically investigate the validity of the proposed criterion
with additional numerical examples on sparse random graphs with different topologies and sign patterns, and for different types of matrices such as adjacency matrices and Jacobian-like matrices.

Aside the numerical evidence, we can also provide an intuitive explanation for the relation between strong local sign stability and the finiteness of the real part of the leading eigenvalue within the context of linear, dynamical system defined on a strongly locally sign stable graph. 
For locally sign stable graphs, when a perturbation is applied to a random node its neighbours react. However typically the finite neighbourhood is, by definition, stable, therefore the perturbation will be damped. This scenario could, in principle, be overturned by cycles of length $O(\log(N))$ that break sign stability, but since they are long we do not expect their effect to be strong enough to destabilise the system in all cases.  The situation is different for large, random graphs that are not  locally sign stable. In such  graphs there exist nodes that are not locally stable, and hence perturbations applied to these nodes will always locally grow and destabilize the system.
In addition, whenever the random graph is locally sign stable but not strongly locally sign stable and short cycles are growing in number with system size, there may exist few nodes that belong to a large number of short cycles and lead again to a local instability.

One very interesting feature of strong local sign stability, following from the concept of sign stability, is that 
it is not affected by the absolute values of the matrix elements and thus it refers to the set of graphs with the same network topology and sign pattern, provided that the distribution of the off-diagonal elements has finite second moment. As a consequence, whenever the interaction matrix $\mathbf{A}$ of an ecosystem is strongly locally sign stable, then also the corresponding interaction-like $\mathbf{B}$ and Jacobian-like $\mathbf{J}$ matrices are strongly locally sign stable, potentially ensuring at once feasibility, structural stability and linear stability. This aspect is particularly interesting in the ecological context where, as anticipated in the Introduction, it is difficult to quantify the strength of interactions between species~\cite{moore2012energetic, jacquet2016no}.
In this regard, Ref.~\cite{solimano1982graph} has shown that for a general predator-prey model defined on a tree, the Jacobian matrix (with negative elements on the diagonal) evaluated on an isolated, feasible equilibrium point has eigenvalues with negative real parts, and hence the equilibrium point is linearly stable for such systems. The concept of strong local sign stability that we have introduced in the present paper generalises the results in Ref.~\cite{solimano1982graph}, as it applies to a broader class of systems, including predator-prey ecosystems defined on random graphs that are locally tree-like, have a finite number of short cycles, and have diagonal elements that are negative and below a finite threshold. The last condition is needed to have all eigenvalues with negative real part despite the finite width of the spectrum.  

Our results challenge the classical notion that complex, dynamical systems  exhibit a trade-off between size and stability, mostly discussed for fully connected models~\cite{may1972will, Allesina2015, jacquet2016no, mccann2000diversity, neutel2014interaction}.
Indeed, in dense graphs the sign pattern does not alter how the leading eigenvalue scales with system size~\cite{Allesina2015, allesina2012, Cure2021}.
Instead, for sparse graphs we find that this trade-off depends on the sign pattern and topology of the graph.
More specifically, Ref.~\cite{mambuca2022dynamical} showed that the real part  of the leading eigenvalue of infinitely large,  antagonistic Erd\H{o}s-R\'{e}nyi graphs with shifted interaction matrix is finite, also by means of theoretical predictions on the boundary of the spectrum in the limit $N\rightarrow\infty$.   The present work generalises this result with numerical evidences by extending it to various matrix structures relevant in ecology and by identifying the common feature %underlying absolute stability. 
the convergence of the real part of the leading eigenvalue to a finite value for large system size.
It will be interesting to extend the theoretical predictions from interaction-like matrices \cite{mambuca2022dynamical} to more general matrices to put on a firmer grounds the extrapolation of numerical results to infinite $N$. 
In general it is also interesting to consider models for which the fraction of sign-symmetric interactions is vanishing with $N$, or antagonistic models defined on graphs with power law degree distribution, which are locally sign stable but not strongly locally sign stable (due to a growing number of finite cycles albeit a vanishing fraction of them), to understand whether the conditions to grant finite leading eigenvalue can be relaxed.  

It is important to emphasise that the strong LSS criterion (\ref{eq:conjecture}) applies to matrices built from graphs as explained in Sec.~\ref{sec:RMTModels}, \emph{i.e.}, without correlations among network topology, sign pattern and interaction weights. In fact, %the prediction of
%our criterion for the asymptotic behavior of the leading eigenvalue relies on the typical scenario which is well described by strong local sign stability. On the other hand, 
including correlations between network topology and interaction weights, the leading eigenvalue can diverse with the system size even though the graph is strongly LSS. 
%can amplify the effect of rare events and make them decisive. 
As an example, consider a sequence $\mathbf{M}_N$ in which the two entries with the largest absolute values concentrate with the same sign on the same edge, say $M_{12} ={\rm max}_{i,j}\left\{|M_{ij}|: i,j\in V \right\}$ and $M_{21} = {\rm max}_{i,j}\left\{|M_{ij}|: i,j\in V \right\}\setminus \left\{M_{12}\right\}$. If the entries are also drawn from a distribution with unbounded support, these largest values $M_{12}$ and $M_{21}$ diverge as a function of $N$. Then in the limit of large $N$, the leading eigenvalue $\lambda_1\approx \sqrt{M_{12}M_{21}}$ and thus it diverges as a function of the system size $N$. At the same time, since we are only imposing the weights of a single edge the sequence $\mathbf{M}_N$ may still be strongly locally sign stable. %However, in this case strong local sign stability is not able to grasp the asymptotic behaviour of the leading eigenvalues since it is determined by a specific local feature.
% Note that in general, unless otherwise explicitly imposed, this realisation will be extremely rare in matrices built as explained in Sec.~\ref{sec:RMTModels}.
Another possible example is a graph in which the node with the largest degree has all sign-symmetric edges. Also in this case, as long as the degree distribution has unbounded support, the leading eigenvalue diverges with the system size $N$, although the graph may be strongly locally sign stable. %, whenever the number of sign-symmetric edges is sub-linear with the system size

Systematic studies of the stability properties of sparse systems are far from being achieved, although potentially relevant for many applications of dynamical systems. In particular new interest is arising on the rich phenomenology of sparse ecosystems \cite{marcus2022local}.
We believe that our paper gives an important contribution by grasping a general criterion for stability of sparse ecosystems.

The second main result presented in this paper concerns the imaginary part  of the leading eigenvalue $\Im[\lambda_1]$ of antagonistic Erd\H{o}s-R\'{e}nyi graphs. The spectra of antagonistic Erd\H{o}s-R\'{e}nyi interaction-like and Jacobian-like matrices display a transition driven by the strength of the diagonal disorder $\sigma_D$ from a low $\sigma_D$-phase in which the leading eigenvalue typically has nonzero imaginary part to a large $\sigma_D$-phase  in which it is real. In the first phase the spectrum is characterised by a reentrance effect around the real axis and this is the cause for it to have a pair of complex conjugate leading eigenvalues. 

This reentrance effect is specific to sparse, low connectivity graphs with sign-antisymmetric weights. In fact, as soon as we move away from the low connectivity limit the spectra tend, under fairly general conditions, to have an elliptic shape.   
It is worth noting that the spectral reentrance observed in sparse random graphs is a qualitatively new feature, which has only been previously observed in the recent work \cite{mambuca2022dynamical}.  That paper  identifies a transition in the spectra of antagonistic Erd\H{o}s-R\'{e}nyi graphs with zero diagonal entries driven by the connectivity $c$, from a large $c$ phase where $\Im[\lambda_1] = 0$ to a small $c$ phase where typically $\Im[\lambda_1] \neq 0$, and which also displayed a reentrance effect.
The present paper generalizes this result by extending it to more elaborate matrix structures. In particular, we have shown that the reentrance effect persists even in the presence of a disordered diagonal or a stripy, Jacobian-like, structure, but only as long as their disorder is not too large.

It is important to stress once more that while the transition described in \cite{mambuca2022dynamical} is continuous, the one discussed in the present paper is discontinuous. In the former case, the reentrance disappears progressively as the connectivity increases. On the other hand, in the latter case, the spectrum is composed of a cloud of complex eigenvalues and of a segment lying on the real axis. In the low $\sigma_D$ phase, the leading eigenvalue is determined by the cloud and is therefore complex. As $\sigma_D$ increases, both of these components gradually change: the cloud reshapes, reducing the reentrance effect, while the real segment elongates. At the transition, the right tip of the real segment reaches the convex hull of the cloud, and the leading eigenvalue starts to be real. This overtaking occurs before the reentrance has completely disappeared, and therefore we observe a jump in the leading eigenvalue imaginary part, giving rise to a discontinuous transition.
At the same time it is possible that, for specific settings, the reentrance effect could disappear completely before (or together with) the overtaking by the segment on the real axis thus making the transition continuous. A more detailed study of the phase diagram describing both the reentrance effect and the transition in $\Im[\lambda_1]$ is left for future work. Moreover, since this reentrance effect is peculiar to sparse graphs, we expect the network topology to be relevant and therefore it would be interesting to investigate its impact.

The transition in $\Im[\lambda_1]$ discussed in this paper, especially with regard to Jacobian-like matrices, is of interest in the context of dynamical systems. In this framework, the imaginary part of the Jacobian leading eigenvalue determines the oscillation frequency of the slowest mode of relaxation towards the related fixed point. Accordingly, the response to a perturbation around a fixed point is oscillatory when the leading eigenvalue has imaginary part different from zero, nonoscillatory otherwise. Our result on the transition indicates that the dynamical response of a nonlinear system defined on antagonistic Erd\H{o}s-R\'{e}nyi is oscillatory only if the strength of the diagonal disorder $\sigma_D$ is small compared to the off-diagonal one $\sigma_{\rm O}$.

In order to appreciate the actual implication of this dynamical transition it would be interesting to derive a phase diagram describing the dynamical behaviors of antagonistic systems defined on sparse graphs. Indeed, in the context of dynamical systems with nonsymmetrical interactions the spectra of Jacobian matrices can play a role only in the case in which the dynamics is attracted by fixed points while chaos and limit cycles are not predominant. In dense ecological models such a phase diagram has been derived and it identifies regions with multiple attractors where solutions are often chaotic \cite{roy2019numerical, roy2020complex, bunin2017ecological, pearce2020stabilization}.
An analogous phase diagram for sparse ecosystems with predator-prey interactions is not known and its derivation is an interesting endeavour left for future work.

Finally, concerning applications, our  findings on the importance of strong local sign stability in enhancing stability are consistent with empirical observations on real food webs, which are graphs that represent predator-prey interactions in ecological systems. According to our results, locally tree-like structures and sign-antisymmetric interactions stabilize large ecosystems, and hence, these are the structures we expect to observe. Empirical studies have shown that food webs are indeed locally tree-like, with a number of cycles that is similar to those found in locally tree-like Erd\H{o}s-R\'{e}nyi graphs \cite{dunne2002food}. Notably, other networks such as social and technological networks have a significantly larger number of cycles, a feature unique to food webs \cite{dunne2002food}.   In addition, 
Ref.~\cite{neutel2002stability}  found that   the weights of long cycles in real food webs are typically smaller than in random matrices, further underlying the importance of locally tree-like structures for large ecosystems.

\section*{Acknowledgments} We thank Andrea Marcello Mambuca for insightful discussions at the initial stage of this work. This work was also supported by the Simons Foundation Grant on Cracking the Glass Problem (\#454935 Giulio Biroli).

\newpage
\appendix

\section{Structural stability to perturbations of various ecological parameters} \label{app:structstab}

In this Appendix we derive the expression for structural stability, defined as the stability of the abundances of surviving species $\vec{N}^*$, as defined in Eq.~(\ref{eq:existence}), with respect to small perturbations of the three different ecological parameters of our model $r_i$, $K_i$ and $\alpha_{ij}$. In particular, we show that in all three cases the susceptibility of $\vec{N}^*$ to perturbations of ecological parameters is related to the inverse of the matrix ${\bf B}$, defined in Eq.~(\ref{eq:Bij_def}) and therefore it is singular when the spectrum of ${\bf B}$ contains the origin of the complex plane.

Let's start with the simplest case of a perturbation applied to the growth rates $r_i\rightarrow r_i+\xi_i$. 
The perturbed equations for the $N_i^*$ then read 
\begin{equation}
    r_i+\xi_i=\frac{N_i^*}{K_i}+\sum^S_{j=1; j\neq i}\alpha_{ij}N_j^* \ ,
\end{equation}
which can be derived with respect to $\xi_k$
\begin{equation}
    \delta_{ik}=\frac{1}{K_i}\frac{\partial N_i^*}{\partial \xi_k}+\sum^S_{j=1; j\neq i}\alpha_{ij} \frac{\partial N_j^*}{\partial \xi_k} = \sum^S_{j=1} B_{ij} \frac{\partial N_j^*}{\partial \xi_k} \ ,
\end{equation}
revealing that the susceptibility of $N_i^*$ to little variations of $r_k$ is directly determined by the inverse of ${\bf B}$:
\begin{equation}
    \frac{\partial N_i^*}{\partial \xi_k}=({\bf B}^{-1})_{ik} \ .
\end{equation}

Let's now consider a perturbation applied to the parameter $K_i\rightarrow K_i+\eta_i$, after which the perturbed equations for the $N_i^*$ read 
\begin{equation}
    r_i=\frac{N_i^*}{K_i+\eta_i}+\sum^S_{j=1; j\neq i}\alpha_{ij}N_j^* \ .
\end{equation}
Deriving with respect to $\eta_k$
%\begin{equation}
%    0=-\delta_{ik}\frac{N_i^*}{(K_i+\eta_i)^2} + \sum^S_{j=1} \left( \frac{\delta_{ij}}{K_i+\eta_i} + \alpha_{ij}\right) \frac{\partial N_j^*}{\partial \eta_k} \ ,
%\end{equation}
and evaluating the derivative at $\vec{\eta}=0$
\begin{equation}
    0=-\delta_{ik}\frac{N_i^*}{K_i^2} + \sum^S_{j=1} \left( \frac{\delta_{ij}}{K_i} + \alpha_{ij}\right) \frac{\partial N_j^*}{\partial \eta_k} = -\delta_{ik}\frac{N_i^*}{K_i^2} + \sum^S_{j=1} B_{ij} \frac{\partial N_j^*}{\partial \eta_k} \ ,
\end{equation}
 we find the susceptibility of $N_i^*$ to little variations of $K_k$ which is determined the inverse of ${\bf B}$ multiply row-by-row by $\frac{N_k^*}{K_k^2}$:
\begin{equation}
    \frac{\partial N_i^*}{\partial \eta_k}\Bigg\rvert_{\vec{\eta}=0}=({\bf B}^{-1})_{ik} \frac{N_k^*}{K_k^2} \ .
\end{equation}

Finally we consider a perturbation applied to the interaction $\alpha_{ij}\rightarrow \alpha_{ij}+\epsilon_{ij}$ which lead to
\begin{equation}
    r_i=\frac{N_i^*}{K_i}+\sum^S_{j=1; j\neq i}\left(\alpha_{ij} + \epsilon_{ij} \right)N_j^* \ .
\end{equation}
Deriving with respect to $\epsilon_{kl}$
%\begin{equation}
%    0 = \frac{\partial N_i^*}{\partial \epsilon_{kl}} \frac{1}{K_i} + \sum^S_{j=1; j\neq i} \left(\alpha_{ij} + \epsilon_{ij} \right) \frac{\partial N_j^*}{\partial \epsilon_{kl}} + N_l^* \delta_{ik} \ ,
%\end{equation}
and evaluating the derivative at $\epsilon_{kl}=0 \;\; \forall \;\; k \neq l$
\begin{equation}
    0 = \sum^S_{j=1} \left( \frac{\delta_{ij}}{K_i} + \alpha_{ij}\right) \frac{\partial N_j^*}{\partial \epsilon_{kl}} + N_l^* \delta_{ik}= \sum^S_{j=1} B_{ij} \frac{\partial N_j^*}{\partial \epsilon_{kl}} + N_l^* \delta_{ik} \ ,
\end{equation}
we find the susceptibility of $N_i^*$ to little variations of $\alpha_{kl}$ which is again related to the inverse of ${\bf B}$:
\begin{equation}
    \frac{\partial N_i^*}{\partial \epsilon_{kl}}\Bigg\rvert_{\epsilon=0} = - ({\bf B}^{-1})_{ik} N_l^* \ .
\end{equation}

In conclusion, a singular behaviour emerges when the spectrum of ${\bf B}$ contains the origin of the complex plane hinting to a large susceptibility of the solution of $\vec{N}^*$ to all three ecological parameters.

\section{The eigenvalues of antagonistic trees have zero real part} \label{app:signstab}
We show that the adjacency matrices $\mathbf{M}\in \mathbb{R}^{N\times N}$ of trees weighted with sign-antisymmetric interactions have purely imaginary eigenvalues, i.e., 
\begin{equation}
    \Re[\lambda_i] = 0 
\end{equation}
for all $i=1,2,\ldots,N$.  As interactions are sign-antisymmetric, it holds that 
\begin{equation}
    M_{ij}M_{ji}<0 
\end{equation}
for all pairs $(i,j)$ for which either $M_{ij} \neq 0$ or $M_{ji}\neq 0$.  The tree condition implies that Eq.~(\ref{eq:cond2}) holds.   We assume that $M_{ii} = 0$.  

The arguments we present are adapted from Ref.~\cite{quirk1965qualitative}, albeit applied to the case of antagonistic, tree graphs.  

First we define a general class of, so-called, strictly quasi-antisymmetric matrices, and we show that these matrices have purely imaginary eigenvalues.   Second, we show that antagonistic trees are strictly quasi-antisymmetric. 

\subsection{Eigenvalues of strictly quasi-antisymmetric matrices are imaginary}
We say that a matrix  $\mathbf{Q}\in \mathbb{R}^{N\times N}$ is strictly quasi-antisymmetric when 
\begin{equation}
\mathbf{Q}^{\rm T}= -\boldeta \mathbf{Q}\boldeta^{-1} , \label{def:strictlyQAS}
\end{equation}
where $\boldeta$ is a symmetric, positive definite matrix; notice that Refs.~\cite{joglekar2011level, feinberg2021pseudo} define strictly quasi-symmetric matrices, which are related to PT-symmetric matrices in quantum mechanics~\cite{bender1999pt}. 

Strictly quasi-antisymmetric matrices have imaginary eigenvalues as they are similar to an antisymmetric matrix $\mathbf{K}$.  Indeed,  since $\boldeta$ is positive and symmetric, it is a diagonalisable matrix with positive eigenvalues.   We define   $\sqrt{\boldeta}$ as the  square root of  $\boldeta$ that is positive definite, which is the (unique) symmetric, matrix that has eigenvalues that are equal to the positive square roots of the eigenvalues of   $\boldeta$.  Consequently, we may define the matrix 
\begin{equation}
\mathbf{K} = \sqrt{\boldeta} \mathbf{Q}\sqrt{\boldeta}^{-1} , \label{eq:K}
\end{equation}
which has real-valued entries as $\sqrt{\boldeta}$ has real-valued entries.  The matrix $\mathbf{K}$ is antisymmetric, as
\begin{equation}
\mathbf{K}=\sqrt{\boldeta}  \mathbf{Q} \sqrt{\boldeta}^{-1}=\sqrt{\boldeta}^{-1} \boldeta  \mathbf{Q}\boldeta^{-1}\sqrt{\boldeta}=-\sqrt{\boldeta}\mathbf{Q}^{\rm T}\sqrt{\boldeta}= \left(\sqrt{\boldeta}\mathbf{Q}\sqrt{\boldeta}\right)^{\rm T} = -\mathbf{K}^{\rm T} \ , \label{eq:KAsym}
\end{equation}
where in the last step we have used that $\sqrt{\boldeta}$ is a symmetric matrix. 
Hence, since $\mathbf{Q}$ is similar to $\mathbf{K}$, both matrices share the same eigenvalues \cite{horn2012matrix}, and since  $\mathbf{K}$ is an antisymmetric with real-valued entries,  the eigenvalues of $\mathbf{Q}$ are purely imaginary. 

\subsection{Adjacency matrices of trees with sign-antisymmetric weights are strictly-quasi-antisymmetric} \label{sec:quasiAnti}

We show that the adjacency matrices  $\mathbf{M}$ of antagonistic trees are strictly quasi-antisymmetric i.e., they satisfy Eq.~(\ref{def:strictlyQAS}).  To this aim, following Ref.~\cite{quirk1965qualitative}, we explicitly construct the matrix  $\boldeta$~.

We  select a random node in the graph, which we call the root node, and we label it as $i_0$.    Subsequently,  we consider (i) the set of nodes $V_1$ that are neighbours of $i_0$; (ii) the set of nodes $V_2$ that are neighbours of nodes in $V_1$ excluding the root node; (iii) the set of nodes $V_3$ that are neighbours of $V_2$ and are not part of $V_1$, etc..   Eventually we obtain a partitioning $V = \left\{i_0\right\}\cup V_1 \cup V_2\cup \ldots V_{\ell}$ of the set of vertices of the tree graph $G=(V,E)$  associated with $\mathbf{M}$, where $\ell$ is the depth of the tree rooted at $i_0$.

\begin{figure}[h]
\begin{center}
\includegraphics[width=0.8\textwidth]{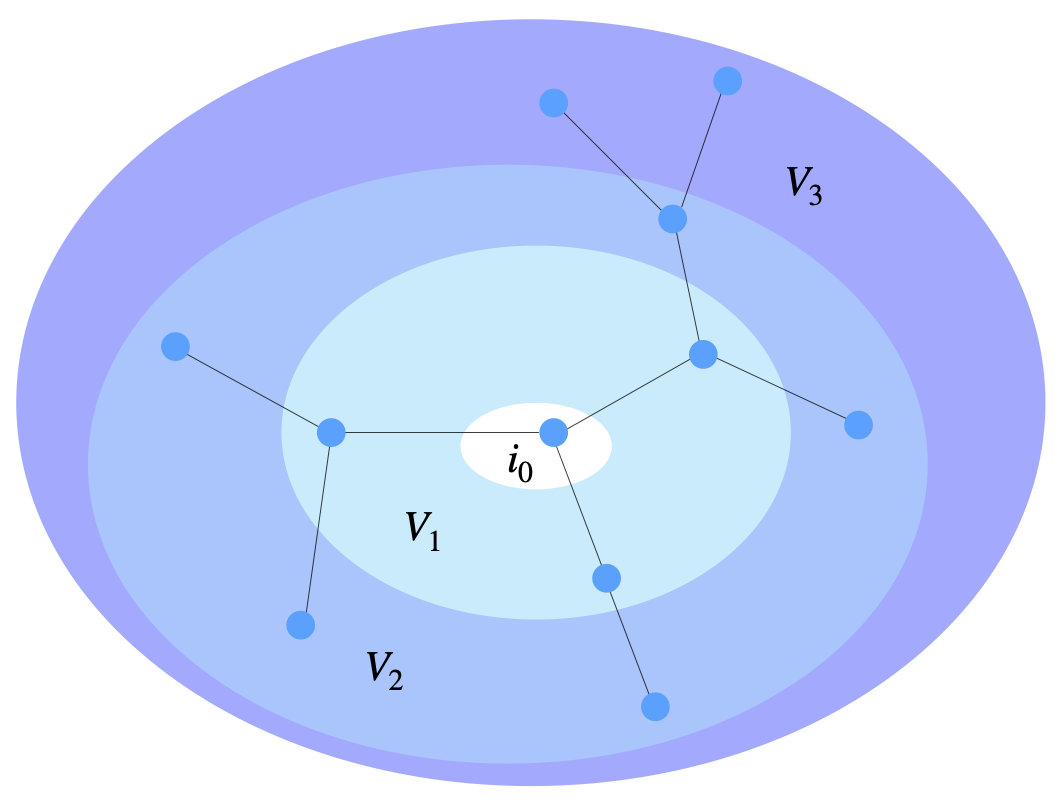} 
\caption{Illustration of the partitioning of nodes in a  tree with root node $i_0$.   The sets $V_1$, $V_2$, and $V_3$ contains nodes that are a distance $1$, $2$, are $3$ separated from the root node $i_0$.}\label{fig:sketchETa}
\end{center}
\end{figure}

The matrix $\boldeta$ is a diagonal matrix with elements on the diagonal defined as follows.   We set the matrix entry associated to the root node to unity, 
\begin{equation}
   \eta_{i_0 i_0}=1,  \label{eq:root1}
\end{equation}
and determine the other nodes through a recursion.   In particular, we set 
\begin{equation}
\eta_{jj} = - \frac{M_{ij}}{M_{ji}}\eta_{ii}, \label{eq:root2}
\end{equation}
for $i\in V_k$ and $j\in V_{k+1}$.   Since  $M_{ij}M_{ji}<0$, the elements $\eta_{ii}>0$, and $\boldeta$ is a symmetric, positive definite matrix.     The procedure of constructing $\boldeta$ is sketched in Fig.~\ref{fig:sketchETa}.  Notice that the  value of each element $\eta_{ii}$ is determined by its  path to the root node $i_0$, and this path is unique, as the graph is a tree. 

Next, we show that
\begin{equation}
    \mathbf{M}^{T} =  -\boldeta \mathbf{M}\boldeta^{-1}, \label{eq:MEta}
\end{equation} 
so that $ \mathbf{M}$ is strictly quasi-antisymmetric.   Components wise, the right-hand side of Eq.~(\ref{eq:MEta}) reads
\begin{equation}
 -[\boldeta \mathbf{M}\boldeta^{-1}]_{ij} = \sum^N_{\ell=1}\sum^N_{k=1}\eta_{i\ell}M_{\ell k}\eta^{-1}_{kj}=\eta_{ii}M_{ij}\eta^{-1}_{jj} \ .
\end{equation} 
If $i=j$, or $i\neq j$ and the nodes are not each other's neighbours, then 
\begin{equation}
 -[\boldeta \mathbf{M}\boldeta^{-1}]_{ij} = -\eta_{ii}M_{ij}\eta^{-1}_{jj}=  0 = M_{ji}. \label{eq:M1}
\end{equation}
On the other hand, when  $i\in V_{k}$ and $j\in V_{k+1}$, then  Eq.~(\ref{eq:root2}) applies, and we obtain 
\begin{equation}
 -[\boldeta \mathbf{M}\boldeta^{-1}]_{ij} = -\eta_{ii}M_{ij}\eta^{-1}_{jj} = M_{ji}.\label{eq:M2}
\end{equation}
Equations (\ref{eq:M1}) together with (\ref{eq:M2}), imply Eq.~(\ref{eq:MEta}), which is what we were meant to show.

\section{Sign stability of  antagonistic trees with nonzero diagonal elements}\label{app:signstabdiag3}
In  \ref{app:signstab}, we have shown that all eigenvalues of the adjacency matrices of weighted, antagonistic tree matrices are purely imaginary.  Now,   we consider matrices of the form 
\begin{equation}
    \mathbf{M}'  = \mathbf{M} - \mathbf{D} \label{eq:a}
\end{equation}
where $\mathbf{M}$ is the adjacency matrix of a weighted, antagonistic tree as in \ref{app:signstab}, and $\mathbf{D}$  is a diagonal matrix with nonnegative diagonal entries, i.e., $[\mathbf{D}]_{ii} = D_{ii}  \geq 0$.    Hence, the distinction with the matrix $\mathbf{M}$ in ~\ref{app:signstab} is that the diagonal entries of $\mathbf{M}'$ can be nonzero, and therefore for clarity  we added the prime.  We show that for this ensemble all eigenvalues have nonpositive  real parts, i.e., 
\begin{equation}
    \Re[\lambda_i(\mathbf{M}')]\leq 0 \label{eq:cond1x} 
\end{equation}
for all $i=1,2,\ldots,N$.

The derivation consists of two parts.   First, in \ref{sec:C1}, we perform a standard linear algebra computation to 
show that Eq.~(\ref{eq:cond1x}) holds for matrices $\mathbf{K}'$ built from subtracting a nonnegative diagonal matrix to an antisymmetric matrix.    Second, in  \ref{sec:C2}, we show that $\mathbf{M}'$ is related to a matrix $\mathbf{K}'$  by a similarity transformation, and hence they share the same eigenvalues.   We end this appendix with a related result: in \ref{sec:C3} we show that the width  spectrum of a diagonal matrix  shrinks when we add to it an antisymmetric matrix.  
  
\subsection{Spectra of antisymmetric matrices with  nonpositive diagonal entries}\label{sec:C1}
Let us consider matrices  of  the form
\begin{equation}
\mathbf{K}'  = \mathbf{K} -  \mathbf{D}, \label{eq:b}
\end{equation}
where  $ \mathbf{K}$ is an antisymmetric matrix (instead of the adjacency matrix of an antagonistic tree in Eq.~(\ref{eq:a})), and  where $\mathbf{D}$ is  a nonnegative, diagonal matrix.   We show that  for matrices of this form, 
\begin{equation}\Re[\lambda_i(\mathbf{K}')] \leq 0  \label{eq:ineq}
\end{equation}
for all $i=1,2,\ldots,N$.

Indeed, in  this case, for any vector $\vec{z}\in\mathbb{C}^N$, it holds that 
\begin{equation}
   \Re( \vec{z}^\dagger\: \mathbf{K}' \vec{z}) =     \Re(\vec{z}^\dagger \:\mathbf{K} \vec{z}) -  \Re(\vec{z}^\dagger\:  \mathbf{D} \vec{z})\leq 0.  \label{eq:final}
\end{equation} 
As $[\mathbf{K}]_{jk} = -[\mathbf{K}]_{kj}\in\mathbb{R}$, it holds that  
\begin{equation}
\sum^N_{k=1}\sum^N_{j=1} z^\ast_j [\mathbf{K}]_{jk} z_k =  - \sum^N_{k=1}\sum^N_{j=1} z_j [\mathbf{K}]_{jk} z^\ast_k , \label{eq:Re1}
\end{equation} 
such that 
\begin{equation}
 \Re(\vec{z}^\dagger \: \mathbf{K} \vec{z}) = 0,\label{eq:Re2}
\end{equation} 
and 
\begin{equation}
\vec{z}^\dagger\:  \mathbf{D} \vec{z} = \sum^N_{j=1}|z_j|^2[\mathbf{D}]_{jj} \geq 0.  \label{eq:Re3}
\end{equation} 
Using Eqs.~(\ref{eq:Re2}) and (\ref{eq:Re3}) in the left-hand side of  (\ref{eq:final}), we obtain the right-hand side of Eq.~(\ref{eq:final}).    Lastly, to obtain the inequalities Eq.~(\ref{eq:ineq}), we set $\vec{z}$ in Eq.~(\ref{eq:final}) equal to a right eigenvector $\vec{r}_i$ of $\mathbf{K}'$, yielding, 
\begin{equation}
    \Re(\vec{r}_i^\dagger \mathbf{K}'\vec{r}_i) = \Re[\lambda_i |\vec{r}_i|^2] = |\vec{r}_i|^2\Re[\lambda_i(\mathbf{K}') ] \leq 0. 
\end{equation}

\subsection{Antagonistic tree matrices with nonpositive  diagonal entries}\label{sec:C2}

We show that antagonistic tree matrices $\mathbf{M}'$ with nonpositive  diagonal entries, as defined in Eq.~(\ref{eq:a}), are related by a similarity transformation to a matrix $\mathbf{K}'$ of the form Eq.~(\ref{eq:b}).     To this aim, we use   $\boldeta$ defined as in Eqs.~(\ref{eq:root1}) and (\ref{eq:root2}); notice that $\boldeta$ is defined with $\mathbf{M}$ and not with $\mathbf{M}$'.    

Indeed, if we define 
\begin{equation}
\mathbf{K}' = \sqrt{\boldeta}\mathbf{M}'\sqrt{\boldeta}^{-1}
\end{equation}
then 
\begin{equation}
\mathbf{K}' = \mathbf{K} - \sqrt{\boldeta}\mathbf{D}\sqrt{\boldeta}^{-1}
\end{equation}
with $\mathbf{K}$ the matrix as defined in Eq.~(\ref{eq:K}), which is antisymmetric as we have shown in Eq.~(\ref{eq:KAsym}).    Since the matrices $\sqrt{\boldeta}$ and  $\mathbf{D}$ are diagonal, 
\begin{equation}
\sqrt{\boldeta}\mathbf{D}\sqrt{\boldeta}^{-1} = \mathbf{D}
\end{equation}
and thus $\mathbf{K}'$ takes the form Eq.~(\ref{eq:b}), as we were meant to show.   

Since $\mathbf{M}'$ is related to $\mathbf{K}'$ by a similarity transformation, they share the same eigenvalues, and since $\mathbf{K}'$ has nonpositive eigenvalues, as we we have shown in Sec.~\ref{sec:C1}, also $\mathbf{M}'$  has  nonpositive eigenvalues.

\subsection{Change in the width of the spectrum after adding an antisymmetric  or  antagonistic tree matrix to a disordered diagonal matrix }\label{sec:C3}
Let $\mathbf{K}'$ be the sum of an antisymmetric matrix $\mathbf{K}$ and a (not necessarily nonpositive) diagonal matrix $-\mathbf{D}$, as defined in Eq.~(\ref{eq:b}).     It then holds that 
\begin{equation}
\Re[\lambda_1(\mathbf{K}')] \leq -D_{\rm min} \label{eq:shrink}
\end{equation}
where $\lambda_1(\mathbf{K}')$ is the leading eigenvalue of the matrix $\mathbf{K}'$,  and where $D_{\rm min}$ is the minimum entry of the $D_i$, as defined in Eq.~(\ref{eq:Dmin}).    Analogously, it holds that 
\begin{equation}
\Re[\lambda_N(\mathbf{K}')] \geq -D_{\rm max}. \label{eq:shrink2}
\end{equation}

Equations (\ref{eq:shrink}) and (\ref{eq:shrink2}) imply that the  width of the spectrum of $-\mathbf{D}$ shrinks when an antisymmetric matrix is added to it, and hence adding antisymmetric interactions to a matrix makes it the matrix more stable.   Note that this result does not extend to the more general case of  sign-antisymmetric interactions, as then the matrix gets more stable in the perturbative regime of small  interactions, but the matrix does not get more stable  for strong  sign-antisymmetric interactions, see \cite{Cure2021}.   

The derivation of the  Eqs.~(\ref{eq:shrink}-\ref{eq:shrink2}) can be seen as an exercise is matrix analysis~\cite{horn2012matrix}.     Nevertheless, for the reader's convenience,  we present here a derivation.  
To derive Eq.~(\ref{eq:shrink}), we consider the matrix 
\begin{equation}
\mathbf{K''} := \mathbf{K'}+ D_{\rm min} \mathbf{1} = \mathbf{K} - \mathbf{D}'
\end{equation}
where $\mathbf{D}' := \mathbf{D} -D_{\rm min} \mathbf{1}  $ is by construction a diagonal matrix with nonnegative diagonal entries.    Therefore, the results of ~\ref{sec:C2} apply to $\mathbf{K''}$, and
\begin{equation}
\Re[\lambda_1(\mathbf{K''})]\leq 0. \label{eq:K''}
\end{equation}
Since,
\begin{equation}
\lambda_1(\mathbf{K''}) =  \lambda_1(\mathbf{K'}) + D_{\rm min} \label{eq:K''2}
\end{equation}
we obtain the Eq.~(\ref{eq:shrink}) from combining Eq.~(\ref{eq:K''}) with (\ref{eq:K''2}).

Using a similar line of reasoning it follows that 
\begin{equation}
\Re[\lambda_1(\mathbf{M}')] \leq -D_{\rm min} ,\label{eq:shrink2x}
\end{equation}
where $\mathbf{M}'$ is now of the form Eq.~(\ref{eq:a}) with $\mathbf{M}$ the adjacency matrix of a weighted, antagonistic tree and $\mathbf{D}$ a diagonal matrix with diagonal entries that can be negative and positive.

\section{All the eigenvalues of the adjacency matrices of weighted, oriented graphs without directed cycles are equal to  zero}\label{app:signstab2}
Let $\mathbf{M}\in \mathbb{R}^{N\times N}$ represent the adjacency matrix of a weighted, directed graph.  We assume that $M_{ii}=0$, so that there are no self-links.   If $M_{ij} = 0$, then  the directed edge from $i$ to $j$ is absent, while if $M_{ij} \in \mathbb{R}\setminus {0}$, then there exists a directed edge from $i$ to $j$ weighted by the value of $M_{ij}$.  

We say that a directed graph is oriented if all of its edges are unidirectional,  i.e., 
\begin{equation}
M_{ij}M_{ji} = 0
\end{equation}
for all pairs of indices $i,j$.  Additionally,  we say that a graph has no directed cycles when Eq.~(\ref{eq:cond2}) holds.   Note that cycles that are nondirected, as for example the feedforward cycles in Panel (c) of Fig.~\ref{fig:sketch_sign_stable}, are allowed.

The characteristic polynomial of an adjacency matrix of an oriented graph without directed cycles  is given by 
\begin{equation}
{\rm det}\left(\lambda \mathbf{1} - \mathbf{M}\right) = \lambda^N, \label{eq:charDir}
\end{equation} 
and consequently all the eigenvalues of $\mathbf{M}$ are equal to zero, i.e., 
\begin{equation}
\lambda_i(\mathbf{A}) = 0   
\end{equation}
for all $i\in \left\{1,2,\ldots,N\right\}$. 

We show that  Eq.~(\ref{eq:charDir}) is true by identifying it as a specific case of the so-called Coefficients Theorem for directed graphs, which we revisit here.  
First, we  present the Coefficients Theorem for unweighted graphs, i.e., for $M_{ij}\in\left\{0,1\right\}$, which is Theorem 1.2 in Ref.~\cite{sachs1980spectra}, and then we present the Coefficients Theorem for weighted graphs, i.e., for $M_{ij}\in \mathbb{R}$.    Before stating the Coefficients Theorem, we need the following definition:  A {\it linear directed graph} is a directed graph for which it holds that all vertices have an indegree and outdegree equal to one.  Hence, linear directed graphs are composed out of one or more directed cycles.

\begin{theorem}[Coefficients Theorem for  directed graphs (Milic \cite{milic1964flow}, Sachs \cite{sachs1964beziehungen}, and Spialter \cite{spialter1964atom})]\label{Theorem12}
Let 
\begin{equation}
{\rm det}\left(\lambda \mathbf{1} - \mathbf{M}\right) = \lambda^N + a_1 \lambda^{N-1} + \ldots + a_N
\end{equation} 
be the characteristic polynomial of an arbitrary directed graph $G$.   Then  
\begin{equation}
    a_n = \sum_{L\in \mathcal{L}_n}(-1)^{p(L)}, \quad {\rm with} \quad n=1,2,\ldots,N,
\end{equation}
where $\mathcal{L}_n$ is the set of all linear directed subgraphs $L$ of $G$ with exactly $n$ vertices; $p(L)$ denotes the number of connected components of $L$ (i.e., the  number of directed cycles of which $L$ is composed). 
\end{theorem}

\begin{theorem}[Coefficients theorem for  weighted directed graphs (Devadas Acharya \cite{acharya1980spectral})]\label{Theorem2}
Let 
\begin{equation}
{\rm det}\left(\lambda \mathbf{1} - \mathbf{M}\right) = \lambda^N + a_1 \lambda^{N-1} + \ldots + a_N
\end{equation} 
be the characteristic polynomial of an arbitrary, weighted, directed graph with adjacency matrix $\mathbf{M}$, then  
\begin{equation}
    a_n = \sum_{L\in \mathcal{L}_n}(-1)^{p(L)}\prod_{(i,j)\in E(L)}M_{ij}, \quad {\rm with} \quad n=1,2,\ldots,N,
\end{equation}
where $E(L)$ is the set of edges in the  linear directed subgraph $L$.  
\end{theorem}

In the case that $G$ has no directed cycles, all coefficients $a_i$  in  Theorem~\ref{Theorem12}  or Theorem~\ref{Theorem2} are equal to zero, and Eq.~(\ref{eq:charDir}) follows as a corollary of the Coefficients Theorem for directed graphs.  

\section{Husimi Plateau in Jacobian-like matrices}
\label{app:HusimiPlateau}
We refine the results in Fig.~\ref{fig:leading_eig_jacobian} by analysing  $\langle \Re[\lambda_1(\mathbf{J})]\rangle$  as a function of $N$ in the limit of $d_{\rm min}\approx 0$, where $d_{\rm min}$ is the smallest value of $d$ that belongs to the support set of $p_D(d)$.     In this limit, the leading eigenvalue of antagonistic, Husimi trees exhibits strong transient effects as a function of $N$, and hence it is important to carefully extrapolate the results to large~$N$.    

\begin{figure}[H]
	\centering
	\includegraphics[width=1.0\textwidth]{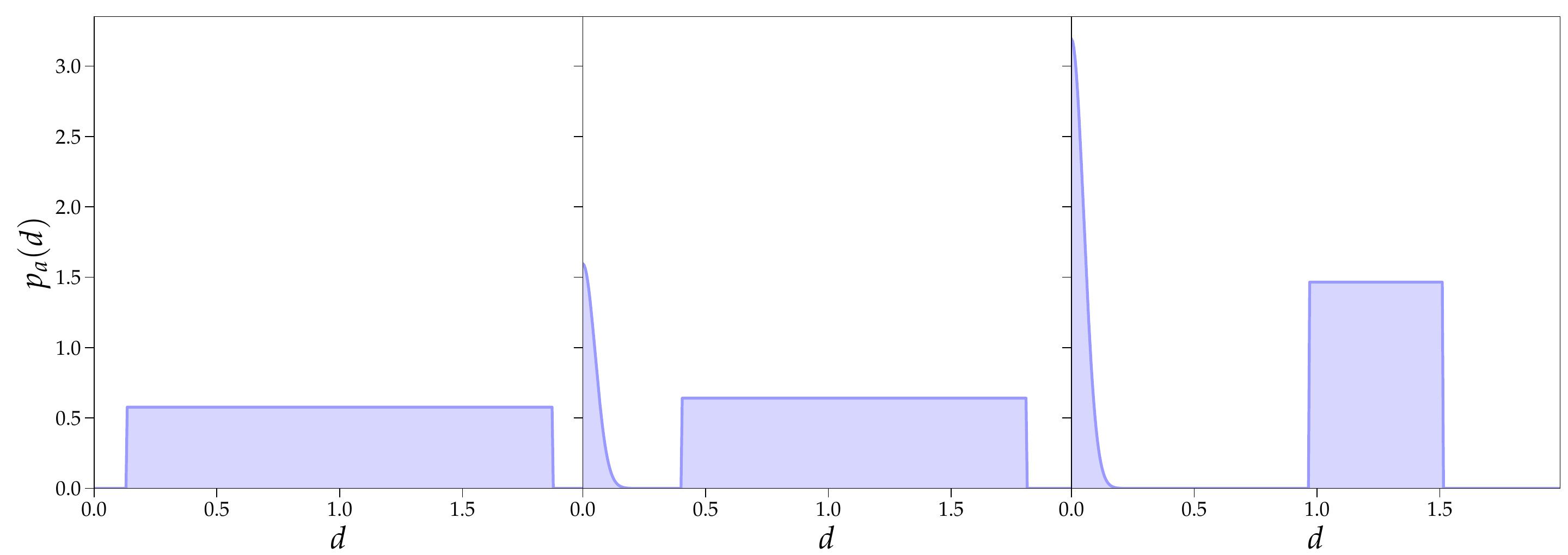}
	\caption[]{The distribution $p_{a}(d)$, as defined in ~\ref{app:HusimiPlateau}, for three values of the parameter: $a=0$ (left), $a=0.1$ (center) and $a=0.2$ (right).}
	\label{fig:plateau_diag_distr}
\end{figure}

To study the influence of a value $d_{\rm min}\approx 0$ on the leading eigenvalue, we  extract the diagonal entries $D_i$ from  a distribution $p_D(d) = p_a(d)$   that is plotted in Fig.~\ref{fig:plateau_diag_distr}. 
As illustrated by Fig.~\ref{fig:plateau_diag_distr}, the distribution  $p_a(d)$  develops a peak around zero, i.e., $d\approx 0$, whose weight increases a function $a$, which is the main reason why we use $p_a(d)$ and not the uniform distribution considered before in Fig.~\ref{fig:leading_eig_jacobian}.     The distribution  $p_a(d)$ is defined by 
\begin{equation} \label{eq:plateau_diag_distr}
   p_a(d):= a \; p_{{\rm HN}}(d) + (1-a) \, p_{\rm U}(d) 
\end{equation}
where $p_{{\rm HN}}(d)$ is a (narrow) half-normal distribution centered at zero, obtained by setting $\mu=0$ and $\sigma=0.05$ on the right-hand side of  Eq.~(\ref{eq:truncated_gaussian}),  and $p_{\rm U}(d)$ is a uniform distribution with a  support that is not   touching zero, and we denote its mean and standard deviation by $\mu_{\rm U}$  and  $\sigma_{\rm U}$, respectively.
We set the parameters $\mu_{\rm U}$ and $\sigma_{\rm U}$  such that, as $a$ varies, the first two cumulants of $p_a(d)$ are identical to those of $p_D(d)$ used in Secs.~\ref{subsec:lss_interaction} and \ref{subsec:lss_jacobian}.

\begin{figure}[h]
	\centering
	\includegraphics[width=1.0\textwidth]{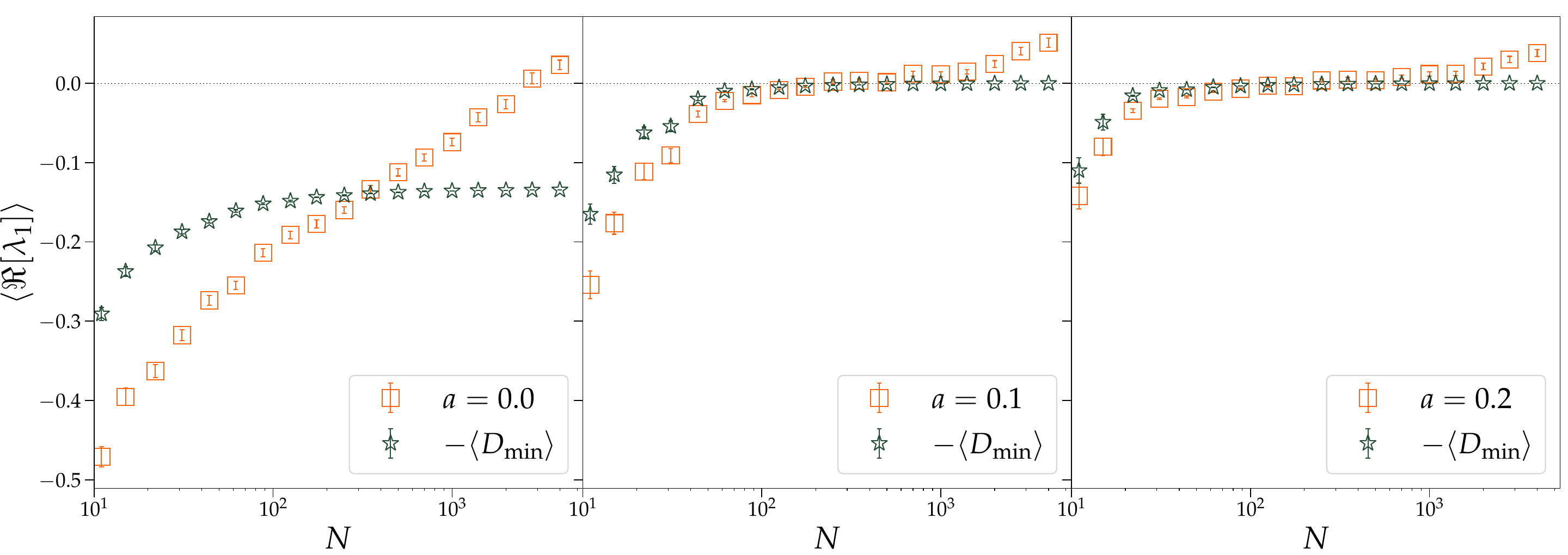}
	\caption[]{Average real part of the leading eigenvalue, $\langle \Re[\lambda_1]\rangle$, as a function of the matrix size $N$ for Jacobian-like matrices $\mathbf{J}$ of antagonistic, Husimi trees.    Numerical  results   obtained from  diagonalising $300$ matrix realisations of antagonistic  Husimi trees (orange squares) are  compared with  the average of the largest element of the diagonal matrix $-D_{\rm min}$ (as defined in Eq.~(\ref{eq:Dmin})), and error bars denote the error on the mean. The  parameters used for the matrix ensembles are detailed in Sec.~\ref{sec:numerical_details}, except for the distribution of diagonal elements which is given by $p_{a}(d)$, defined in Eq.~(\ref{eq:plateau_diag_distr}).}
\label{fig:husimi_plateau}
\end{figure}

Figure \ref{fig:husimi_plateau} plots $\langle\Re[\lambda_1(\mathbf{J})]\rangle$ as a function of $N$ for antagonistic Husimi trees  with diagonal elements drawn from  $p_{a}(d)$ and  for three values of $a$.  
In the left panel, the support of the diagonal distribution $p_{a}(d)$ does not include  $d=0$,   and hence the functional behaviour of $\langle\Re[\lambda_1(\mathbf{J})]\rangle$ is analogous to the one shown in Fig.\ref{fig:leading_eig_jacobian}.  On the other hand, in the middle and right panels  $p_{a}(d)$ develops a peak around $d=0$, and consequently  the monotonic increase of $\langle\Re[\lambda_1(\mathbf{J})]\rangle$  as a function of  $N$  slows down significantly at intermediate values of $N$, leading to the appearance of a    plateau.  Note that the plateau is a transient effect, as for large enough values of $N$ the steady increase of  $\langle\Re[\lambda_1(\mathbf{J})]\rangle$ continues.    Observe that  increasing $a$    only widens the plateau  from the left side,  making it   appear   at smaller sizes  of f$N$.    Hence, also when $p_{a}(d)$ is peaked around $d=0$, $\langle\Re[\lambda_1(\mathbf{J})]\rangle$  diverges as a function of $N$, albeit with a strong, transient, plateau effect.   

We end with some final remarks.  Due to the Husimi plateau, we should  carefully assess finite size effects in Husimi trees.   In particular, we could wrongly conclude that $\langle \Re[\lambda_1]\rangle$ converges to a finite value when not considering large enough values of $N$.    The Husimi plateau only occurs in Jacobian-like matrices when the support of the diagonal distribution contains $d=0$, and hence we conjecture that it  is  related to the stripy structure  of Jacobian-like matrices.

\section{Finite size effect for the histograms of the imaginary part distribution of nonreal leading eigenvalues} \label{app:finite_size_scaling}
\begin{figure}[hb]
	\centering
	\includegraphics[width=\textwidth]{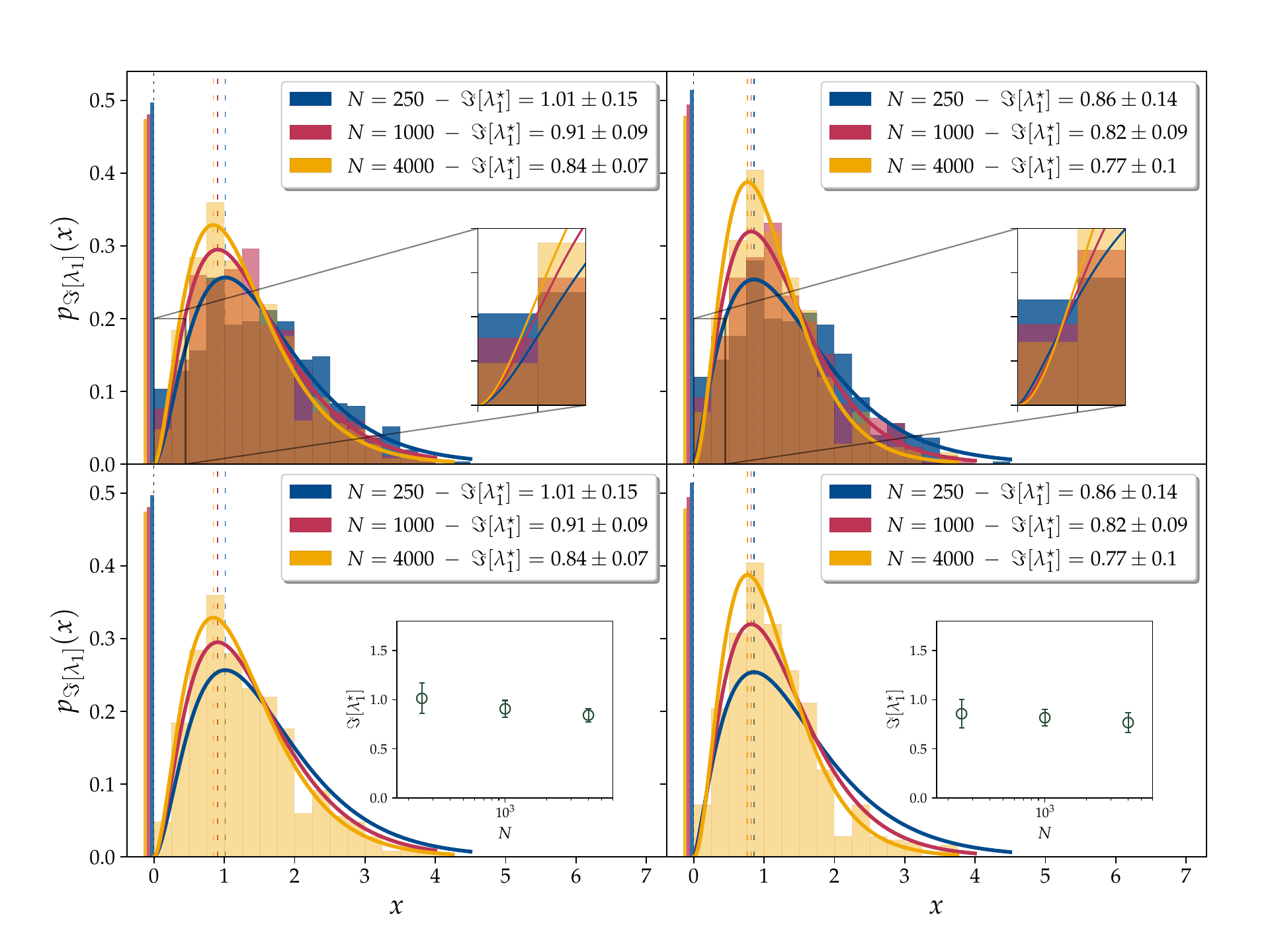}
	\caption[]{Histograms of the imaginary part of the leading eigenvalues of interaction-like (left panels) and Jacobian-like (right panels) matrices  of antagonistic  Erd\"{o}s-R\'{e}nyi graphs.  The data  shown corresponds with the parameter value $\sigma_D=0.10$  at which the dynamical transition takes place in Fig.~\ref{fig:reentrances_hist}, albeit now for different matrix sizes $N$ as indicated in the legend. Additionally, the insets show the histograms for the range $[0,0.45]$ on the $x$-axis. The other parameters are the same as in  Fig.~\ref{fig:reentrances_hist}. The figures at the bottom show, for reasons of clarity, the same data as the figures at the top, except for the absence of the histograms for $N=250$ and $N=1000$. The insets show the typical value of the leading eigenvalue imaginary part $\Im[\lambda^\star_1]$ as a function of $N$.
 } 
	\label{fig:reentrances_hist_size_scaling}
\end{figure}
In this Appendix, we determine  the effect of  a finite size $N$ on   the distribution  $p_{\Im[\lambda_1]}(x)$ plotted in Fig.~\ref{fig:reentrances_hist}.   As we show, increasing the system size $N$, the discontinuity of the transition becomes more pronounced. Figure~\ref{fig:reentrances_hist_size_scaling} plots the distribution $p_{\Im[\lambda_1]}(x)$  for different sizes  $N$  when the control parameter $s$ is roughly at the transition $s_{\rm d}$ (in particular we set $s \simeq 0.08$, in correspondence with $\sigma_D=0.10$, $\sigma_{\rm G}=0.6$ and $\mu_{\rm G}=1.0$).   

As shown in Fig.~\ref{fig:reentrances_hist_size_scaling}, the  typical value  $\Im[\lambda_1^{\star}]$ of the distribution $p_{\Im[\lambda_1]}(x)$ presents in one instance only a very mild trend (although it is unclear whether it is statistically significant) due to finite size correction, still consistent with saturation to a finite value and therefore compatible with the discontinuous nature of the transition found in Fig.~\ref{fig:reentrances_hist} and strongly supported by the reentrant behaviour of the support of the spectrum shown in Fig.~\ref{fig:reentrances_spectra_interaction_with_diagonal} and Fig.~\ref{fig:reentrances_spectra_jacobian}.  Interestingly,  as the size $N$ increases, the  $\gamma$-distribution  peaked at $\Im[\lambda_1^{\star}]$ gets more narrow, indicating that in the infinite size limit  $\tilde{p}_{\Im[\lambda_1]}(x)$, as defined in Eq.~(\ref{eq:pInm}), possibly converges to the delta distribution $\delta(x-\Im[\lambda_1^{\star}])$.   Moreover, focusing on the behaviour of the nonreal histogram near the zero (see the insets in the top row of Fig.~\ref{fig:reentrances_hist_size_scaling}), we see find  that the bin closest to the real delta peak decreases as a function of $N$, while the second bin increases as a function of $N$.    

\newpage

\section*{References}
\bibliographystyle{unsrt} % {plain}
\bibliography{Biblio_aps}

\end{document}